\documentclass[preprint]{aastex}

\shorttitle{New Orbital Analysis of Stars at the Galactic Center}
\shortauthors{Boehle et al.}
\usepackage{apjfonts}
\usepackage{lscape}
\usepackage[utf8]{inputenc}
\usepackage{amsmath}
\usepackage{mathrsfs}

\begin{document}

\title{An Improved Distance and Mass Estimate for Sgr A* from a Multistar Orbit Analysis}

\author{A. Boehle\altaffilmark{1}, A. M. Ghez\altaffilmark{1}, R. Sch{\"o}del\altaffilmark{2}, L. Meyer\altaffilmark{1}, S. Yelda\altaffilmark{1}, S. Albers\altaffilmark{1}, G. D. Martinez\altaffilmark{1}, E. E. Becklin\altaffilmark{1}, T. Do\altaffilmark{1}, J. R. Lu\altaffilmark{3}, K. Matthews\altaffilmark{4}, M. R. Morris\altaffilmark{1}, B. Sitarski\altaffilmark{1}, G. Witzel\altaffilmark{1}}


\altaffiltext{1}{UCLA, Department of Physics and Astronomy, Los Angeles, CA 90095, USA}
\altaffiltext{2}{Instituto de Astrof\'{i}sica de Andaluc\'{i}a (CSIC), Glorieta de la Astronom\'{i}a S/N, 18008 Granada, Spain}
\altaffiltext{3}{Institute for Astronomy, University of Hawaii, Honolulu, HI 96822, USA}
\altaffiltext{4}{Division of Physics, Mathematics, and Astronomy, California Institute of Technology, MC 301-17, Pasadena, CA 91125, USA}

\email{aboehle@astro.ucla.edu}

\begin{abstract}

We present new, more precise measurements of the mass and distance of our Galaxy's central supermassive black hole, Sgr A*. These results stem from a new analysis that more than doubles the time baseline for astrometry of faint stars orbiting Sgr A*, combining two decades of speckle imaging and adaptive optics data.  Specifically, we improve our analysis of the speckle images by using information about a star's orbit from the deep adaptive optics data (2005 - 2013) to inform the search for the star in the speckle years (1995 - 2005).  When this new analysis technique is combined with the first complete re-reduction of Keck Galactic Center speckle images using speckle holography, we are able to track the short-period star S0-38 (K-band magnitude = 17, orbital period = 19 years) through the speckle years. We use the kinematic measurements from speckle holography and adaptive optics to estimate the orbits of S0-38 and S0-2 and thereby improve our constraints of the mass ($M_{bh}$) and distance ($R_o$) of Sgr A*: $M_{bh} = 4.02\pm0.16\pm0.04\times10^6~M_{\odot}$ and $7.86\pm0.14\pm0.04$ kpc.  The uncertainties in $M_{bh}$ and $R_o$ as determined by the combined orbital fit of S0-2 and S0-38 are improved by a factor of 2 and 2.5, respectively, compared to an orbital fit of S0-2 alone and a factor of $\sim$2.5 compared to previous results from stellar orbits. This analysis also limits the extended dark mass within 0.01 pc to less than $0.13\times10^{6}~M_{\odot}$ at 99.7\% confidence, a factor of 3 lower compared to prior work. 

\end{abstract}

\section{Introduction} \label{intro}

Following the motions of stars in the center of our Galaxy has given many insights into the properties of the gravitational potential in which they move.  
The measurement of the high proper motions, and later, accelerations of these stars implies that they move in the gravitational potential of a concentrated dark mass (\citealt{1997MNRAS.284..576E}; \citealt{1998Ghez}; \citealt{2000Ghez}; \citealt{2002MNRAS.331..917E}).  
With further observations, these stellar motions have provided strong evidence for the presence of a supermassive black hole (SMBH) at the Galactic Center (Sgr A*) with a mass of about $4 \times 10^{6}$ $M_{\odot}$.  
Once the star S0-2 went through closest approach in 2002, it was possible to fit its motion with a Keplerian orbit (\citealt{2002Natur.419..694S}; \citealt{2003ApJ...586L.127G}).  
In addition to the mass of the SMBH, stellar orbits with measured radial velocities have been used to determine the distance to the Galactic Center ($R_o$; \citealt{2003ApJ...586L.127G}; \citealt{2003ApJ...597L.121E}).  
With S0-2's short orbital period, this star provides the best constraint on the mass of the central black hole and $R_o$ from stellar orbits to date (e.g., \citealt{2008Ghez}, \citealt{2009Gillessen}).  
The focus of recent work has been to continue assessing the central black hole's properties as well as exploring the potential for using the measurement of stellar motions to test general relativity.

The mass of Sgr A* ($M_{bh}$) and the distance to the Galactic Center ($R_{o}$)  are both important ways of characterizing this unique region of our Galaxy and putting it in context with others galaxies.  
Measuring the mass of the central supermassive black hole allows it to be compared to supermassive black holes in the centers of other galaxies.  With the mass of Sgr A*, the Milky Way can be added to observed correlations between the mass of the central SMBH and other galactic properties, such as velocity dispersion of stars and bulge luminosity (\citealt{2000ApJ...539L...9F}; \citealt{2002ApJ...574..740T}; see \citealt{2013ARA&A..51..511K} for a review).  
$R_o$ is a key parameter characterizing our galaxy's size, mass and kinematics.  
The adopted value of $R_o$ affects estimates of the Milky Way's rotation curve and thereby also measurements of the mass and shape of the dark matter distribution (e.g., \citealt{2000MNRAS.311..361O}).
An independent, very accurate measurement of $R_o$ could possibly be used to calibrate stellar distance indicators, such as RR Lyrae and Cepheids, which are important steps on the cosmic distance ladder (see \citealt{1993ARA&A..31..345R}).  
$R_o$ additionally serves to calibrate the extinction towards the Galactic Center (e.g., \citealt{2010A&A...511A..18S}).  
The mass-to-distance ratio of Sgr A* as derived from stellar orbits is also necessary to determine the predicted size of the black hole shadow that will be observed by the upcoming Event Horizon Telescope, which can be used as a null hypothesis test of general relativity (see \citealt{2015Psaltis}).  
Finally, future tests of general relativistic effects on the motion of S0-2 depend on accurate measurements of the gravitational potential due to the SMBH in the Newtonian regime.

Until now, our group has used only S0-2 to constrain $R_o$ and the mass of Sgr A*.  
This is because S0-2 is unique: it is bright (14.2 in the K band) and has a short orbital period (16.2 years).  
We therefore have been able to track its motion since Keck observations of the Galactic Center began in 1995 such that our observations now cover more than one full orbit of this star.  We would like to also use other short-period stars in the Galactic Center to help determine the gravitational potential, but ideally only those stars with high orbital phase coverage like S0-2.  
It has been shown for visual binary stars that if less than 40 - 50\% of a body's orbit is covered by astrometric observations, the orbital parameters estimated from the data are systematically biased from their true values (\citealt{2014Lucy}).  
We therefore only use additional short-period stars to constrain the black hole mass and $R_o$ that conservatively have at least this minimum orbital phase coverage.  

Achieving this minimum orbital coverage for other short-period stars is a challenge because those stars are fainter than S0-2 by more than an order of magnitude, making them difficult to track in the first 10 years of speckle imaging data.  
In this work, we present a method to increase the orbital phase coverage of these faint stars at the Galactic Center through a complete reanalysis of this data set.  
In all past analyses of stellar orbital parameters using the speckle data, the data from each observation run has been treated independently.  Stars are blindly searched for in the summed image from a given epoch of observation, as if the Galactic Center had never been observed before.  No information from other observations is used in this search.  
In this work, we present a new methodology for analyzing the speckle images that does use information from the much deeper adaptive optics images and the vast improvement that has been made in the knowledge of the central black hole's properties.  We use constraints on a star's orbit from the deep adaptive optics data to inform the search for the star in the earlier speckle years.  

As a pilot study for this new methodology, we apply this technique to S0-38, one of the three stars at the Galactic Center with an orbital period of less than 20 years (in addition to S0-2 and S0-102; see \citealt{Meyer12}), with the ultimate goal of using this star as an additional constraint on the black hole mass and $R_o$.  At a magnitude of K=17.0, S0-38 is consistently detected in our deep adaptive optics data taken from 2005 - 2013 but it has not previously been detected in our speckle imaging data taken from 1995 - 2005.  S0-38 is an ideal star for the application of this methodology because it is consistently detected in all 21 adaptive optics images and its radial velocity has been measured (\citealt{2009Gillessen} and this work), so its orbit is well known even with just over 40\% of its orbit covered by AO observations. 
Our results are also made possible by 
a new reduction of the speckle data using the more sophisticated reconstruction algorithm called speckle holography (\citealt{2013SchoedelHolo}).  This is the first work that includes the speckle holography re-reduction of all Keck Galactic Center speckle data.

The paper is organized as follows:  
Section \ref{datasets} describes the data sets used in this work, including the results of the new speckle holography reduction on the full set of Keck Galactic Center speckle imaging data.  Section \ref{forcemethod} describes the new methodology of analyzing the speckle images as applied to S0-38.  Section \ref{results} contains the results of the S0-38 orbital analysis, including improved constraints on mass and $R_{o}$ as well as extended mass.

\section{Data Sets}
\label{datasets} 

This paper is based on three different types of data sets, which are briefly summarized here.  The first type is previously reported speckle imaging data that, in this work, are re-reduced with the speckle holography technique (\citealt{2013SchoedelHolo}).  
The result is a higher-quality, deeper final image for each observation epoch (Section \ref{speckledata} and Table \ref{tab:speckle_obs}). 
The second is new and previously reported adaptive optics (AO) astrometry data (Section \ref{AOimagedata}).  
New AO imaging observations (Table \ref{tab:AOobs}) are analyzed by methods used in prior work by our group to produce star lists of relative astrometric measurements.  
Initial star lists of relative astrometric measurements from re-analyzed speckle data and star lists from new and existing AO data are transformed to an absolute reference frame using updated measurements for the astrometric standards (see Appendix \ref{app:ref_frame}).  
The third data type is new and previously reported AO spectroscopic data.  These data are analyzed by standard techniques to extract radial velocity measurements (Section \ref{AOspecdata}), which are combined with radial velocity measurements from the literature.  No astrometric measurements from the literature are used due to the difficulty in consistently transforming these measurements to our reference frame.  These data and initial data reduction steps allow us to extend the time baseline for S0-38 by a factor of two through analysis steps that are described in Section \ref{forcemethod}.

\subsection{Speckle Imaging}
\label{speckledata}

\subsubsection{Existing Observations}
The speckle data sets used for this study provide astrometric measurements of the central $\sim$ 5"x5" of the Milky Way between 1995 and 2005.  While the details of these observations can be found in earlier papers from our group (\citealt{1998Ghez}; \citealt{2000Ghez}; \citealt{2005aGhez}; \citealt{2005Lu}; \citealt{2007Rafelski}), we provide a brief summary. Individual K-band (2.2 $\mu$m) frames were obtained with NIRC (\citealt{1994ExA.....3...77M}; \citealt{1996PASP..108..615M}) with a pixel scale of 20 milliarcseconds on the W. M. Keck I telescope.
During each epoch of observation, between 2,000 and 20,000 frames were obtained using very short exposure times (0.1 sec) to freeze the distorting effects of the Earth's atmosphere. 
We begin our analysis of these data with individual frames that have had the instrumental effects removed (i.e., sky subtracted, flat fielded, bad pixel corrected, distortion corrected).  Table \ref{tab:speckle_obs} provides the dates and number of frames for each of the 27 epochs of observations.

\subsubsection{New Image Reconstruction and Initial Star Lists}
\label{initstarlists}
\begin{figure}[ht]  
   \centering
   \includegraphics[width=1.0\columnwidth]{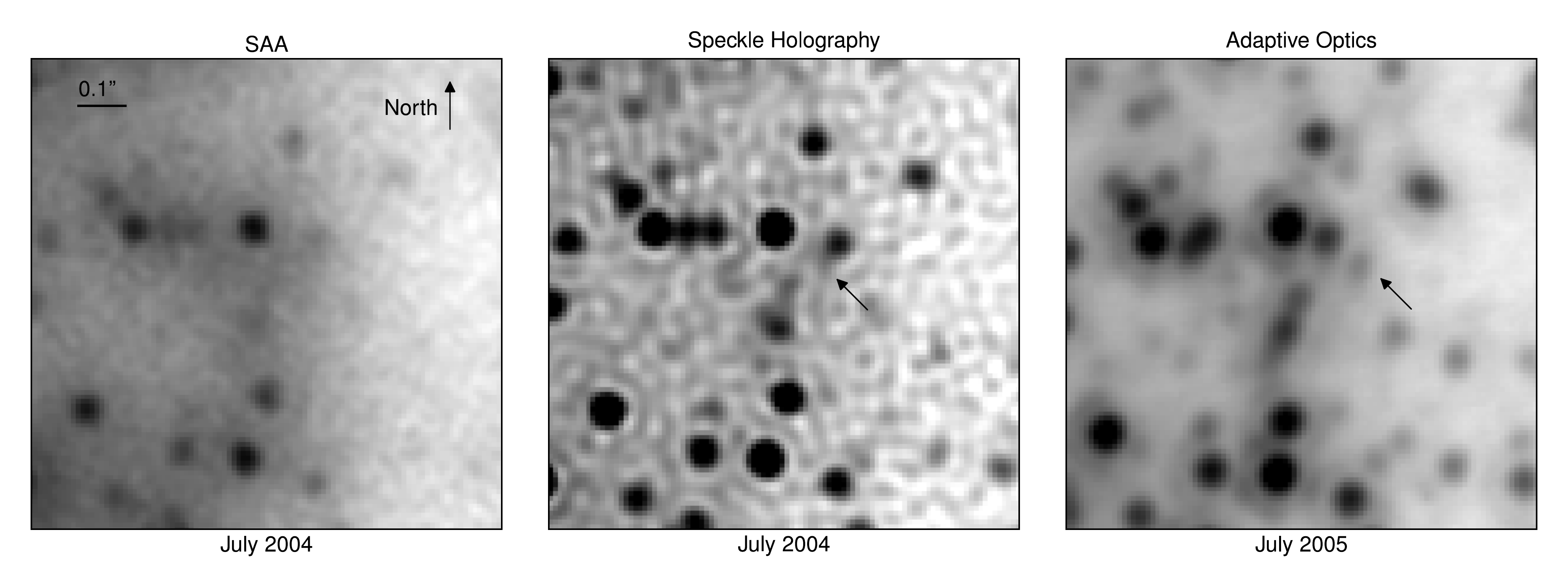}
      \caption[Comparison of Adaptive Optics and Speckle Holography Images]{Comparison of a speckle shift-and-add final image (\emph{left}), a speckle holography final image (\emph{middle}), and an adaptive optics final image (\emph{right}).  The speckle holography and SAA images are made from the same observed frames from July 2004 and the AO image is from July 2005.  Note that between these two observations dates the relative positions of the stars are significantly different.  Each image is 1 arcsecond on a side.  S0-38 is directly detected in both the AO and speckle holography images, but not in the shift-and-add image.  Its detected position is shown with the black arrow.  With a K-band magnitude of 17.0, S0-38 is easily detected in the deep AO images but it is at the direct detection limit for the speckle holography images.}
         \label{fig:images}  
\end{figure}

The individual short-exposure frames are combined via post-processing techniques that compensate for the blurring effects of the Earth's atmosphere.  
The result is a diffraction-limited image whose final Strehl ratio and depth depend on the algorithm used to make the combined image.  
Originally, our speckle images were reduced using the shift-and-add algorithm (SAA; \citealt{1998Ghez}; \citealt{2000Ghez}; \citealt{2002ApJ...577L...9H}, \citealt{2005aGhez}; \citealt{2005Lu}; \citealt{2007Rafelski}).  In this technique, the individual frames are shifted so that the brightest pixel of each frame is at the same position before the highest quality frames 
 are averaged together (see \citealt{1991PASP..103.1040C}).  
With SAA, only the brightest speckle from each frame contributes to the diffraction-limited core of the final point spread function (PSF) and all other speckles are part of a substantial halo.  
In the present work, we employ a more sophisticated speckle reconstruction technique called speckle holography (\citealt{1973A&A....22..319B}; \citealt{1990JOSAA...7.1598P}; \citealt{1998ApJ...500..825P}), as implemented by \cite{2013SchoedelHolo}. 
This technique involves the deconvolution of the observed, distorted images with the instantaneous PSF as measured from a set of reference sources.  
In this approach, the information from all the speckles contributes to the final diffraction-limited core.  In the implementation of speckle holography presented here, the diffraction-limited core of the final PSF contains $\sim6.5$ times more flux than the case of the SAA PSFs (Strehl ratio in speckle holography $\sim0.4$ versus Strehl ratio in SAA $\sim0.06$)\footnote{We note that these Strehl ratios are computed from the post-processed speckle holography and shift-and-add images.  These Strehl ratios differ from those of the raw frames, with the SAA ratios being a closer representation.} and a much greater fraction of the data obtained is used in the analysis (average fraction of frames used in speckle holography $\sim92\%$ versus average fraction of frames used in SAA $\sim17\%$).  See Figure \ref{fig:images} for a comparison between SAA, speckle holography, and AO images.  
For each epoch, both a final image and three subset images of similar quality are constructed.  
An early version of our analysis of these data was reported in our recent work on S0-102 (\citealt{Meyer12}).  
A subset of the epochs presented in that early analysis used a preliminary version of the algorithm described in \cite{2013SchoedelHolo}, with a smaller field of view and a different treatment of the PSF reference sources.  
In the present work, the full implementation of the speckle holography algorithm from \cite{2013SchoedelHolo} was used.  Image quality was further improved by rebinning the speckle frames from the original 20 milliarcsecond pixel scale down to 10 milliarcseconds before applying speckle holography as was done in the original SAA analysis.  
This is the first complete application of speckle holography to the Galactic Center speckle imaging data set from Keck Observatory.

After the final speckle holography images are constructed for each observation epoch, an initial search for stars in these images is performed.  In this preliminary analysis, each observation epoch is treated independently.  The positions and fluxes of stars in the field are determined by the PSF fitting program \emph{StarFinder} (\citealt{2000SPIE.4007..879D}).  
In the running of \emph{StarFinder} to detect sources in both the main and subset images, we choose to set the correlation threshold to 0.8 instead of the value of 0.7 used in \cite{Meyer12}.  
This more conservative correlation threshold value is chosen to minimize the number of fake sources detected in this preliminary analysis of the images.  
This correlation threshold is also motivated by the fact that we will later go back and look for sources that this initial analysis missed (see Section \ref{forcemethod}).  
To further ensure that each detection is robust and to estimate the
measurement uncertainties, \emph{StarFinder} is run in the same way on the three subset images.  
We keep a detection in our initial star list if it was detected in all three of these subset images as well the main image.  The astrometry and photometry measurements are taken from the main image.  The error on our astrometric and photometric measurements is taken as the error of the mean (standard deviation divided by $\sqrt{3}$) of the positions and fluxes determined by \emph{StarFinder} in the three subset images.  

\begin{figure}[ht]  
   \centering
   \includegraphics[width=10.0cm]{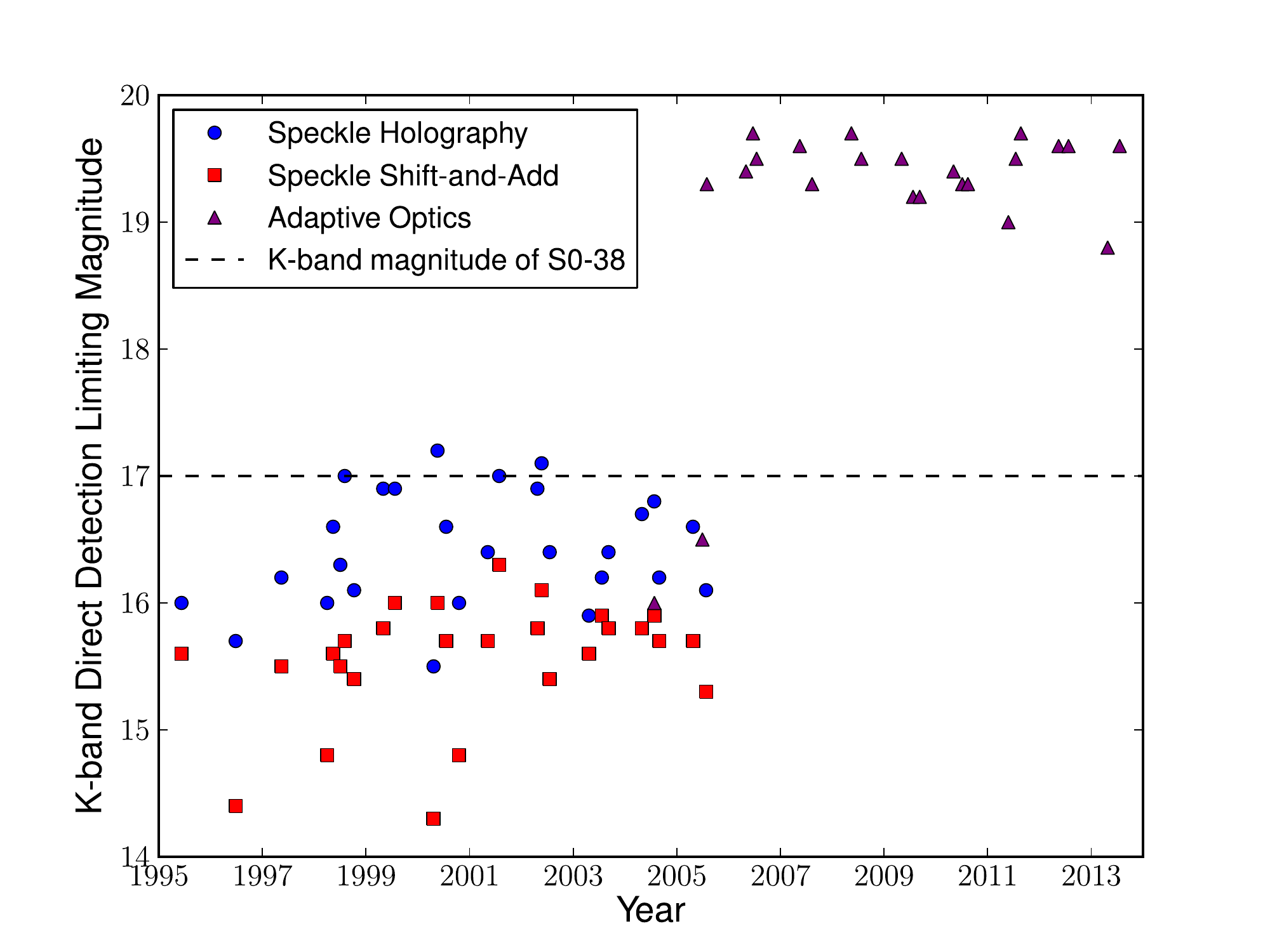}
      \caption[K-band direct detection limiting magnitude]{The direct detection K-band limiting magnitude of the imaging data used in this work (from Table \ref{tab:speckle_obs}).  The limiting magnitude is defined as the magnitude at which the cumulative distribution function of the K magnitudes of the directly detected sources reaches 95\% of the total sample size.  The red squares show this limiting magnitude as derived for the original shift and add reduction of the speckle data and the blue circles show the improved depth that comes from the new speckle holography reduction of the same data.  The purple triangles show the direct detection limiting magnitudes of the AO images for comparison.  The dashed line shows the average magnitude of S0-38 derived from the AO data.  This shows that S0-38 is consistently detected in the deep AO images but is near the limit of direct detection in the speckle holography images.  
Note that the limiting magnitudes of the first two AO epochs are similar to those of the speckle holography epochs.  This is due to the small number of frames going into these AO images (10 frames used versus the typical 100-200 frames).  This fact also explains why S0-38 is detected in 21 out of 23 AO epochs.}
         \label{fig:Klims}  
\end{figure}

We use the initial star lists to determine the K-band limiting magnitude for "direct detections" in each speckle holography image.  The K limiting magnitude is defined as the magnitude at which the cumulative distribution function of the observed K magnitudes of stars in the initial list reaches 95\% of the total sample size.  
Figure \ref{fig:Klims} shows the K-band limiting magnitude of the SAA and speckle holography images for each observation epoch, as well as the limiting magnitude of the AO images for comparison.  In our data, the median SAA K limiting magnitude is 15.7 and the median speckle holography K limiting magnitude is 16.4.  For comparison, the median K limiting magnitude of our AO data is 19.4.  Table \ref{tab:speckle_obs} summarizes and compares the SAA and speckle holography analyses of the 27 epochs of speckle data used in this work. 

\subsection{Adaptive Optics}
\label{AOdata}

\subsubsection{Imaging Data and Astrometric Reference Frame}
\label{AOimagedata}
New high-resolution images of the central $\sim10^{\prime\prime}\times10^{\prime\prime}$ of our galaxy were taken in 2012 and 2013 using the laser guide star adaptive optics system (\citealt{2006PASP..118..297W}; \citealt{2006PASP..118..310V}) on the Keck II telescope.
The images were taken with NIRC2 (PI: Keith Matthews) in the K$^{\prime}$-band (2.1 $\mu$m) and have a plate scale of 9.952 milliarcseconds per pixel (\citealt{2010Yelda}).
The AO observations of this field from 2004 through 2011 have been previously reported (\citealt{2005bGhez}; \citealt{2008Ghez}; \citealt{2009Lu}; \citealt{2010Yelda}; \citealt{2012Yelda}; \citealt{2014Yelda}).  
The new AO observations from the years 2012 and 2013 were collected in the same manner as earlier observations.  In this setup, each frame was taken with an exposure time of 2.8 seconds with 10 coadds (280 times longer than the speckle imaging frames).  
Astrometry and photometry are extracted using the same techniques reported in our previous work.  Uncertainties in these measurements are also determined by running the same analysis on 3 subset images constructed with 1/3 of the frames used in the deep image.  Table \ref{tab:AOobs} summarizes the results from the new AO observations.  Together with existing star lists from earlier Keck AO data, these 3 new star lists bring the total number of epochs of relative AO astrometry available for this study to 23 epochs, from which 21 AO detections of S0-38 are unambiguously made (see Figure \ref{fig:images}).  The average magnitude of S0-38 as derived from these detections is $17.0 \pm 0.1$.  S0-38 is well above the direct detection limit of the AO data, but comparable to the direct detection limit of the speckle holography data (see Figure \ref{fig:Klims}).

In addition to the central field, we also take observations designed to measure the astrometry in the near infrared (NIR) of a set of 7 SiO masers.
We tie the NIR measurements of these masers to astrometric and velocity measurements made in the radio to construct an absolute reference frame with Sgr A* at rest at the origin (see \citealt{2007ApJ...659..378R}; \citealt{2008Ghez}; \citealt{2009Gillessen}; \citealt{2010Yelda}; \citealt{2014Yelda}).
Here we repeat the exact procedure detailed in \cite{2014Yelda} to construct the absolute reference frame, adding three new epochs of NIR maser observations from 2011 to 2013.  These new maser observations are summarized in Table \ref{tab:maserobs} in Appendix \ref{app:ref_frame}.  Appendix \ref{app:ref_frame} also provides the resulting absolute astrometry for the IR secondary standards that are used to combine all the speckle holography and AO star lists of relative positions together into a common absolute reference frame.

\subsubsection{Spectroscopic Data}
\label{AOspecdata}

New spectroscopic observations for S0-38 were obtained using the integral field spectrograph OSIRIS with the laser guide star AO system on the Keck 1 telescope (\citealt{2006SPIE.6269E..42L}) on 2013 May 11-13.  
The S0-38 data were taken with the K broadband filter (Kbb; 1.965 - 2.382 $\mu m$) in the 35 milliarcsecond (mas) plate scale.  This observational setup is designed to measure the CO bandheads ($2.3 \mu m$) of the short-period, late-type star S0-38.  The resulting spectrum has a resolution of $R \sim 3,600$.  Thirty four 900-second exposures  were taken, each dithered by a small amount ($\sim$ 0.2 arcseconds).  In 6 of these frames, the AO correction was such that S0-38 is confused with a neighboring source $\sim100$ milliarcseconds away.  The remaining 28 frames are used to obtain the spectrum of S0-38, providing a total on-source time of 7 hours.  
For calibration purposes, we observed skies with the same exposure time (900 s).  
The AO performance was excellent during these observations, providing a PSF with a full width at half maximum (FWHM) of $\sim$ 70 mas.  

\begin{figure}[ht]
   \centering
   \includegraphics[width=8.0cm]{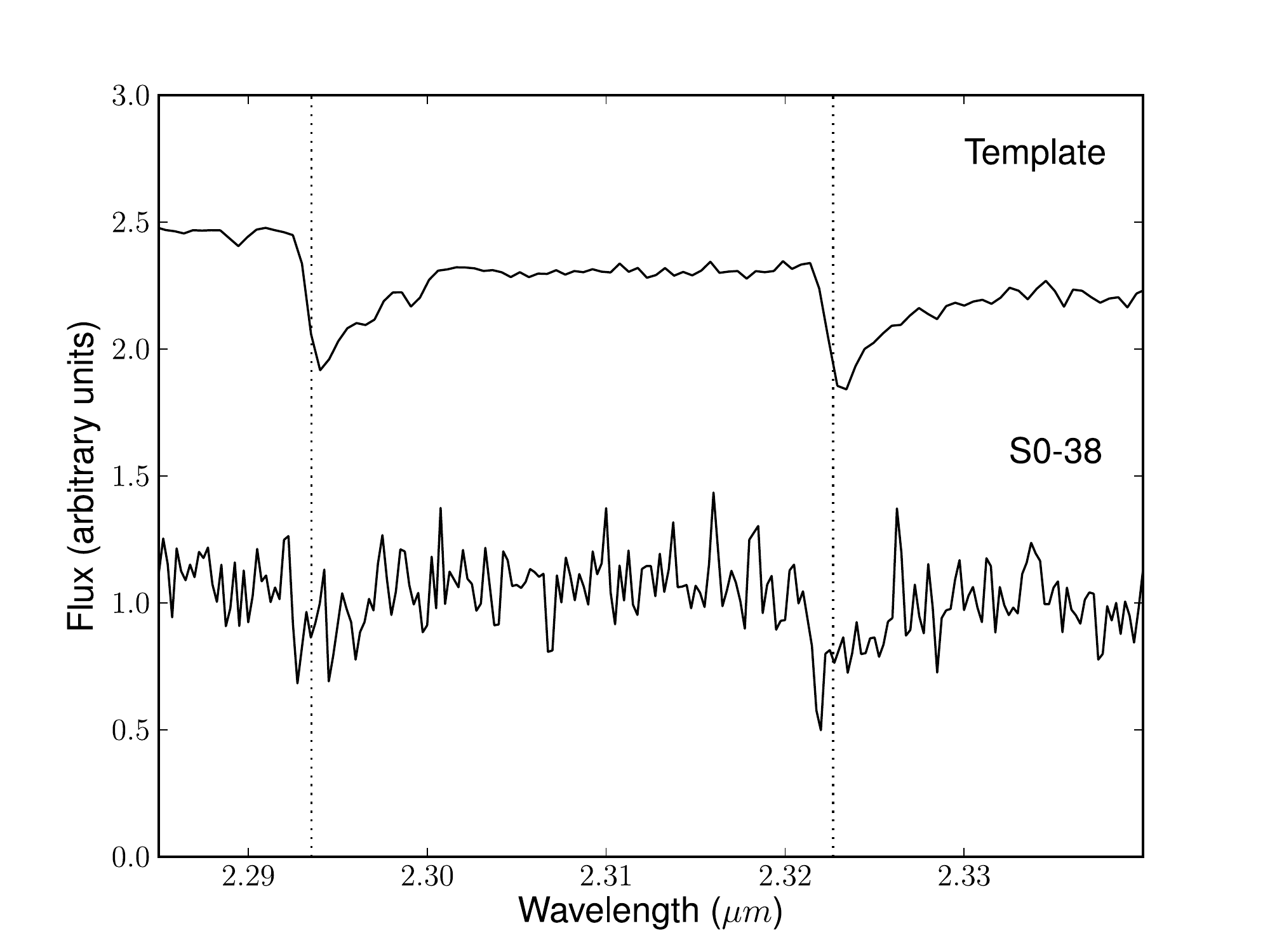}
   \includegraphics[width=8.0cm]{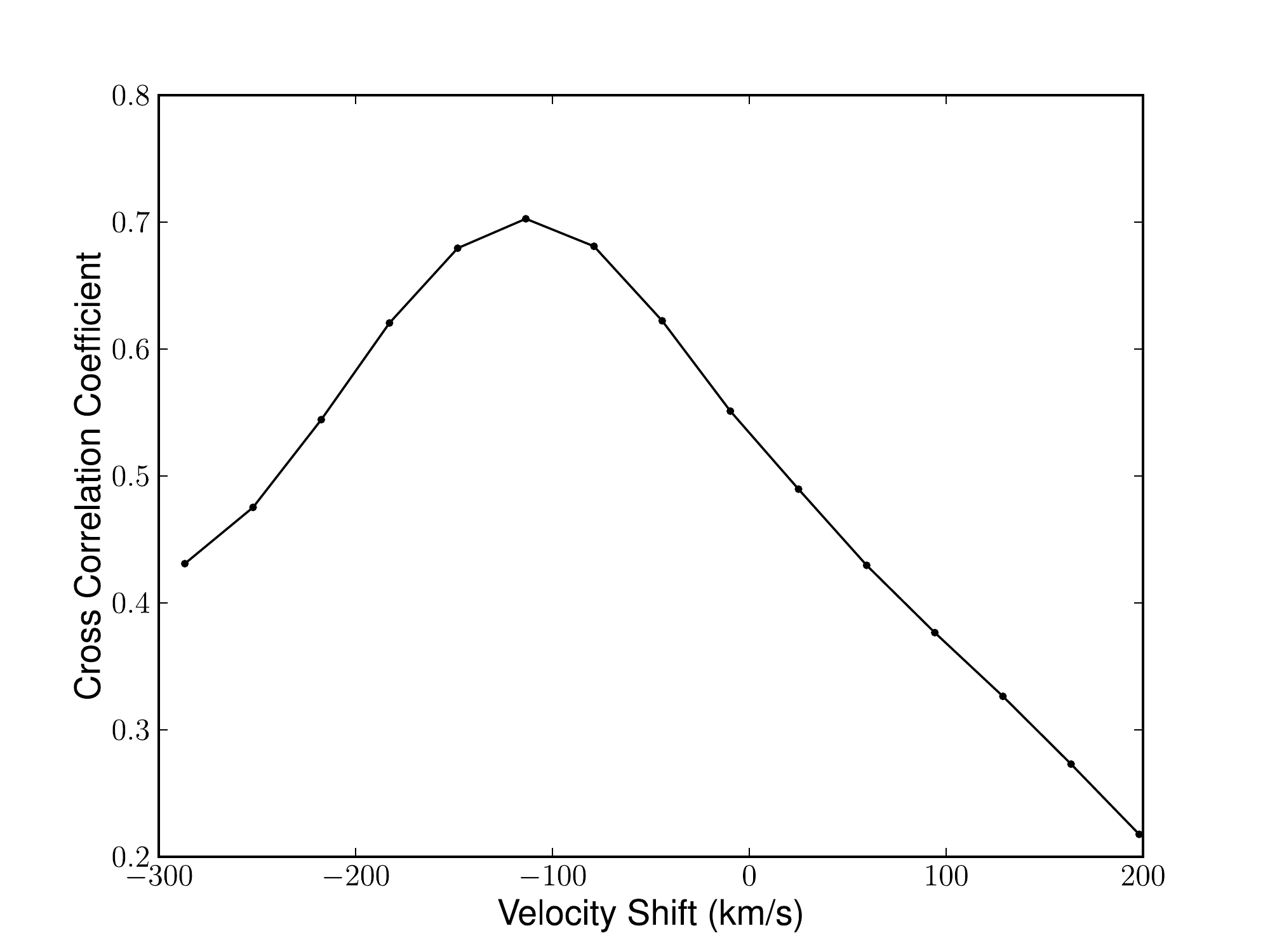}
      \caption[Spectrum of S0-38]{\emph{Left}: Observed spectrum of S0-38 as well as the template spectrum (shifted to a radial velocity of 0 km s$^{-1}$; \citealt{2009ApJS..185..289R}) used for the cross correlation showing the wavelength range over which this analysis is performed.  Dotted vertical lines indicate the rest wavelengths of the CO bandhead absorption features seen in both the template and in S0-38's spectrum.  \emph{Right}: The measured cross correlation coefficient between the template spectrum and S0-38's spectrum over a range of radial velocity shifts of S0-38.  These velocity shifts are in the local standard of rest reference frame.}
         \label{fig:spectrum}
\end{figure}

The analysis used to extract spectra closely follows previous analyses (\citealt{2013Do}).  For the new Kbb data, we extract S0-38's spectrum using a 35 milliarcsecond aperture.  The resulting spectrum is calibrated by subtracting an annulus around the extraction aperture with a width of 70 milliarcseconds.  The calibrated and normalized spectrum is shown in Figure \ref{fig:spectrum}.  This spectrum of S0-38 has a signal-to-noise ratio of 10, computed over a small, featureless wavelength range (between 2.212 and 2.218 microns).  The radial velocity (RV) of S0-38 is then determined by measuring the cross-correlation of the extracted spectrum and a template spectrum in the wavelength region around the CO bandhead absorption features (from 2.285 to 2.340 $\mu m$).  We use the spectrum of the M3II giant HD40239 observed by the SPEX telescope (\citealt{2009ApJS..185..289R}) as the template spectrum.  
\cite{2009Gillessen} first reported the detection of the CO bandheads in this source, and the RV reported in that paper is also used for our orbital analysis.  
The observed radial velocity is finally transformed to the local standard of rest reference frame by adding a correction of 27 km s$^{-1}$.  
The error on the RV measurements is taken as the root mean square of the RV measurements made on three subset spectra that are each made with $\sim$1/3 of the individual frames.  
From this we obtain an LSR-corrected RV of -111 $\pm$ 25 km s$^{-1}$ for S0-38 in May 2013.  

\begin{deluxetable}{lrrrrrcccccccc}
\tabletypesize{\scriptsize}
\tablewidth{0pt}
\tablecaption{Summary of Speckle Imaging Observations}
\tablehead{
  \multicolumn{2}{c}{Date} & 
  \colhead{Frames} & 
  \multicolumn{2}{c}{Frames Used} & 
  \multicolumn{2}{c}{$N_{stars}$} & 
  \multicolumn{2}{c}{Strehl Ratio} & 
  \multicolumn{2}{c}{$K_{lim}\tablenotemark{a}$ (mag)} & 
  \multicolumn{2}{c}{$\sigma_{pos}\tablenotemark{b}$ (mas)} & 
  \colhead{} \\ 
  \colhead{(UT)} & 
  \colhead{(Decimal)} & 
  \colhead{Available} & 
  \colhead{Holo} & 
  \colhead{SAA} & 
  \colhead{Holo} & 
  \colhead{SAA} & 
  \colhead{Holo} & 
  \colhead{SAA} & 
  \colhead{Holo} & 
  \colhead{SAA} & 
  \colhead{Holo} & 
  \colhead{SAA} & 
  \colhead{Data Source\tablenotemark{c}} 
}
\startdata
      1995 June 9-12  & 1995.439  & 7532 & 4201 & 1800 &  199 &  151 & 0.23 & 0.06 & 16.0 & 15.6 &  0.5 &  1.1 &                (1)\\ 
     1996 June 26-27  & 1996.485  & 5391 & 4287 &  865 &  136 &   77 & 0.23 & 0.03 & 15.7 & 14.4 &  1.9 &  1.8 &                (1)\\ 
         1997 May 14  & 1997.367  & 3400 & 3400 & 1837 &  229 &  139 & 0.47 & 0.04 & 16.2 & 15.5 &  0.8 &  1.3 &                (1)\\ 
      1998 April 2-3  & 1998.251  & 2744 & 2712 & 1639 &  170 &   83 & 0.31 & 0.04 & 16.0 & 14.8 &  1.3 &  1.5 &                (2)\\ 
      1998 May 14-15  & 1998.366  & 9708 & 9708 & 2102 &  231 &  126 & 0.41 & 0.04 & 16.6 & 15.6 &  0.5 &  1.3 &                (2)\\ 
       1998 July 3-5  & 1998.505  & 2352 & 2352 &  933 &  172 &  127 & 0.46 & 0.06 & 16.3 & 15.5 &  0.5 &  1.2 &                (2)\\ 
        1998 Aug 4-6  & 1998.590  & 19741 & 12061 & 1933 &  235 &  172 & 0.48 & 0.06 & 17.0 & 15.7 &  0.3 &  0.8 &                (2)\\ 
          1998 Oct 9  & 1998.771  & 2548 & 2166 & 1082 &  161 &  120 & 0.58 & 0.07 & 16.1 & 15.4 &  0.5 &  1.5 &                (2)\\ 
        1999 May 2-4  & 1999.333  & 9800 & 9775 & 1857 &  263 &  183 & 0.51 & 0.07 & 16.9 & 15.8 &  0.4 &  1.1 &                (2)\\ 
     1999 July 24-25  & 1999.559  & 5684 & 5684 & 2108 &  273 &  232 & 0.50 & 0.09 & 16.9 & 16.0 &  0.3 &  0.9 &                (2)\\ 
       2000 April 21  & 2000.305  & 2940 & 2940 &  805 &  122 &   63 & 0.18 & 0.04 & 15.5 & 14.3 &  1.4 &  2.1 &                (3)\\ 
      2000 May 19-20  & 2000.381  & 15680 & 15680 & 2492 &  293 &  242 & 0.47 & 0.08 & 17.2 & 16.0 &  0.3 &  0.7 &                (3)\\ 
     2000 July 19-20  & 2000.548  & 11172 & 10637 & 1581 &  269 &  194 & 0.34 & 0.07 & 16.6 & 15.7 &  0.7 &  1.1 &                (3)\\ 
         2000 Oct 18  & 2000.797  & 2352 & 2352 & 1517 &  140 &   77 & 0.39 & 0.04 & 16.0 & 14.8 &  1.1 &  1.3 &                (3)\\ 
        2001 May 7-9  & 2001.351  & 7306 & 7306 & 1994 &  225 &  175 & 0.39 & 0.07 & 16.4 & 15.7 &  0.5 &  1.0 &                (3)\\ 
     2001 July 28-29  & 2001.572  & 6860 & 6565 & 1695 &  290 &  239 & 0.54 & 0.11 & 17.0 & 16.3 &  0.3 &  0.8 &                (3)\\ 
    2002 April 23-24  & 2002.309  & 13620 & 13501 & 1958 &  267 &  183 & 0.41 & 0.05 & 16.9 & 15.8 &  0.4 &  1.2 &                (3)\\ 
      2002 May 23-24  & 2002.391  & 18052 & 11800 & 1443 &  290 &  252 & 0.53 & 0.08 & 17.1 & 16.1 &  0.3 &  0.9 &                (3)\\ 
     2002 July 19-20  & 2002.547  & 8081 & 5518 & 1118 &  208 &  125 & 0.36 & 0.06 & 16.4 & 15.4 &  0.9 &  1.4 &                (3)\\ 
    2003 April 21-22  & 2003.303  & 9392 & 9392 & 1841 &  144 &  121 & 0.36 & 0.04 & 15.9 & 15.6 &  0.6 &  1.1 &                (3)\\ 
     2003 July 22-23  & 2003.554  & 5028 & 5028 & 1703 &  190 &  180 & 0.58 & 0.07 & 16.2 & 15.9 &  0.6 &  1.2 &                (3)\\ 
       2003 Sept 7-8  & 2003.682  & 6270 & 6228 & 1723 &  233 &  182 & 0.35 & 0.07 & 16.4 & 15.8 &  0.6 &  1.2 &                (3)\\ 
    2004 April 29-30  & 2004.327  & 9289 & 9289 & 1423 &  236 &  185 & 0.46 & 0.08 & 16.7 & 15.8 &  0.4 &  0.8 &                (4)\\ 
     2004 July 25-26  & 2004.564  & 13110 & 12920 & 2161 &  293 &  200 & 0.43 & 0.08 & 16.8 & 15.9 &  0.4 &  0.9 &                (4)\\ 
         2004 Aug 29  & 2004.660  & 2850 & 2850 & 1301 &  203 &  167 & 0.31 & 0.08 & 16.2 & 15.7 &  0.5 &  1.3 &                (4)\\ 
    2005 April 24-25  & 2005.312  & 10553 & 10553 & 1679 &  226 &  162 & 0.37 & 0.06 & 16.6 & 15.7 &  0.4 &  1.0 &                (5)\\ 
     2005 July 26-27  & 2005.566  & 6080 & 5683 & 1331 &  171 &  111 & 0.32 & 0.05 & 16.1 & 15.3 &  0.5 &  1.2 &                (5)\\ 
\enddata 
\label{tab:speckle_obs}

\tablenotetext{a}{K$_{lim}$ is the magnitude at which the cumulative distribution function of the observed K magnitudes reaches 95\% of the total sample size.}
\tablenotetext{b}{Positional error taken as error on the mean from the three sub-images in each epoch and includes stars with $K <$ 15.}
\tablenotetext{c}{Data originally reported in (1) \citet{1998Ghez}, (2) \citet{2000Ghez}, (3) \citet{2005aGhez}, (4) \citet{2005Lu}, and (5) \citet{2007Rafelski}}

\end{deluxetable}
\clearpage

\begin{deluxetable}{lrrrrrrrrr}
\tabletypesize{\scriptsize}
\tablewidth{0pt}
\tablecaption{Summary of New AO Imaging Observations}
\tablehead{
  \colhead{Date} & 
  \colhead{Date} &  
  \colhead{Frames} & 
  \colhead{Frames} & 
  \colhead{FWHM} & 
  \colhead{Strehl} & 
  \colhead{N$_{stars}$} & 
  \colhead{K$_{lim}\tablenotemark{a}$} & 
  \colhead{$\sigma_{pos}\tablenotemark{b}$} \\
  \colhead{(UT)} & 
  \colhead{(Decimal)} & 
  \colhead{Obtained} & 
  \colhead{Used} & 
  \colhead{(mas)} & 
  \colhead{} & 
  \colhead{} & 
  \colhead{(mag)} & 
  \colhead{(mas)} & 
}
\startdata
        2012 July 24  & 2012.562  &  224 &  162 & 58 & 0.29 & 2344 & 19.6 & 0.06\\ 
    2013 April 25-26  & 2013.318  &  268 &  140 & 67 & 0.17 & 1432 & 18.8 & 0.11\\ 
        2013 July 20  & 2013.550  &  239 &  193 & 58 & 0.29 & 2492 & 19.6 & 0.08 \\ 
\enddata 
\label{tab:AOobs}

\tablenotetext{a}{K$_{lim}$ is the magnitude at which the cumulative distribution function of the observed K magnitudes reaches 95\% of the total sample size.}
\tablenotetext{b}{Positional error taken as error on the mean from the three sub-images in each epoch and includes stars with $K <$ 15 and with an average distance < 4 arseconds from the black hole.}

\end{deluxetable}
\clearpage

\section{Data Analysis}

\label{forcemethod} 

The goal of the work presented here is to increase the time baseline of observations of S0-38 so that more than $\sim$40\% of its orbit is covered and its information can be combined with that of S0-2 to get further constraints on the gravitational potential in the Galactic Center.  While the procedure described in Section 2 results in unambiguous detections of S0-38 in all epochs of deep AO data (2005-2013), the same is not true for all epochs of speckle holography data (1995-2005).  
We refer to the AO and speckle holography detections from Section 2 as "direct detections."  
In order to add more speckle holography detections of S0-38, we use a new methodology that takes advantage of the knowledge of (1) the black hole properties from S0-2 and (2) S0-38's orbital motion from the AO epochs to search for S0-38 in the speckle holography maps.  
We refer to the detections that result from this procedure as "prior-assisted detections," in contrast to the "direct detections."

The data analysis associated with establishing and using the prior-assisted detections of S0-38 is described in the following four sections.  
In Section \ref{S038selection}, we discuss why S0-38 is chosen as the first star to which the new methodology is applied.  
In Section \ref{sampleselection}, we identify stars that are similar in magnitude to S0-38 and that may have overlapped with S0-38 on the plane of the sky between 1995 and 2005.  By tracking these stars along with S0-38, we can better determine to which star any new speckle holography detections should be assigned.  In Section \ref{priormethod}, we create positional predictions for S0-38, in addition to the stars in the sample, based on existing knowledge of the orbital parameters, and use these predictions to search for S0-38 in the speckle epochs.  In Section \ref{combofits}, all the AO and speckle holography astrometry along with the RV measured for S0-38 are simultaneously fit with S0-2's measurements to further constrain the gravitational potential through which these stars move.

\subsection{Selecting First Star to Apply New Methodology}
\label{S038selection}

In this work, we chose a single star as a pilot test for our new methodology of searching for new detections in speckle holography data.  We limit the possible stars to those for which full orbital phase coverage may be acheived, including both turning points, by adding the 11-year speckle holography time baseline to the 9-year AO time baseline.  This puts an upper limit of 20 years on the orbital period of candidate stars.  The goal of this orbital period criteria is to select a star that would give the most information about the central gravitational potential once new speckle holography detections are included in its orbital fit.  There are two stars not consistently directly detected in the speckle holography images that meet this constraint: S0-38 ($P = 19$ years) and S0-102 ($P = 11.5$ years; \citealt{Meyer12}).  We chose to apply the new methodology to S0-38 for the following reasons.  
First, S0-38 is consistently detected in every one of our deep adaptive optics data sets (23 astrometric measurements) while S0-102 is detected in just over half (14 astrometric measurements).  The reason that S0-102 is often lost in the adaptive optics data is that its orbit moves through areas that are very crowded by other stars and by the NIR component of Sgr A* itself.   S0-38 also moves through a similar area on the plane of the sky as it goes through closest approach, but it spends the majority of its orbit away from the very central area in which S0-102 moves, thanks to its longer orbital period/semi-major axis and its projection on the plane of the sky.  In the speckle years specifically, S0-38 is predicted overall to be in a less crowded region of the sky than S0-102, which makes finding additional prior-assisted speckle holography detections more feasible.  
Additionally, the radial velocity of S0-38 has been successfully measured (this work and \citealt{2009Gillessen}), whereas the radial velocity of S0-102 has not.  
This is largely thanks to the CO band heads present in S0-38's late-type star spectrum.  The distinct shape and depth of these features make them easier to detect than the Br-$\gamma$ line present in the spectra of early-type stars.  
In the specific case of the K broadband spectroscopic data used to measure S0-38's RV in this work (Section \ref{AOspecdata}), we are not able to measure the spectrum of S0-102 because it is confused with nearby sources in most frames.  
S0-38 was near furthest approach and in a more isolated region at the time of its spectroscopic measurement, making the extraction of its spectrum possible.  
We therefore choose to apply the new methodology to S0-38.

\subsection{Identifying Sources Potentially Confused with S0-38}
\label{sampleselection}

We construct a sample of stars that may have overlapped with S0-38 on the plane of the sky between 1995 and 2005.  The orbits of these stars are tracked along with that of S0-38 to avoid mismatching a star with any new speckle holography detection.  An initial sample set of possibly confusing sources is identified through a radius and magnitude cut, where the radius used here is with respect to Sgr A*.  The radius is set by how far away a star could be in the AO epochs from 2005 - 2013 and 
still have traveled into the central $0.^{\prime\prime}3$ (0.012 pc for $R_o$ = 8 kpc), where S0-38's orbit lies, during the 1995-2005 speckle epochs.  In order to set a radius cut, we assume a representative velocity for the other stars in this central region. We make a conservative estimate of this velocity by taking 3 times the measured velocity dispersion at the projected distance of $0.^{\prime\prime}3$ from the central black hole  ($\sigma$ = 400 km/s; \citealt{2013Do}).
While this assumed velocity is not an exact calculation of the velocities of stars in this region, it is a conservative value for the velocities of possibly confusing stars.

We then assume that the star is constantly moving at this velocity and that this velocity is fully in the plane of the sky in order to calculate the maximum radius of possible confusing stars.  The maximum radius is then: 
\begin{equation}
r_{max}\, =\, 3\; \times\; \sigma(\textrm{at } 0.^{\prime\prime}3) \times 10~yrs \; + \; 0.^{\prime\prime}3 \,=\, 0.^{\prime\prime}62
\end{equation}
This equation says that a star starting at a distance $r_{max}$ from Sgr A* and moving at a constant velocity of $3 \times \sigma(0.^{\prime\prime}3)$ towards the black hole on the plane of the sky would be at a distance of $0.^{\prime\prime}3$ from Sgr A* after 10 years.  
Any star that is within $r_{max}$ at some point in the AO epochs could have been inside the $0.^{\prime\prime}3$-radius circle in which S0-38 is orbiting at some point in the speckle epochs.  

From the stars within this radius, we then eliminate those sources whose magnitude is not consistent with S0-38's within 3 sigma ($K_{mag} = 17.00 \pm 0.24$).  Finally, we also remove sources whose velocities are not within $90^{\circ}$ of pointing to Sgr A* on the plane of the sky.  The final set of possible confusing sources is: S0-104, S0-102, S0-103, S0-37, S0-40, S0-42, S0-45, S0-49, and S0-23.

\subsection{Extending S0-38's Astrometric Orbital Phase Coverage using Orbital Priors}
\label{priormethod}

We apply a new methodology of analyzing the speckle holography data to find sources not detected in the initial, blind analysis.  Our technique uses the information in the adaptive optics astrometric and radial velocity measurements of S0-38 as a starting point of where to look for this star in the speckle holography images.  
The orbits of the sample of possibly confusing stars described in Section \ref{sampleselection} are tracked along with that of S0-38 so that any new speckle holography detection is not misidentified.   
The basic steps of this new methodology as applied to S0-38 are as follows:
\begin{enumerate}

\item Perform an orbital fit of S0-2's direct astrometric detections from AO and speckle holography epochs and radial velocities, with all 6 orbital parameters and all 7 gravitational potential parameters left free.

\item Perform an orbital fit of the direct astrometric detections and radial velocity measurements (if available) of S0-38 and the possibly confusing stars (Section \ref{orbitalfits}).  In these fits, the gravitational potential is fixed to the best-fit solution from S0-2's orbital fit.  For the sample stars, both speckle holography and AO direct detections are included in this fit.  For S0-38, only AO direct detections are included.  

\item From the probability distributions of the orbital parameters found in each orbital fit, determine the probability distribution of the (X, Y) (on-the-sky) position of each star in the 
next earlier speckle holography epoch relative to those epochs included in S0-38's orbital fit (i.e., start with the most recent speckle holography epoch and move backwards in time in  subsequent steps).

\item Search for a speckle holography detection of S0-38 in the image from this epoch.  We first check whether S0-38 is detected in the initial analysis of the image.  
If it is not in the initial star list, then apply the newly-developed \emph{StarFinder} Force 1 algorithm to the image (Section \ref{force1}).  This algorithm takes as input the predicted pixel position of the star from the probability distribution of X and Y and searches the speckle holography image for a point source around that position that had not previously been detected in the initial star list.

\item If a new detection is found that meets a minimum correlation (0.3) and signal-to-noise threshold (3$\sigma$) and it is not blended with another known star, add that point to the list of detections of S0-38 and refit the orbit with that point (Section \ref{orbitalfits}).  If no detection is found (i.e., the PSF fit in \emph{Starfinder} Force 1 does not converge), then search again in the next earlier speckle holography epoch.  Note that the potentially confusing stars are not searched for in the speckle holography images and their orbital fits are therefore not updated.  The majority of these stars are typically directly detected in the speckle holography images because their orbital periods are longer than S0-38's and they move through less crowded regions on the sky.

\item Repeat steps 3 - 5 until all speckle holography images are searched for S0-38.  

\item After all new prior-assisted detections are identified, the black hole parameter probability distributions are determined by a simultaneous fit with S0-2 (Section \ref{combofits}).

\end{enumerate}

\subsubsection{6-Dimensional Orbital Fits and Positional Priors}
\label{orbitalfits}
Orbital fits are carried out several times in our analysis.  
The initial task in the new analysis is to fit the orbits of all stars in the sample.  
For each new speckle holography detection, the orbit of S0-38 is also fit again to improve the prediction for earlier speckle epochs.  
All the orbital fits performed in steps 2 - 5 assume that the star moves in the gravitational potential of a single point mass that is the central supermassive black hole (BH), Sgr A*.  
In all orbital fits used to inform the search for the star in speckle holography images, we use the information from S0-2 to determine the probability distributions of the 7 parameters that describe the gravitational potential: the black hole's mass ($M_{bh}$), position on the sky ($x_{o}$, $y_o$), 3D velocity ($V_x$, $V_y$, $V_z$), and the line of sight distance ($R_o$).  
The details of how S0-2 is used to determine the BH parameter probability distributions as well as our new S0-2 measurements are presented in Appendix \ref{appendix:S0-2}.  
We then use the measurements of the star's position and radial velocity along with the constraints from S0-2 on the black hole parameters to determine probability distributions of the star's 6 Keplerian orbital parameters: orbital period ($P$), time of closest approach ($T_o$), eccentricity ($e$), inclination ($i$), argument of periapse ($\omega$), and angle to the ascending node ($\Omega$).  

Our approach of using externally constrained black hole parameters speeds up the process of determining positional priors for S0-38 and the other sample stars by a factor of $\sim5$, allowing these orbits to be fit in a few hours instead of 1 day.  The external constraints are implemented by stepping through a grid in the 7D BH parameter space with 5 steps in each dimension, placed at -1, -0.5, 0, 0.5, and 1 sigma from the mean in that dimension.\footnote{We choose to use a 1 sigma range in this grid spacing to make the fitting process computationally feasible.}  
Each of the $5^{7} = 78,125$ points in this 7D grid is a single combination of values of the 7 BH parameters.  These combinations are assigned weights based on the probability of those values as determined by S0-2
\footnote{The 7D joint probability distribution of the BH parameters as determined by S0-2 is approximated by a multivariate gaussian.  This approximation represents the distribution well (see Figure\ref{fig:BHparams})}.  
The weights of these combinations are very unevenly distributed because some sets of BH parameters are highly correlated.  
In order to not include combinations of BH parameter values that have a very low probability due to these correlations, we keep only the combinations that together make up 95\% of the total cumulative weight.  
This weight cut reduces the number of individual combinations by a factor of $\sim$14 to 5,500.  

We then perform a series of individual fits to the star's detections in astrometry and radial velocity, one for each BH parameter value combination that made it past the weight cut.  
In each of these individual fits to the star's data, the 7 BH parameters are fixed to the values in each combination and the 6 orbital parameters are left free.  Each individual fit then results in a set of samples of the posterior distribution of the star's 6 orbital parameters.  
These individual fits are performed using the Bayesian multimodal nested sampling algorithm {\sc MultiNest} (\citealt{2008Feroz}; \citealt{2009Feroz}), which was implemented in our code by Leo Meyer.  This algorithm samples the posterior distribution more efficiently than traditional Markov chain Monte Carlo (MCMC) sampling schemes, especially when the distribution is multimodal.  
The {\sc MultiNest} algorithm has the advantage over our previous orbit-fitting method (see \citealt{2008Ghez}) in that the best-fit orbital parameters and their probability distributions are found simultaneously and with much less total computing time.  
When the {\sc MultiNest} is run, uniform priors are applied to every free parameter, with ranges chosen such that the full range of the posterior distribution is probed.  
The results of the individual fits are then combined in order to get the final probability distributions for the orbital parameters of the star.  This is achieved by combining the sets of samples of the posterior distributions found in each individual fit with the weight assigned to that fit's unique BH parameter value combination.  Finally, the probability distributions of the individual orbital parameters are determined by constructing histograms of the weighted posterior samples.  

The result is a probability distribution for each of the 6 orbital parameters determined by the star's detections, with the BH parameter information determined only by S0-2's detections.
From the final probability distributions of all the BH and orbital parameters we can then derive the probability distribution of the (X, Y) position on the sky of the star in the speckle holography observation epochs.  These probability distributions of X and Y constitute our prior knowledge about the position of this star in all earlier epochs and are used as inputs into the \emph{StarFinder} Force 1 algorithm (Section \ref{force1}).

\begin{figure}[h!]
\centering
 \includegraphics[width=6in]{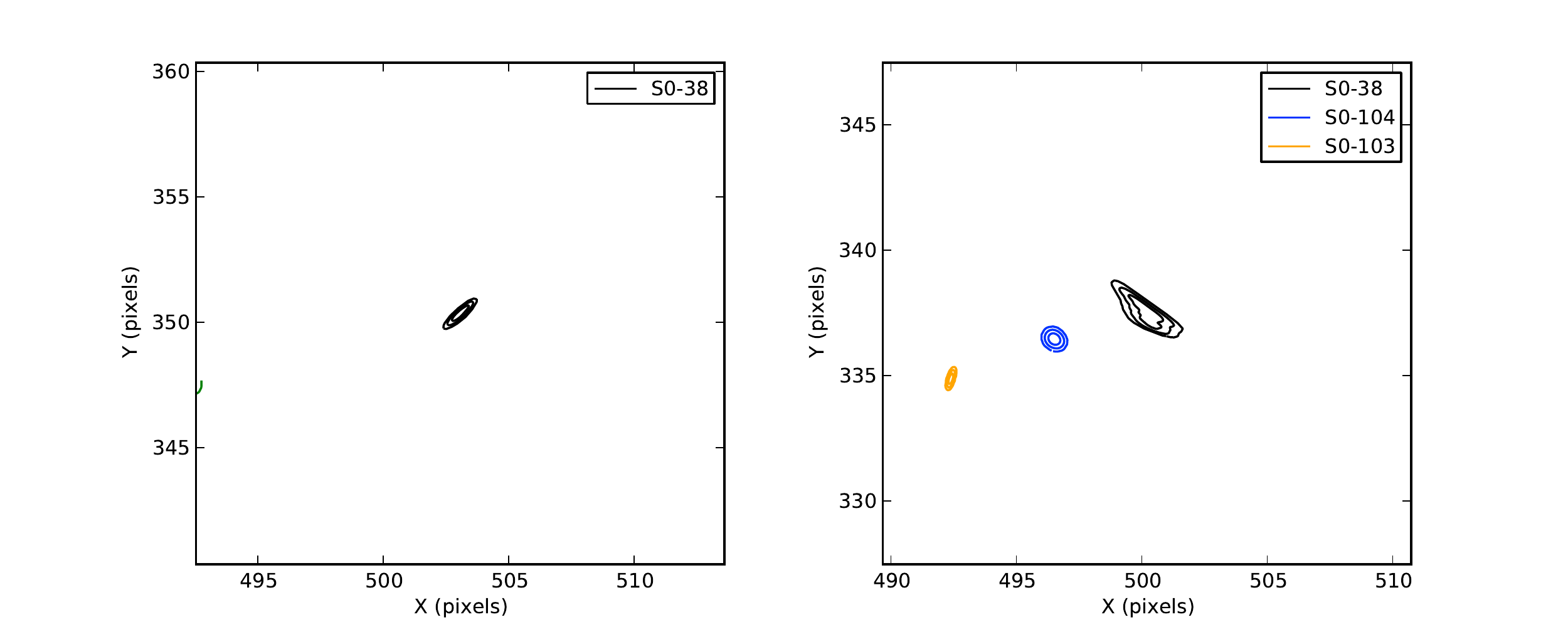} 
 \caption{\label{confusingsources}Probability distributions of X and Y for S0-38 and nearby sample stars in two example epochs: September 2003 (\emph{left}) and July 2002 (\emph{right}).  The pixel position is shown on the image, which has a plate scale of 10 milliarcseconds per pixel.  In September 2003, the prediction of the (X, Y) position of S0-38 is well-isolated from predicted positions of all the other stars in the sample.  This level of isolation is typical throughout the speckle epochs.  In July 2002, however, S0-38 is predicted to be within 70 milliarcseconds of S0-104.  These two sources are not separately detected in the 2002 speckle epochs due to their close proximity, so no new detection is assigned to S0-38 in these epochs.  
}
\end{figure}

Although orbital fits are performed for all the stars in the sample, there are only two that came close enough to be potentially confused with S0-38: S0-104 (in 2002) and S0-49 (in 1998).  
Figure \ref{confusingsources} shows the probability distributions of X and Y for S0-38 and these confusing sources in one of the epochs in which these stars are close on the plane of the sky (July 2002), as well as a more typical epoch where S0-38 is isolated from the possibly confusing sources (September 2003).  
Any new detection that is found in the images from these observation epochs is only used in the fit of S0-38 if both S0-38 and the nearby source are separately detected.  

The initial orbital fit of S0-38 includes only the AO direct astrometric detections and radial velocity measurements, while the orbital fits of the possibly confusing stars include both AO and speckle holography direct detections.  S0-38 moved so quickly in the speckle years compared to other stars in the sample that its direct speckle holography detections are generally too far away to associate with S0-38 by eye.  
An initial fit of S0-38's orbit with the gravitational potential fixed to S0-2's best-fit solution gives a reduced chi-squared of 12.2, implying that the AO errors determined by the 3 subset image method are underestimated.  
In order to correct this, we add in quadrature to the 3 subset image errors an additive error term constant for all AO epochs so that the reduced chi-squared of the AO-only fit equaled 1.0.  
This additive term is set to 1.7 mas, compared to an average 3 subset image error of 2.9 mas for S0-38's AO direct detections.  
This same procedure is not required for S0-2 because the reduced chi-squared of the orbital fit of its AO and speckle holography direct detections is already $\sim1$.

\subsubsection{\emph{StarFinder} Force 1 Algorithm}
\label{force1}
We can now use the probability distributions of S0-38's position on the sky in a given speckle epoch to inform our search for the star in that image, while also tracking the orbits of the potentially confusing stars included in our sample.  
We first check whether there is a direct detection in the initial star list for that epoch that coincides within uncertainties with the predicted position of S0-38.  If a detection is not found in the initial analysis, then the newly-developed \emph{StarFinder} Force 1 algorithm is applied.  

The \emph{StarFinder} Force 1 algorithm is designed to search for any additional point source that is not in the initial star list.  
The main inputs to the algorithm are the best guess whole pixel position of the source (Xo, Yo), the size of the search box over which to initially look for the source, and the positions and fluxes of the stars in the initial list for that epoch.  (Xo, Yo) and the search box size are chosen so that the initial search is done over the whole 3 sigma range of the probability distribution of X and Y, but in the end new detections are allowed to be found outside the search box.

\emph{StarFinder} Force 1 uses many of the sub-routines of the original \emph{StarFinder} code and follows the same basic steps, outlined here: 

\begin{enumerate}
\item{First, the code takes a subset of the main image centered on the input position (Xo, Yo).  The size of this subset depends on the full width at half maximum (FWHM) of the point spread function (PSF).  For a typical PSF FWHM, the size of the subset image in this correlation step is 5 pixels on a side.  It then finds the position at which the correlation between this small subset image and the inputted PSF is maximized.  The search box input parameter sets the offsets over which the PSF and the subset image are cross correlated.}

\item{Next, the code takes a larger subset of the image (for a typical PSF FWHM, this subset image is 9 pixels on a side) and performs the initial fit of that subset image.  The free parameters of this fit are the positions and fluxes of all stars within the subset image (including any sources found in the initial star list plus the additional source).  The initial guess for the position of the additional source is the position of maximum correlation found in the previous step.  
The position of the additional point source is allowed to move away from this initial guess during the fit and is not constrained by the inputted search box or the previously predicted (X,Y) position.  
Any contributions of flux in the subset image from stars with positions outside the subset image are considered fixed and are subtracted from the subset image before the fit is performed.}

\item{Finally, a series of 2 iterative fits is performed over a slightly larger subset image (for a typical PSF FWHM, this is 11 pixels on a side).  
If these iterative fits converge, then the best-fit position and flux of the newly-detected source are returned.  To assign this new detection to S0-38, we require that the new detection (1) has a correlation above 0.3, (2) has a flux more than 3$\sigma$ above the noise, and that (3) S0-38 is not confused with any other source in the image.
}
\end{enumerate}

We generally cannot assign errors to the points detected with \emph{StarFinder} Force 1 using the three subset image method that is used in the construction of the initial star lists.  Since the magnitude of S0-38 is close to the detection limit in the deep speckle holography images of each epoch, it is not consistently detected in the three subset images that use only 1/3 of the frames.  We therefore assign the error of each \emph{StarFinder} Force 1 astrometric point so that the reduced chi-squared of the orbital fit with the additional point is approximately 1.  This is done on an epoch-by-epoch basis, so the errors of each new point are set independently.  In this way, an updated orbital fit with the new detection can be done before the search for the star in the next earlier speckle holography epoch is performed.  If S0-38 is detected in the initial analysis of that holography epoch (and therefore is already detected in all three subset images), then the errors determined by the three subset image method are used in the subsequent orbital fit.  

\subsection{Orbital Fits That Use S0-38 to Constrain the Gravitational Potential}
\label{combofits}

The result of applying the method outlined above to all speckle holography images is a final set of astrometric measurements of S0-38, including both the direct detections and the prior-assisted detections.  With a higher orbital phase coverage that covers both closest and furthest approach, we now use S0-38 to constrain the gravitational potential.  

Before performing these orbital fits, we reassign the errors of S0-38's direct and prior-assisted speckle holography detections to a single value for all epochs.  The single value is chosen so that the reduced chi-squared of an orbital fit of S0-38 with the black hole potential fixed to the best-fit solution from S0-2 equals 1.  This reassignment is done so that the errors on the position of S0-38 are not underestimated in the epochs of the direct detections, whose errors were originally assigned by the three subset image method.  
The errors determined by the error of the mean of the measurements on three subset images are expected to be on-average slightly underestimated 
and are also a noisy estimator of the true error of the main map measurement. 
The reassignment therefore ensures that the errors on S0-38's positions, as well as the resulting errors on the black hole parameters, are not underestimated.  This issue of underestimated error bars is already addressed in the AO epochs by the inclusion of an additive error term so that the AO-only orbital fit gives a reduced chi-squared of 1.  Additionally, with this reassignment the errors of the direct and prior-assisted detections are treated consistently in that both are set by requiring the reduced chi-squared to equal 1.0.  

Once S0-38's speckle holography errors have been reassigned, we perform two Keplerian orbital fits in which S0-38 is used to constrain the gravitational potential parameters: (1) a 13-dimensional fit of only the orbit of S0-38 and (2) a 19-dimensional fit of the orbits of both S0-2 and S0-38.  The 13 free parameters in the first fit are the 7 black hole potential parameters and the 6 Keplerian orbital parameters that describe the motion of S0-38.  This fit is used to compare the gravitational potential solutions determined by S0-38 alone with those determined by S0-2 alone.  
The 19 free parameters in the second fit are the 7 black hole potential parameters and a set of 6 Keplerian orbital parameters for each star.  This simultaneous fit of the orbits of S0-2 and S0-38 is used to determine the final best-fit values and errors of the black hole potential parameters.  
Both the 13- and 19-dimensional fits are performed using {\sc MultiNest}.

We finally perform a simultaneous fit of the orbits of S0-2 and S0-38 that includes deviations from a pure Keplerian orbit due to a distribution of extended dark mass.  From this fit, we find an upper limit for the amount of extended mass within the orbits of these stars.  
Following \cite{2008Ghez}, we set the extended mass distribution to follow a power-law density profile such that the total enclosed mass is given by:
\begin{equation}
M(<r) = M_{bh} + M_{ext}(<r_o)\left(\frac{r}{r_o}\right)^{3-\gamma}
\end{equation}
The power law slope, $\gamma$, is fixed to 1.5 in the orbital fits presented here, though \cite{2008Ghez} showed that the extended mass upper limit within the small radius enclosed by stellar orbits did not depend strongly on the value chosen for $\gamma$ in the range from 0.5 to 3.  We set the outer radius cutoff of the extended mass distribution to 3.4 $\times 10^{11}$ km = 0.011 pc, such that it encloses the apoapse distances of S0-2 and S0-38 within uncertainties.  Our data is not sensitive to the mass outside of the orbits of these stars.  
For this fit, we also set the characteristic radius in the equation above, $r_o$, to the same value of 0.011 pc.  We then use the extended mass parameters to find the extended dark mass within the apoapse of S0-2's orbit (0.01 pc), to compare with results from previous works.  
This orbital fit is performed using {\sc MultiNest} and has 20 free parameters: 7 black hole potential parameters, a set of 6 Keplerian orbital parameters for each star, and the total extended dark mass within $r_o$ ($M_{ext}(<r_o)$).  We also perform an extended mass fit of the orbit of S0-2 alone for comparison purposes, which is discussed in Appendix \ref{appendix:S0-2}.

As the final part of these analyses, we explore the additional uncertainties and biases arising from the construction of the absolute reference frame.  In Appendix \ref{appendix:jackknife}, we describe the details of a jack knife analysis performed on the 7 SiO masers used to tie our NIR observations to the absolute reference frame.  The results of this jack knife analysis are used to correct the statistical bias (which is small compared to the orbital fitting uncertainties for $M_{bh}$ and $R_o$) and add an additional source of error for all the black hole potential parameters.  These bias shift and the additional errors are only significant for the parameters describing the position and velocity of Sgr A* on the plane of the sky.

\subsection{Simulations to Investigate Statistical and Systematic Errors}
To confirm that the errors on the black hole parameters resulting from the simultaneous fit of S0-2 and S0-38 are reasonable given our measurement uncertainties, we simulate mock S0-2 and S0-38 data.  We then fit these mock data and compare the widths of the resulting $M_{bh}$ and $R_o$ probability distributions to the results from the real data.  

For every speckle holography, AO, and RV observation epoch, a simulated measurement is created by drawing from a gaussian distribution.  This gaussian distribution has an average equal to the best-fit model from the combined S0-2 and S0-38 orbital fit and a standard deviation equal to that epoch's measurement error.  The mock data are then fit using {\sc MultiNest}, in the same way as the real S0-2 and S0-38 data.  The widths of the resulting $M_{bh}$ and $R_o$ probability distributions are quantified as the standard deviation of the posterior samples generated by the {\sc MultiNest} fit.  One hundred mock data sets are created and fit in order to sample the possible widths of the $M_{bh}$ and $R_o$ distributions well while not requiring large amounts of computation time.

\section{Results}
\label{results}

\subsection{New Speckle Holography Detections of S0-38}
\label{detections_S038alone}

\begin{figure}[h]
\begin{center}
\includegraphics[width=0.7\columnwidth]{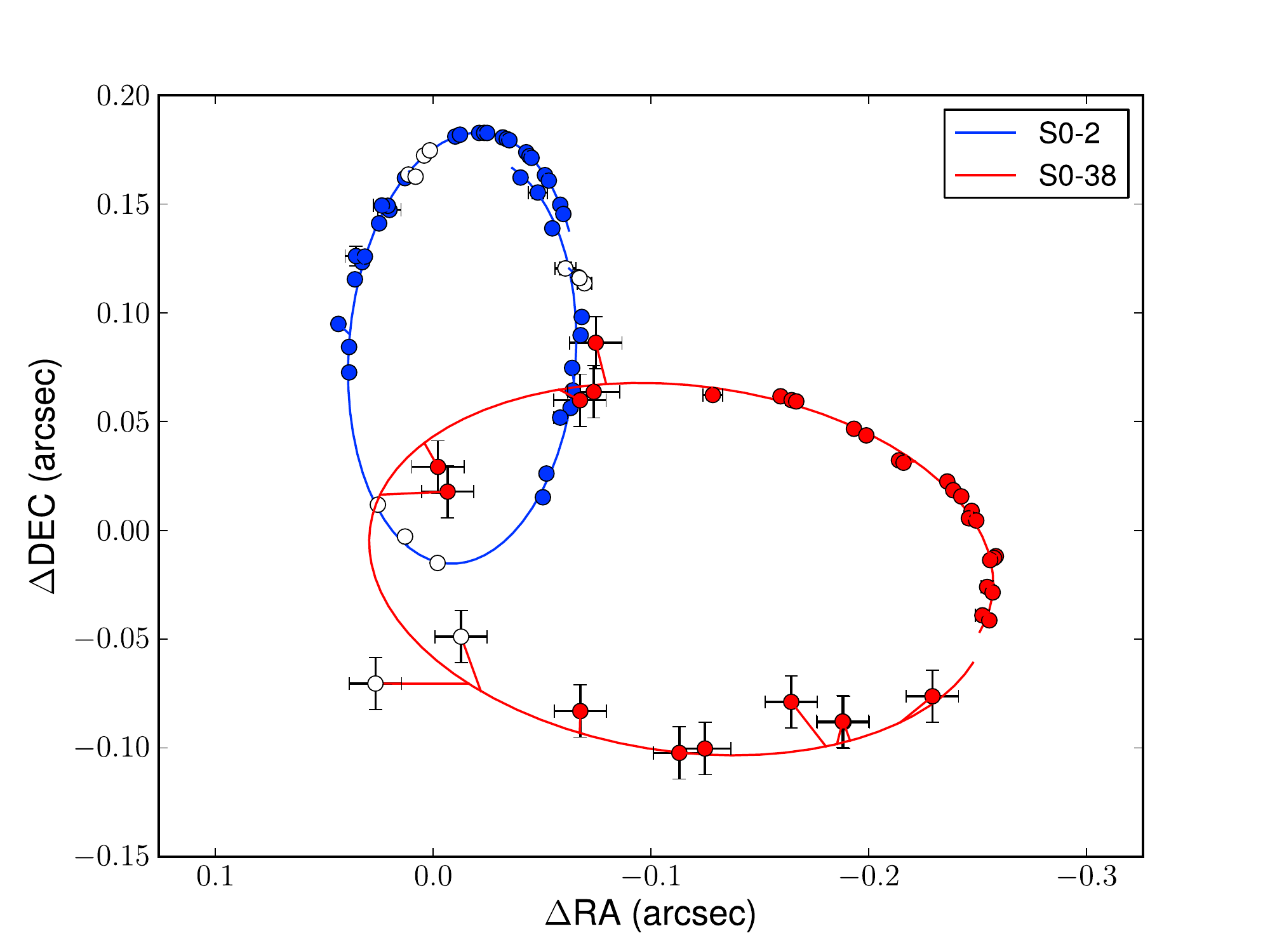}
\caption{\label{fig:XY_S0-2_S0-38}
The best-fit orbit for S0-2 (blue line) and for S0-38 (red line) on the plane of the sky.  These model orbit lines show the positions of these stars from 1995 to 2014.  Both stars orbit clockwise on the plane of the sky.  Closed circles indicate astrometric detections that were used in the orbital fits.  Open circles indicate points that were not used in the fits because these astrometric detections are biased due to the proximity of other known sources on the plane of the sky. 
For S0-38, this consists of the two epochs of May and June 2002 in which the position of S0-38 shows a bias because it is blended with S0-104; in all other confusion epochs \emph{Starfinder} Force 1 can only recover a position close to the neighboring source and S0-38 is undetected.  Note in particular the paucity of detections around the point of closest approach at the bottom of S0-2's orbit.  The implication of the absence of useable detections in this part of S0-2's orbit is that S0-2 constrains the horizontal position of Sgr A* in our reference frame ($x_o$) better than its vertical position ($y_{o}$).  S0-38's orbit on the plane of the sky is perpendicular to that of S0-2, so it can provide additional constraints on $y_{o}$.}
\end{center}
\end{figure}

The application of the new methodology of analyzing speckle holography images on S0-38 results in 12 new astrometric detections of this star.  These new detections go back to May 1997, spanning nearly the entirety of our time baseline for speckle holography.  Of these 12 new speckle holography detections, 6 are direct detections found in the initial analysis of the speckle holography data and 6 are prior-assisted detections found using the force method.  The 6 direct detections are in the 6 deepest epochs in which S0-38 was not confused with another source.  All the prior-assisted detections have K magnitudes that are consistent with S0-38's average AO magnitude within $\pm$0.5, are 3 - 20 sigma above the noise, and have correlations ranging from 0.7 - 0.85 (well above the 0.3 minimum value). 
Adding these new speckle holography detections to the 21 adaptive optics detections of S0-38 brings the total number of astrometric measurements to 33 and almost doubles the time baseline that these measurement cover.  
Table \ref{tab:S0-38_measurements} lists all the astrometry and radial velocity measurements of S0-38 used in this study.  
Appendix \ref{appendix:S0-2} lists the measurements of S0-2 used in the orbital modelling.  
With the new speckle holography detections, S0-38 has now been observed for over 80\% of its 19-year orbit (see Figure \ref{fig:XY_S0-2_S0-38}).

S0-38 is not detected in the other 15 speckle epochs for a variety of reasons.  S0-38 is confused with another star in 7 speckle epochs, either with a star of similar brightness (with S0-104 in April and May 2002 and with S0-49 in July 1998) or with a much brighter star (with S0-1, K = 14.7, in October 2000 and with S0-20, K = 15.8, in July 2005 and May and July 1999).  S0-38 was not detected in 6 other speckle epochs because the PSF fit performed by \emph{Starfinder} Force 1 did not converge (April 2005, September 2003, July 2002, May 2001, April 2000, and April 1998).  Finally, only 2 detections were rejected because they did not meet either the correlation or signal-to-noise threshold: June 1996 (signal-to-noise = 1.7) and June 1995 (correlation = 0.04).

\begin{deluxetable}{lccccccccr}
\tabletypesize{\scriptsize}
\tablewidth{0pt}
\tablecaption{S0-38 Astrometric and Radial Velocity Measurements}
\tablehead{
  \colhead{Date} & 
  \colhead{$K^{\prime}$} & 
  \colhead{$\Delta$R.A.} & 
  \colhead{$\Delta$Dec.} & 
  \colhead{$\Delta$R.A. Error} & 
  \colhead{$\Delta$Dec. Error} & 
  \colhead{$V_z$\tablenotemark{b}} & 
  \colhead{$V_z$ Error} & 
  \colhead{Reference\tablenotemark{a}} & 
  \colhead{Detection} \\
  \colhead{(Decimal)} & 
  \colhead{(mag)} & 
  \colhead{(arcsec)} & 
  \colhead{(arcsec)} & 
  \colhead{(arcsec)} & 
  \colhead{(arcsec)} & 
  \colhead{(km s$^{-1}$)} & 
  \colhead{(km s$^{-1}$)} & 
  \colhead{} & 
  \colhead{Method} \\
}

\startdata
  1997.367 & 16.9 & -0.229 & -0.076 & 0.012 & 0.012 & - & - & (1) & Speckle (prior)\\ 
  1998.366 & 16.8 & -0.188 & -0.088 & 0.012 & 0.012 & - & - & (1) & Speckle (prior)\\   
  1998.590 & 17.9 & -0.188 & -0.088 & 0.012 & 0.012 & - & - & (1) & Speckle (direct)\\ 
1998.771 & 17.0 & -0.164 & -0.079 & 0.012 & 0.012 & - & - & (1) & Speckle (prior)\\ 
2000.381 & 17.3 & -0.125 & -0.100 & 0.012 & 0.012 & - & - & (1) & Speckle (direct)\\ 
2000.548 & 16.5 & -0.113 & -0.102 & 0.012 & 0.012 & - & - & (1) & Speckle (direct)\\ 
2001.572 & 17.0 & -0.068 & -0.083 & 0.012 & 0.012 & - & - & (1) & Speckle (direct)\\ 
2003.303 & 17.0 & -0.007 & 0.018 & 0.012 & 0.012 & - & - & (1) & Speckle (prior)\\ 
2003.554 & 16.8 & -0.002 & 0.029 & 0.012 & 0.012 & - & - & (1) & Speckle (prior)\\ 
2004.327 & 17.0 & -0.067 & 0.060 & 0.012 & 0.012 & - & - & (1) & Speckle (direct)\\ 
2004.564 & 17.0 & -0.074 & 0.064 & 0.012 & 0.012 & - & - & (1) & Speckle (direct)\\ 
2004.660 & 16.6 & -0.075 & 0.086 & 0.012 & 0.012 & - & - & (1) & Speckle (prior)\\ 
2005.580 & 17.1 & -0.1285 & 0.0622 & 0.0045 & 0.0023 & - & - & (1) &           AO\\ 
2006.336 & 16.9 & -0.1596 & 0.0616 & 0.0017 & 0.0017 & - & - & (1) &           AO\\ 
2006.470 & 17.0 & -0.1647 & 0.0599 & 0.0020 & 0.0025 & - & - & (1) &           AO\\ 
2006.541 & 16.9 & -0.1668 & 0.0593 & 0.0017 & 0.0017 & - & - & (1) &           AO\\ 
2007.374 & 17.2 & -0.1933 & 0.0467 & 0.0017 & 0.0018 & - & - & (1) &           AO\\ 
2007.612 & 16.9 & -0.1990 & 0.0437 & 0.0018 & 0.0017 & - & - & (1) &           AO\\ 
2008.260 & - & - & - & - & - & -185 &  70 & (2) &           AO\\ 
2008.371 & 16.9 & -0.2140 & 0.0322 & 0.0019 & 0.0018 & - & - & (1) &           AO\\ 
2008.562 & 16.9 & -0.2161 & 0.0311 & 0.0022 & 0.0017 & - & - & (1) &           AO\\ 
2009.340 & 17.1 & -0.2361 & 0.0225 & 0.0020 & 0.0019 & - & - & (1) &           AO\\ 
2009.561 & 17.1 & -0.2389 & 0.0185 & 0.0018 & 0.0018 & - & - & (1) &           AO\\ 
2009.689 & 17.1 & -0.2426 & 0.0156 & 0.0019 & 0.0021 & - & - & (1) &           AO\\ 
2010.342 & 17.0 & -0.2473 & 0.0090 & 0.0017 & 0.0017 & - & - & (1) &           AO\\ 
2010.511 & 16.9 & -0.2461 & 0.0056 & 0.0024 & 0.0022 & - & - & (1) &           AO\\ 
2010.620 & 17.1 & -0.2495 & 0.0045 & 0.0019 & 0.0019 & - & - & (1) &           AO\\ 
2011.401 & 17.0 & -0.2585 & -0.0119 & 0.0022 & 0.0018 & - & - & (1) &           AO\\ 
2011.543 & 16.9 & -0.2576 & -0.0127 & 0.0017 & 0.0017 & - & - & (1) &           AO\\ 
2011.642 & 17.0 & -0.2558 & -0.0136 & 0.0018 & 0.0018 & - & - & (1) &           AO\\ 
2012.371 & 17.0 & -0.2545 & -0.0260 & 0.0028 & 0.0020 & - & - & (1) &           AO\\ 
2012.562 & 17.0 & -0.2570 & -0.0285 & 0.0018 & 0.0017 & - & - & (1) &           AO\\ 
2013.318 & 16.7 & -0.2523 & -0.0391 & 0.0033 & 0.0020 & - & - & (1) &           AO\\ 
2013.360 & - & - & - & - & - & -111 &  25 & (1) &           AO\\ 
2013.550 & 16.8 & -0.2555 & -0.0413 & 0.0021 & 0.0020 & - & - & (1) &           AO\\ 
\enddata 
\label{tab:S0-38_measurements}

\tablenotetext{a}{Measurement originally reported in (1) this work and (2) \citet{2009Gillessen}.}
\tablenotetext{b}{$V_z$ values listed here are in the local standard of rest reference frame.}

\end{deluxetable} 
\clearpage

\subsection{Gravitational Potential Parameters}

\subsubsection{Values Derived from the Orbital Fit of S0-38 Alone}

\begin{figure}[h!]
\begin{center}
\includegraphics[width=1.0\columnwidth]{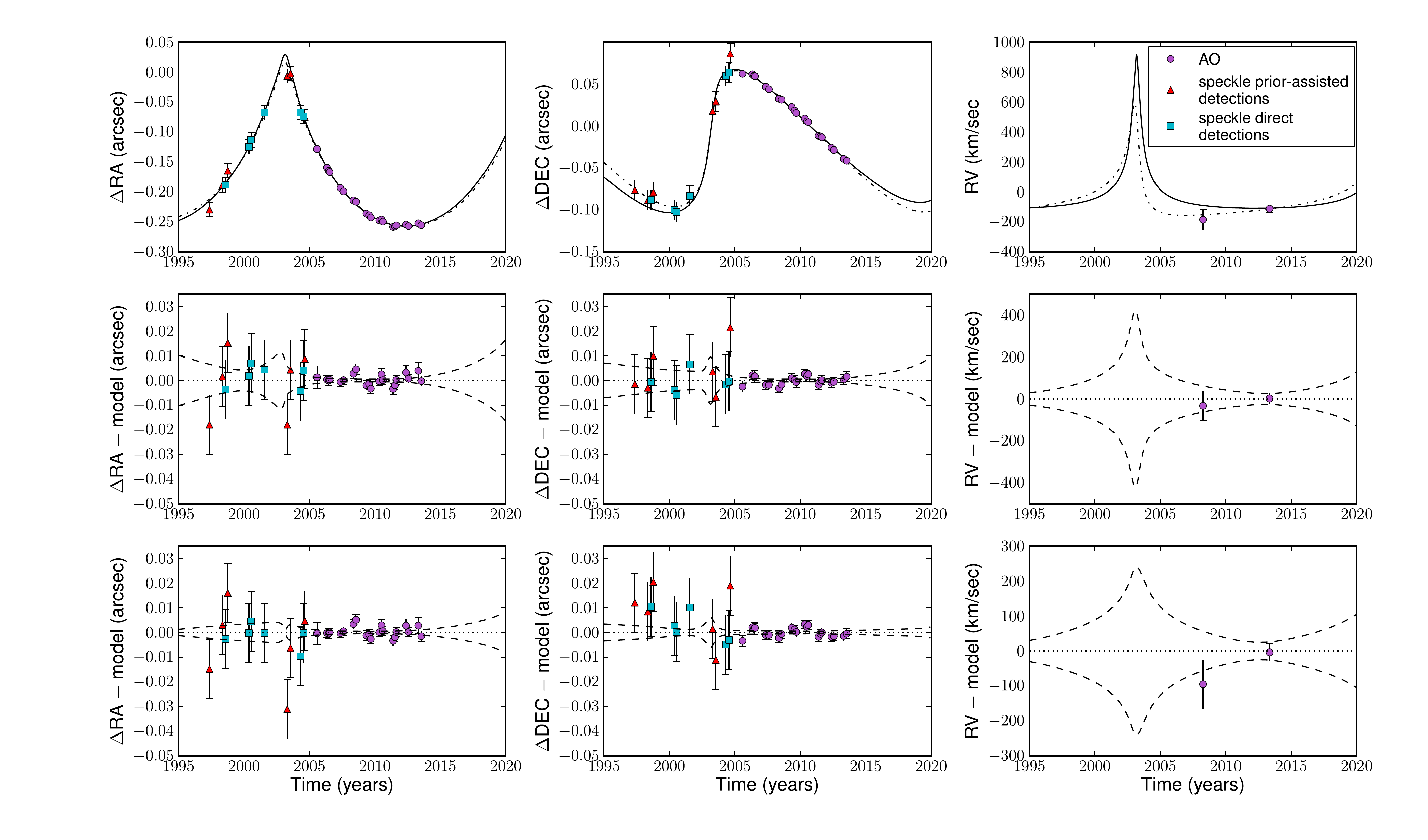}
\caption{\label{fig:S0-38_finalorbit}
Orbital fits of all speckle holography and AO astrometry and radial velocity measurements of the short-period star S0-38.  The top row shows the measurements of S0-38's position and radial velocity along with the best-fit models of S0-38's orbit as determined by the fit of S0-38 alone (dash-dotted line) and the simultaneous fit of S0-38 and S0-2 (solid line).  
The middle and bottom rows show the residuals of the position and radial velocity measurements from the best-fit models.  
The middle row shows the residuals from the fit of S0-38 alone and the bottom rows shows the residuals from the simultaneous fit of S0-2 and S0-38.  
The dashed lines in these rows indicate the 1-sigma uncertainties in the best fit models.  
In all plots, the detections of S0-38 are shown in different colors and shapes depending on the method in which they were detected: AO detections are purple circles, direct detections from the speckle holography images are teal squares, and prior-assisted detections from the speckle holography images are red triangles.  The astrometry of S0-38 is plotted as the offset from the position of Sgr A*-radio, which is defined as the origin of our absolute reference frame.}
\end{center}
\end{figure}

\begin{figure}[h!]
\begin{center}
\includegraphics[width=1.0\columnwidth]{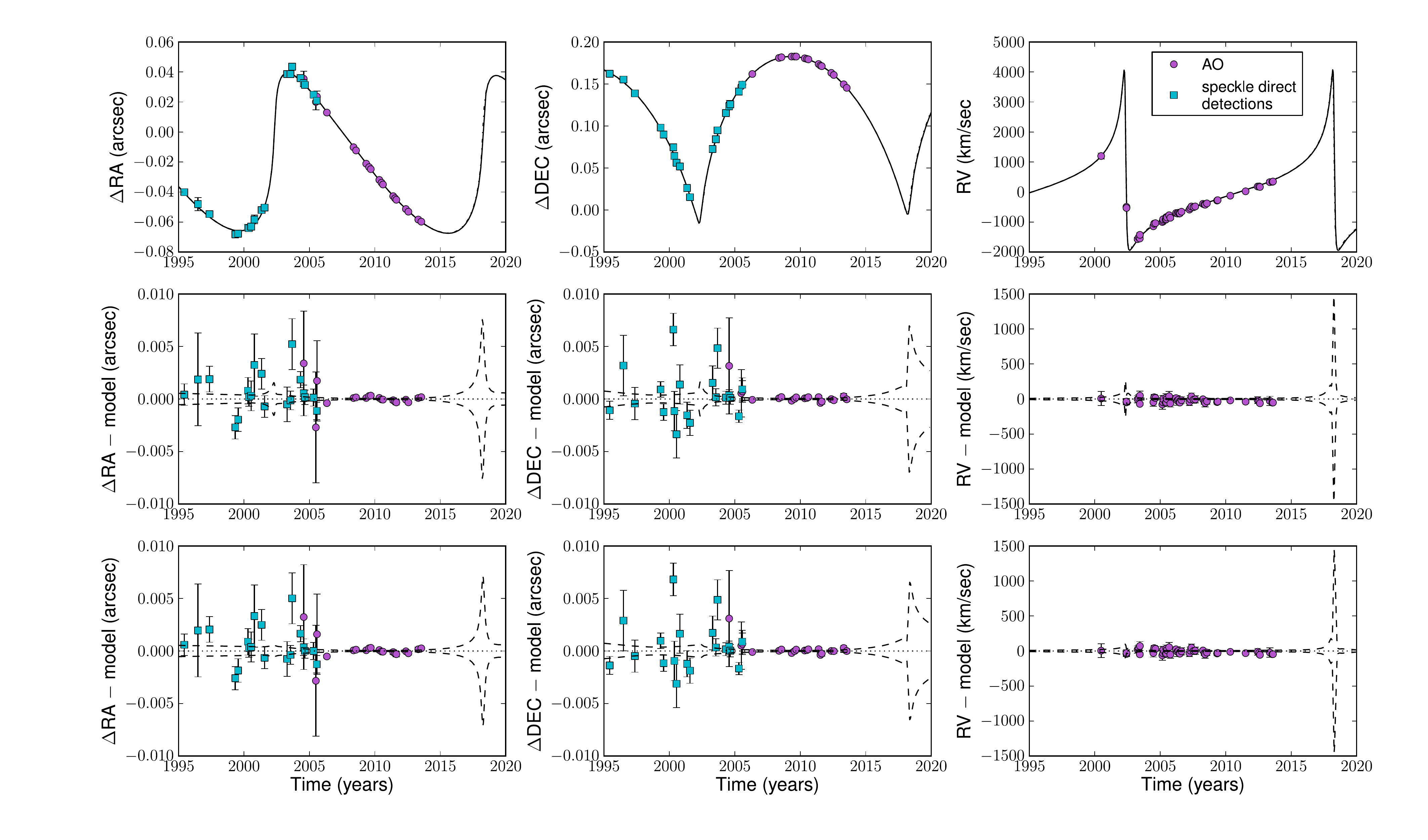}
\caption{\label{fig:S0-2_finalorbit}
Orbital fits of all speckle holography and AO astrometry and radial velocity measurements of the short-period star S0-2.  The top row shows the measurements of S0-2's position and radial velocity along with the best-fit model of S0-2's orbit as determined by the fit of S0-2 alone (dash-dotted line) and the simultaneous fit of S0-38 and S0-2 (solid line).  Note that these twzo best-fit models are very similar on the scale of the plots shown here.  
The middle and bottom rows show the residuals of the position and radial velocity measurements from the best-fit models.  
The middle row shows the residuals from the fit of S0-2 alone and the bottom rows shows the residuals from the simultaneous fit of S0-2 and S0-38.  
In all plots, the detections of S0-2 are shown in different colors and shapes depending on the method in which they were detected: AO detections are purple circles and direct detections from the speckle holography images are teal squares.  The astrometry of S0-2 is plotted as the offset from the position of Sgr A*-radio, which is defined as the origin of our absolute reference frame.  Note that the uncertainty in the prediction of the radial velocity of S0-2 around its next time of closest approach in 2018 exceeds the limits of the y axis of the lower right plot.  The uncertainty in the RV at this time ranges between $\pm1400$ km s$^{-1}$ relative to the best-fit model.}
\end{center}
\end{figure}

We fit the orbit of S0-38 alone, leaving the orbital parameters as well as the black hole parameters free, to compare with the fit of S0-2's orbit alone (discussed in Appendix \ref{appendix:S0-2}).  The motion of S0-38 alone is well fit by a Keplerian orbital model.  Comparisons of the best-fit model from the fit S0-38's data alone are shown in Figure \ref{fig:S0-38_finalorbit} (middle and bottom rows).  Similarly, the best fit to S0-2's data alone is displayed in Figure \ref{fig:S0-2_finalorbit}.  

\begin{figure}[h]
\begin{center}
\includegraphics[width=0.8\columnwidth]{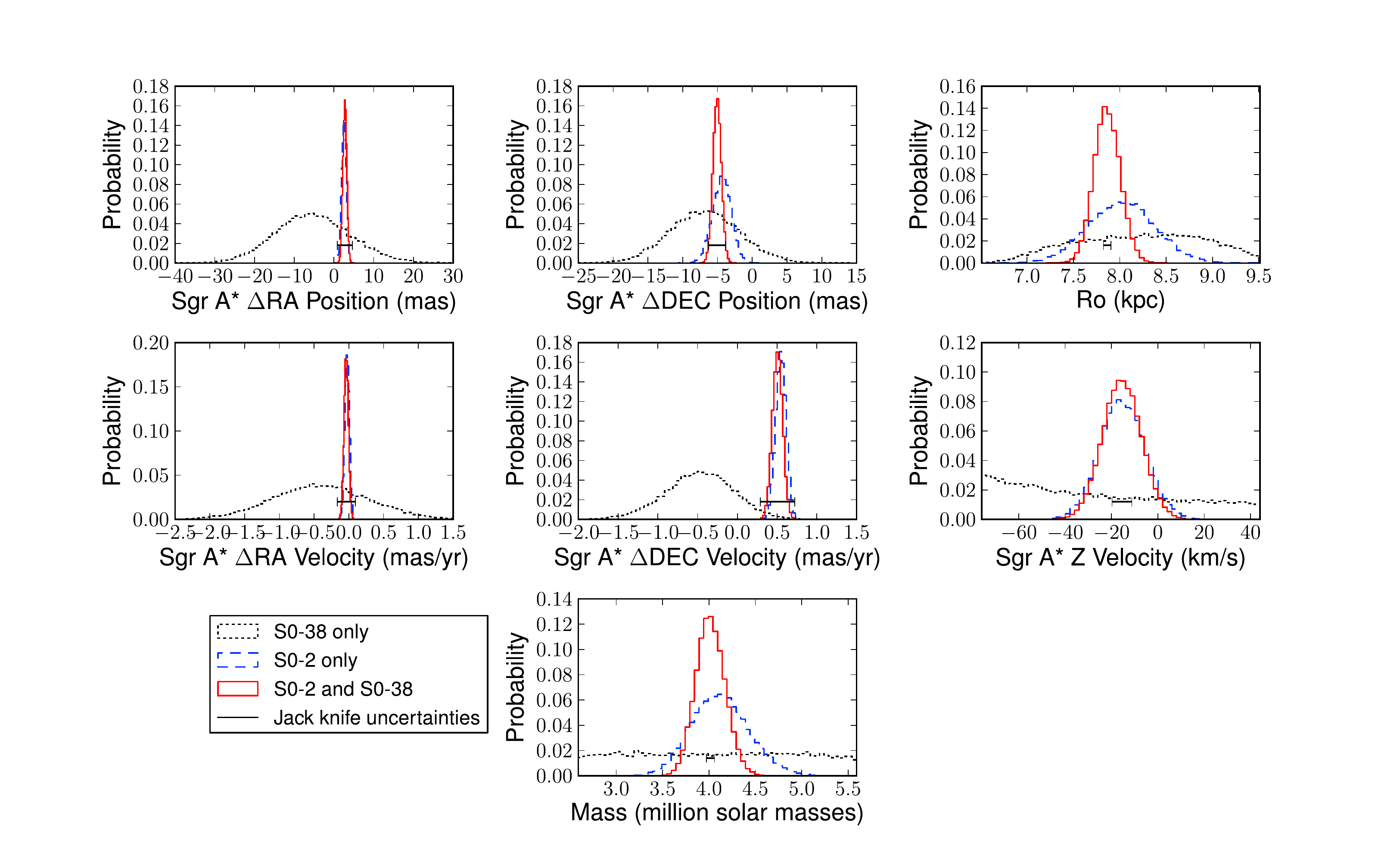}
\caption{\label{fig:BHparams}Probability distributions of the 7 black hole parameters as determined by an orbital fit of S0-2 alone (blue dashed lines), a fit of S0-38 alone (black dotted lines), and a combined orbital fit of S0-2 and S0-38 (red solid lines).  All three probability distributions in each panel are shifted by the bias determined in the jack knife analysis on the construction of the absolute reference frame (see Appendix \ref{appendix:jackknife}).  This shift is also applied in all subsequent probability distribution plots.  The black, horizontal error bars show the uncertainties in this bias shift.  The addition of S0-38 to the orbital fit increases the constraints on the black hole parameters, most notably the $R_o$ and the mass of Sgr A*.}
\end{center}
\end{figure}

\begin{figure}[h]
\begin{center}
\includegraphics[width=0.4\columnwidth]{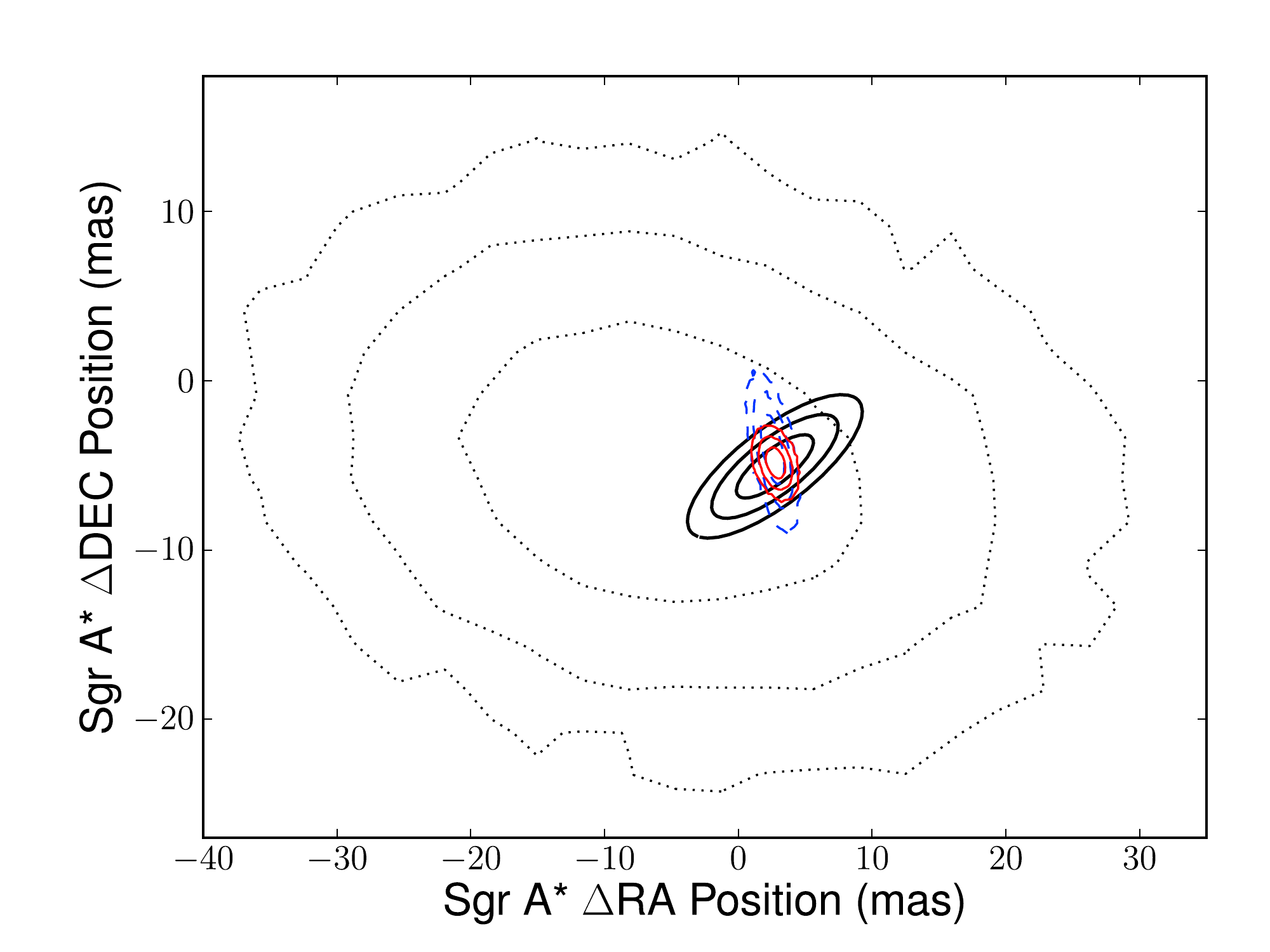}
\includegraphics[width=0.4\columnwidth]{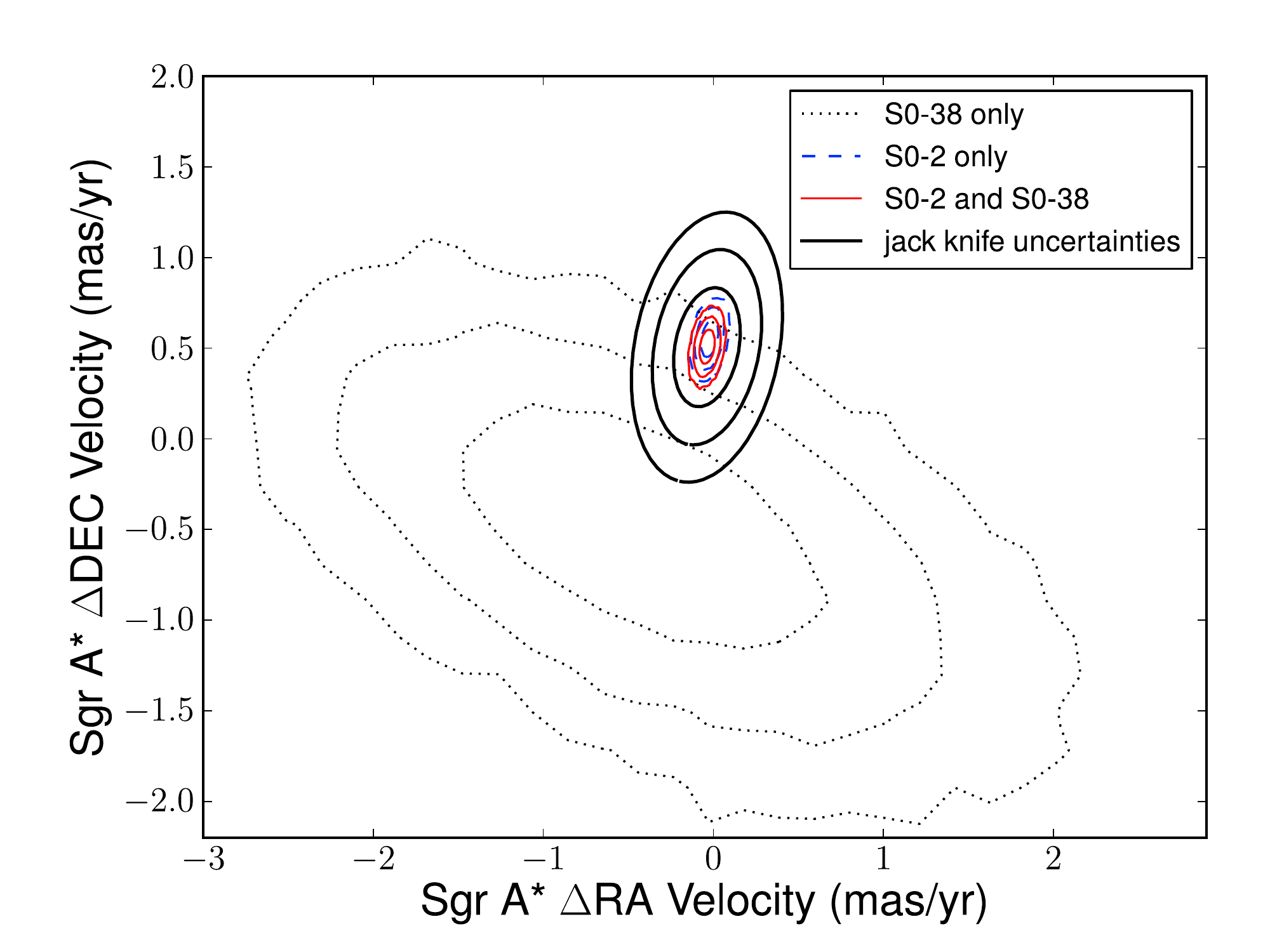}
\caption{\label{fig:SgrAPosVel}Joint probability distribution of $x_o$ and $y_o$ (\emph{left}), the position of Sgr A* on the plane of the sky, and the joint probability of $V_x$ and $V_y$, the velocity of Sgr A* on the plane of the sky (\emph{right}).  The joint probability distributions as determined by three orbit fits are shown: S0-2 alone (blue, dashed line), S0-38 alone (black, dotted line), and S0-2 and S0-38 together (red, solid line).  All the probability distributions are shifted by the bias determined in the jack knife analysis (see Appendix \ref{appendix:jackknife}).  The black, solid lines show the 1, 2, and 3-sigma uncertainties in this bias shift.  
}
\end{center}
\end{figure}

The orbital model fit of S0-38's motion alone provides independent estimates the black hole potential parameters $x_o$, $y_o$, $V_x$, and $V_y$ (see Figure \ref{fig:BHparams} and Table \ref{tab:bestfit_BHandorbitparams}).  The other black hole parameters ($V_z$, $R_o$, $M_{bh}$) rely on radial velocity measurement  and are not well determined.  This is because S0-38 has only two radial velocity measurements and these RV values are close to 0 km s$^{-1}$.  These three parameters are still left free in this orbital fit of S0-38, but the uniform priors applied to these parameters are set to ranges that fully encompass at least the $\pm4\sigma$ uncertainties derived from the fit of S0-2 alone (see Table \ref{tab:bestfit_BHandorbitparams}). Those that are independently well fit are those that only require astrometric measurements ($x_o$, $y_o$, $V_x$, $V_y$; see Figure \ref{fig:SgrAPosVel}).  The value for these parameters preferred by S0-38 agree with the values given by S0-2 within 1 sigma, except for the $V_y$ parameter which differs by 2 sigma.  Note that reference frame uncetainties are a significant additional error source for $x_o$, $y_o$, $V_x$, and $V_y$ (see Figure \ref{fig:SgrAPosVel}), but are negligible for $M_{bh}$ and $R_o$ (see Appendix \ref{appendix:jackknife}).

\begin{deluxetable}{lcrrr}
\tabletypesize{\scriptsize}
\tablewidth{0pt}
\tablecaption{Best-Fit Black Hole and Orbital Parameters as Derived From the Fit of S0-2 alone, S0-38 alone, and the Simultaneous Fit of S0-2 and S0-38}
\tablehead{
  \colhead{} & 
  \colhead{} & 
  \multicolumn{3}{c}{Best-Fit Parameter Values from Orbital Fits\tablenotemark{a}} \\
  \colhead{Model Parameter (units)} & 
  \colhead{} & 
  \colhead{S0-2 only} & 
  \colhead{S0-38 only} & 
  \colhead{S0-2 and S0-38}\\
}
\startdata
Black Hole Properties:& & & \\ 
                     Distance (kpc) &&$ 8.02  \pm 0.36  \pm0.04  $&$ [6.5, 9.5]\tablenotemark{b} $&$ 7.86  \pm 0.14  \pm0.04   $\\ 
             Mass ($10^6M_{\odot}$) &&$ 4.12  \pm 0.31  \pm0.04  $&$ [2.5, 5.5]\tablenotemark{b}$&$ 4.02  \pm 0.16  \pm0.04   $\\ 
       $X$ Position of Sgr A* (mas) &&$ 2.52  \pm 0.56  \pm1.90  $&$ -5.25  \pm 9.41  \pm1.90  $&$ 2.74  \pm 0.50  \pm1.90   $\\ 
       $Y$ Position of Sgr A* (mas) &&$ -4.37  \pm 1.34  \pm1.23  $&$ -6.85  \pm 5.00  \pm1.23  $&$ -5.06  \pm 0.60  \pm1.23   $\\ 
              $X$ Velocity (mas/yr) &&$ -0.02  \pm 0.03  \pm0.13  $&$ -0.40  \pm 0.70  \pm0.13  $&$ -0.04  \pm 0.03  \pm0.13   $\\ 
              $Y$ Velocity (mas/yr) &&$ 0.55  \pm 0.07  \pm0.22  $&$ -0.48  \pm 0.43  \pm0.22  $&$ 0.51  \pm 0.06  \pm0.22   $\\ 
              $Z$ Velocity (km/sec) &&$ -15  \pm 10  \pm4  $&$ [-80, 40]\tablenotemark{b}$&$ -15.48  \pm 8.36  \pm4.28   $\\ 
\\ 
S0-2 Properties:& & & \\ 
                        Period (yr)  &&$ 15.90 \pm 0.04 $&&$ 15.92 \pm 0.04  $\\ 
      Time of Closest Approach (yr)  &&$ 2002.343 \pm 0.008 $&&$ 2002.347 \pm 0.003  $\\ 
                       Eccentricity  &&$ 0.890 \pm 0.005 $&&$ 0.892 \pm 0.002  $\\ 
                  Inclination (deg)  &&$ 134.7 \pm 0.9 $&&$ 134.2 \pm 0.3  $\\ 
         Argument of Periapse (deg)  &&$ 66.5 \pm 0.9 $&&$ 66.8 \pm 0.5  $\\ 
  Angle to the ascending node (deg)  &&$ 227.9 \pm 0.8 $&&$ 228.0 \pm 0.5  $\\ 
\\
S0-38 Properties:& & & \\ 
                        Period (yr)  &&& $ 19.1 \pm 0.6 $&$ 19.2 \pm 0.2  $\\ 
      Time of Closest Approach (yr)  &&& $ 2003.1 \pm 0.1 $&$ 2003.19 \pm 0.04  $\\ 
                       Eccentricity  &&& $ 0.80 \pm 0.02 $&$ 0.810 \pm 0.004  $\\ 
                  Inclination (deg)  &&& $ 170 \pm 4 $&$ 170 \pm 3  $\\ 
         Argument of Periapse (deg)  &&& $ 20 \pm 30 $&$ 12 \pm 21  $\\ 
  Angle to the ascending node (deg)  &&& $ 110 \pm 30 $&$ 95 \pm 20  $\\ 
\\ 
                Reduced Chi-Squared   &&  1.5   &  0.9    &  1.3\\ 
\enddata 
\label{tab:bestfit_BHandorbitparams}
\tablenotetext{a}{All best-fit values are shifted by the bias determined by the jack knife analysis of the reference frame (see Appendix \ref{appendix:jackknife}).  The first error term for each best fit value corresponds to the error determined by the orbital fit.  For the black hole parameters, the second error term corresponds to jack knife uncertainty from the reference frame.}
\tablenotetext{b}{These parameters are not well-determined by S0-38 alone, so instead of a best fit the uniform prior ranges that were used in the fit are reported.}

\end{deluxetable}
\clearpage

\subsubsection{Values Derived from the Simultaneous Orbital Fit of S0-38 and S0-2}
\label{BHconstraints}
We fit the orbits of S0-38 and S0-2 simultaneously to constrain the gravitational potential.  Using the information from both of these stars gives the final values of $M_{bh}$ and $R_o$.  
Figure \ref{fig:S0-38_finalorbit} and Figure \ref{fig:S0-2_finalorbit} (top and bottom rows) show the resulting best-fit model orbits for S0-38 and S0-2 respectively in $\Delta$R.A., $\Delta$Decl., and radial velocity versus time.  The astrometric data are shown in an absolute reference frame that defines Sgr A*-radio to be at rest at the origin.  These figures also give the residuals of these data from the model.  The residuals show that orbital models fit the data well within the uncertainties, with a reduced chi-squared of 1.3.  
Table \ref{tab:bestfit_BHandorbitparams} shows the best fit and errors on the black hole and orbital parameters of S0-2 and S0-38 as determined by the simultaneous fit of these stars.  

The best-fit black hole parameters presented in Table \ref{tab:bestfit_BHandorbitparams} from the fits of S0-2 alone, S0-38 alone, and S0-2 and S0-38 simultaneously are taken as the weighted average of the {\sc MultiNest} samples for each parameter.  The fitting error is then taken as the weighted standard deviation of the {\sc MultiNest} samples.  The best-fit values are then shifted by the bias determined by a jack knife analysis on the 7 SiO masers used to construct the absolute reference frame (see Appendix \ref{appendix:jackknife} for details).  The error in this bias shift is also presented as a second error term in Table \ref{tab:bestfit_BHandorbitparams}.  All probability distributions of the 7 black hole parameters presented in this work include the bias shift.  Note that the jack knife uncertainties due to the absolute reference construction are small compared to the fitting errors for $M_{bh}$ and $R_o$, so systematics in the reference frame do not greatly affect the values of these two key black hole parameters (see Figure \ref{fig:BHparams}).  Additionally, the black hole position and velocity on the sky ($x_o$, $y_o$, $V_x$, and $V_y$) as determined by the fit of S0-2 and S0-38 are offset from 0.0, but they become consistent with 0.0 when the reference frame errors are included (see Figure \ref{fig:SgrAPosVel}).

\begin{figure}[ht]
\centering
\includegraphics[width=0.4\columnwidth]{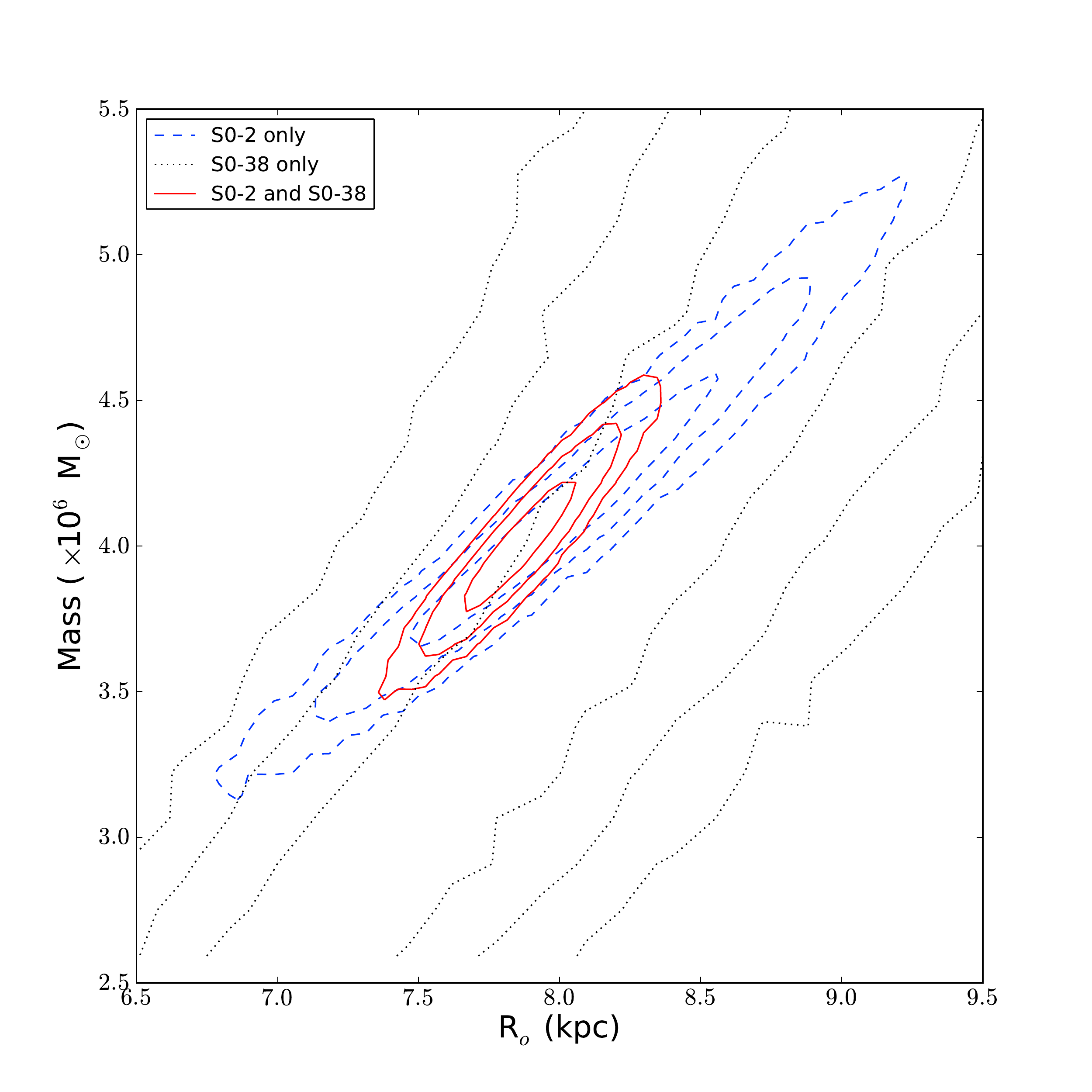}
\includegraphics[width=0.4\columnwidth]{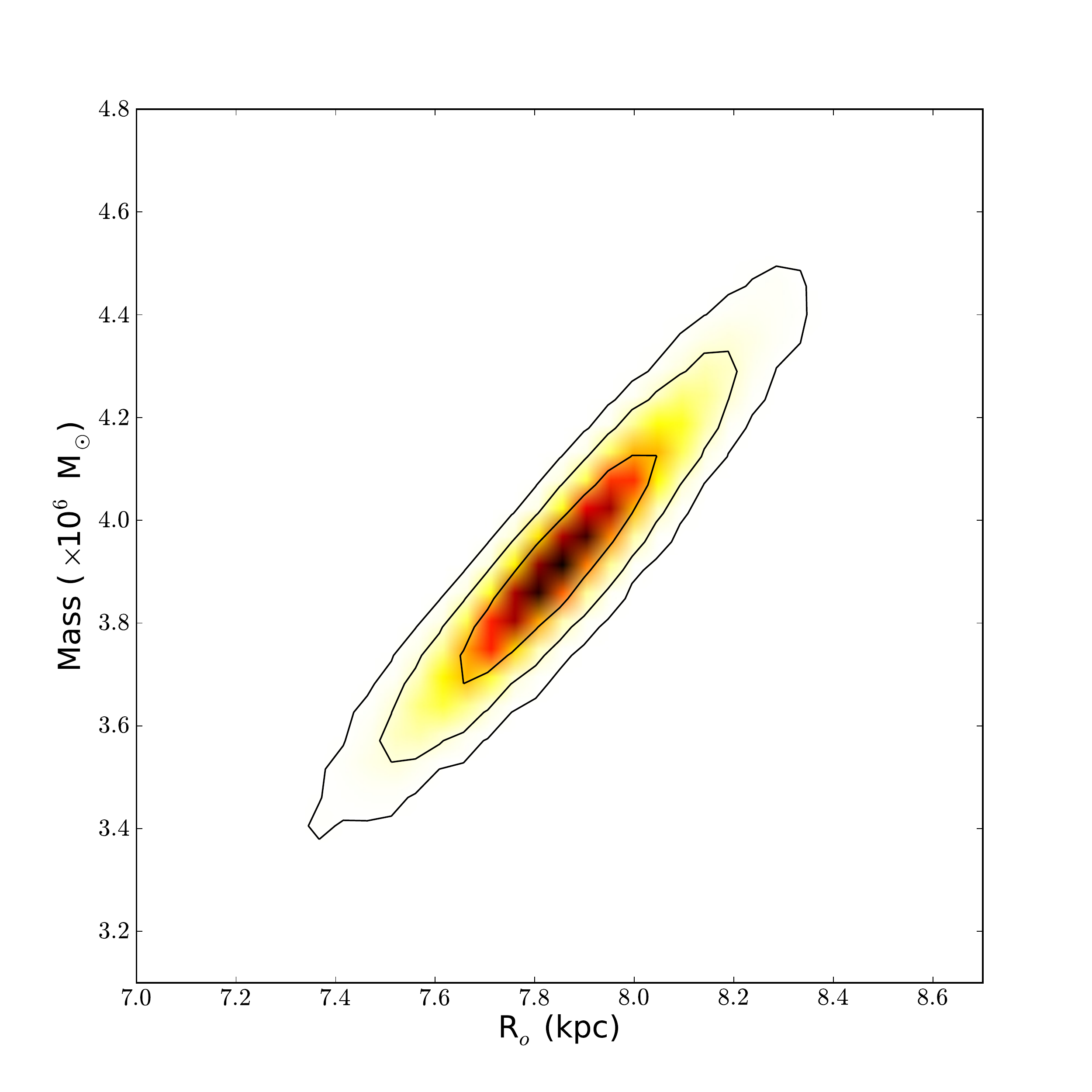}
 \caption{\label{fig:MassRo}
\emph{Left}: 2D joint probability distribution with 1-, 2-, and 3-sigma contours of $M_{bh}$ and R$_{o}$ as derived by the orbital fit of S0-2 alone (blue dashed lines), S0-38 alone (black dotted line), and a simultaneous fit of S0-2 and S0-38, with the new speckle holography detections of S0-38 included (red solid lines).  The orientation of S0-38's orbit on the plane of the sky leads to a different correlation of $M_{bh}$ and $R_o$, resulting in small errors on these parameters when S0-2 and S0-38 are fit simultaneously.  The precision of our measurements of $M_{bh}$ and R$_{o}$ has increased by a factor of $\sim$2 and 2.5 respectively compared to S0-2 alone.  Note that the $y_o$ parameter (not visualized here) is also highly correlated with $M_{bh}$ and $R_o$ (see Section \ref{discussion1}).  
\emph{Right}: Final joint probability distribution of $M_{bh}$ and $R_o$ as derived by the simultaneous fit of S0-2 and S0-38.}
\end{figure}

The addition of S0-38 leads to a better constraint of the black hole parameters compared to using the information from S0-2 alone.  The best-fit black hole parameters and the errors on these values for the simultaneous fit of S0-2 and S0-38 as well as the fit of S0-2 alone are shown in Table \ref{tab:bestfit_BHandorbitparams}.  The errors on the mass of Sgr A* and the distance to the Galactic Center decrease by a factor of $\sim$2 and 2.5 respectively.  The best fit values of $M_{bh}$ and $R_o$ as determined from the combined S0-2 and S0-38 orbital fit are: 
$M_{bh} = 4.02 \pm 0.16 \pm 0.04$ $\times 10^6 M_{\odot}$ and $7.86  \pm 0.14  \pm0.04$ kpc.
Figure \ref{fig:MassRo} shows the joint probability distributions of $M_{bh}$ and $R_o$ as determined by S0-2 alone, S0-38 alone, and by S0-2 and S0-38 together.

To confirm that the estimated uncertainties on $M_{bh}$ and $R_o$ from the simultaneous fit of S0-2 and S0-38 are reasonable given the measurement uncertainties, we compare our estimated uncertainties to the distributions of the uncertainties on $M_{bh}$ and $R_o$ derived from the 100 mock data sets.  These mock data sets are simulated assuming the measurement uncertainties are only statistical.  The estimated uncertainties on $M_{bh}$ and $R_o$ from the real data are well within the distributions of uncertainties derived from the simulated data sets, with values smaller on average than the widths resulting from the mock data by only 1- and 0.7-sigma respectively.  In addition, the $R_o$ uncertainty derived from the real data falls in the most likely bin of the simulated uncertainties.  We therefore conclude that the estimated uncertainties on $M_{bh}$ and $R_o$ are reasonable given the measurement uncertainties.  

\begin{figure}[ht]
\begin{center}
\includegraphics[width=0.7\columnwidth]{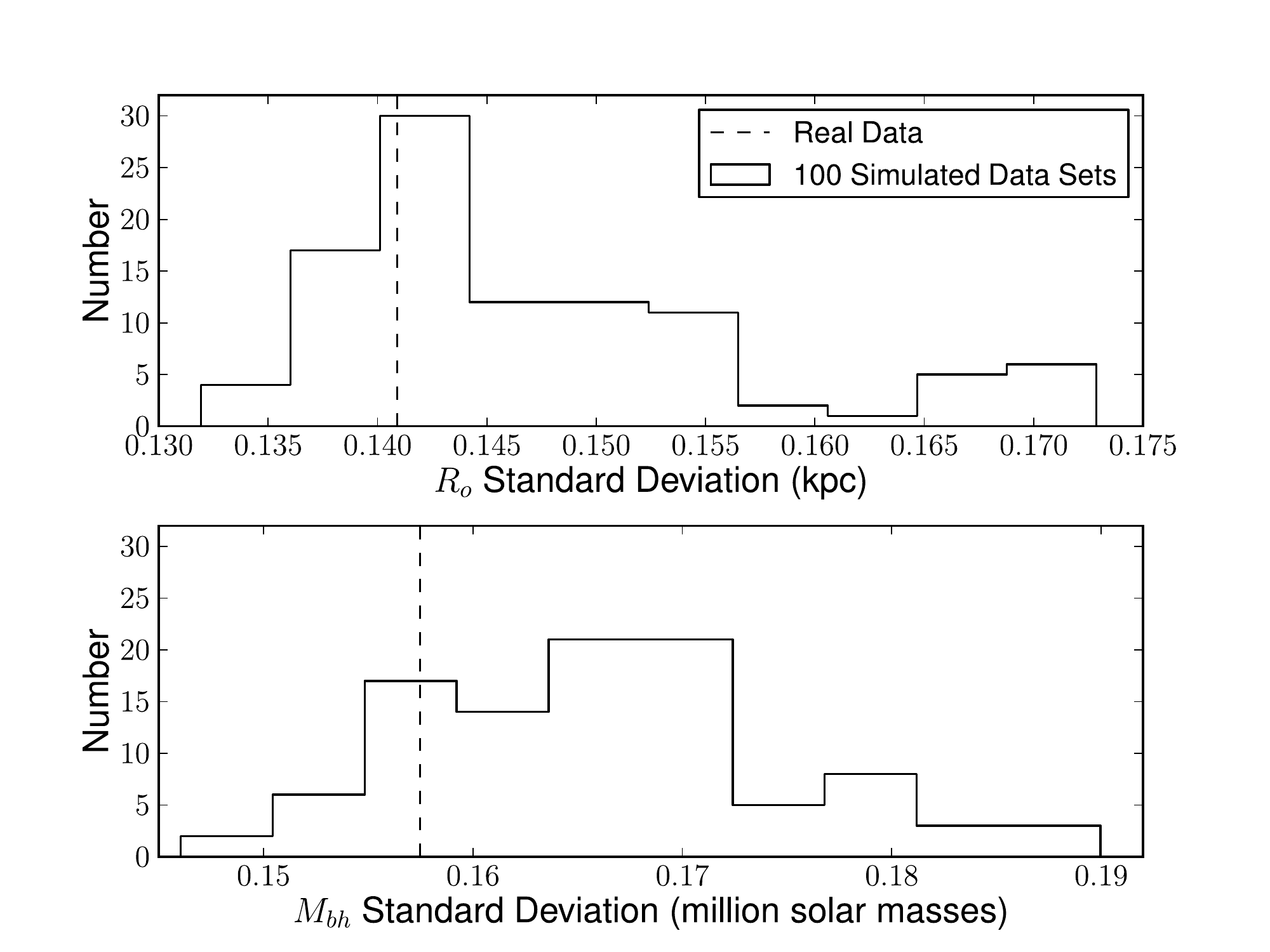}
\caption{\label{fig:simulations}
Histogram of standard deviation of posterior samples of $M_{bh}$ and $R_o$ generated by the {\sc MultiNest} fit of 100 fake sets of S0-2 and S0-38 data.  The standard deviation quantifies the width of the $M_{bh}$ and $R_o$ probability distributions.  The 100 data sets were created by assuming only statistical measurement uncertainties.  The vertical dashed lines show standard deviations of $M_{bh}$ and $R_o$ samples generated by the fit of the real S0-2 and S0-38 data, which are smaller on average than the simulated standard deviations by 1- and 0.7-sigma respectively.  
}
\end{center}
\end{figure}

\begin{figure}[h]
\centering
\includegraphics[width=0.5\columnwidth]{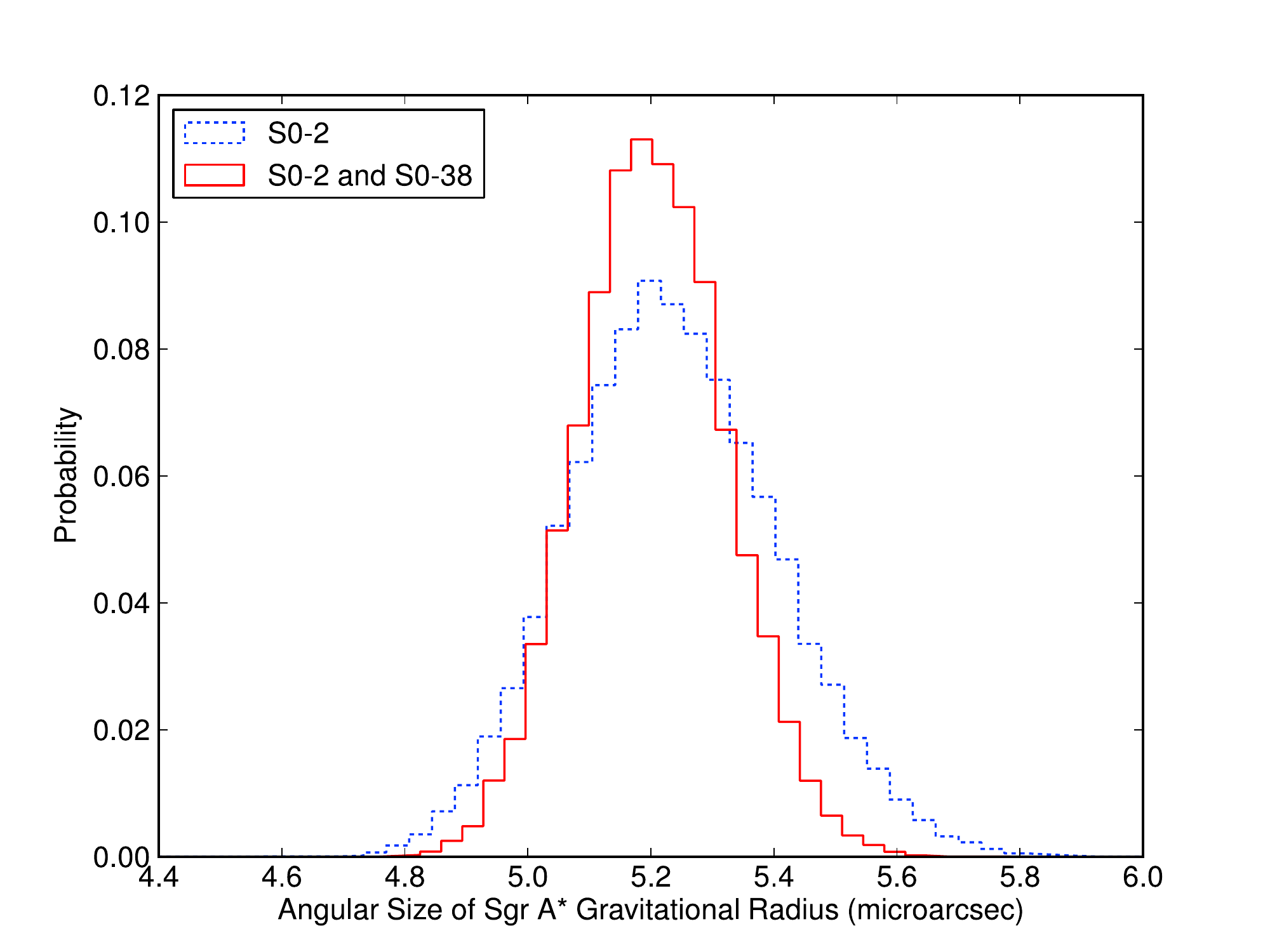}
 \caption{\label{fig:GravRadius}
Probability distribution of the angular size of the gravitational radius of Sgr A*: $GM_{bh}/c^2R_o$.  The distribution is calculated from the $M_{bh}$ and $R_o$ distributions that are shifted by the negligible bias determined in the jack knife analysis (see Appendix \ref{appendix:jackknife}).}
\end{figure}

The mass-to-distance ratio of Sgr A*, $M_{bh}/R_o$
sets the apparent size of the black hole's gravitational radius at the distance of $R_o$ to Sgr A*: $r_g = GM_{bh}/c^2R_o$.  
The probability distribution of this value as determined by the orbital fit of S0-2 and S0-38 is shown in Figure \ref{fig:GravRadius}.  
We find that the best fit angular size of the gravitational radius of Sgr A* is $5.20 \pm 0.12 \pm 0.06$, where the second error term is the estimated uncertainty in the reference frame, jack knife bias shifts of $M_{bh}$ and $R_o$ propagated to the gravitational radius.

\begin{figure}[h]
\centering
\includegraphics[width=0.5\columnwidth]{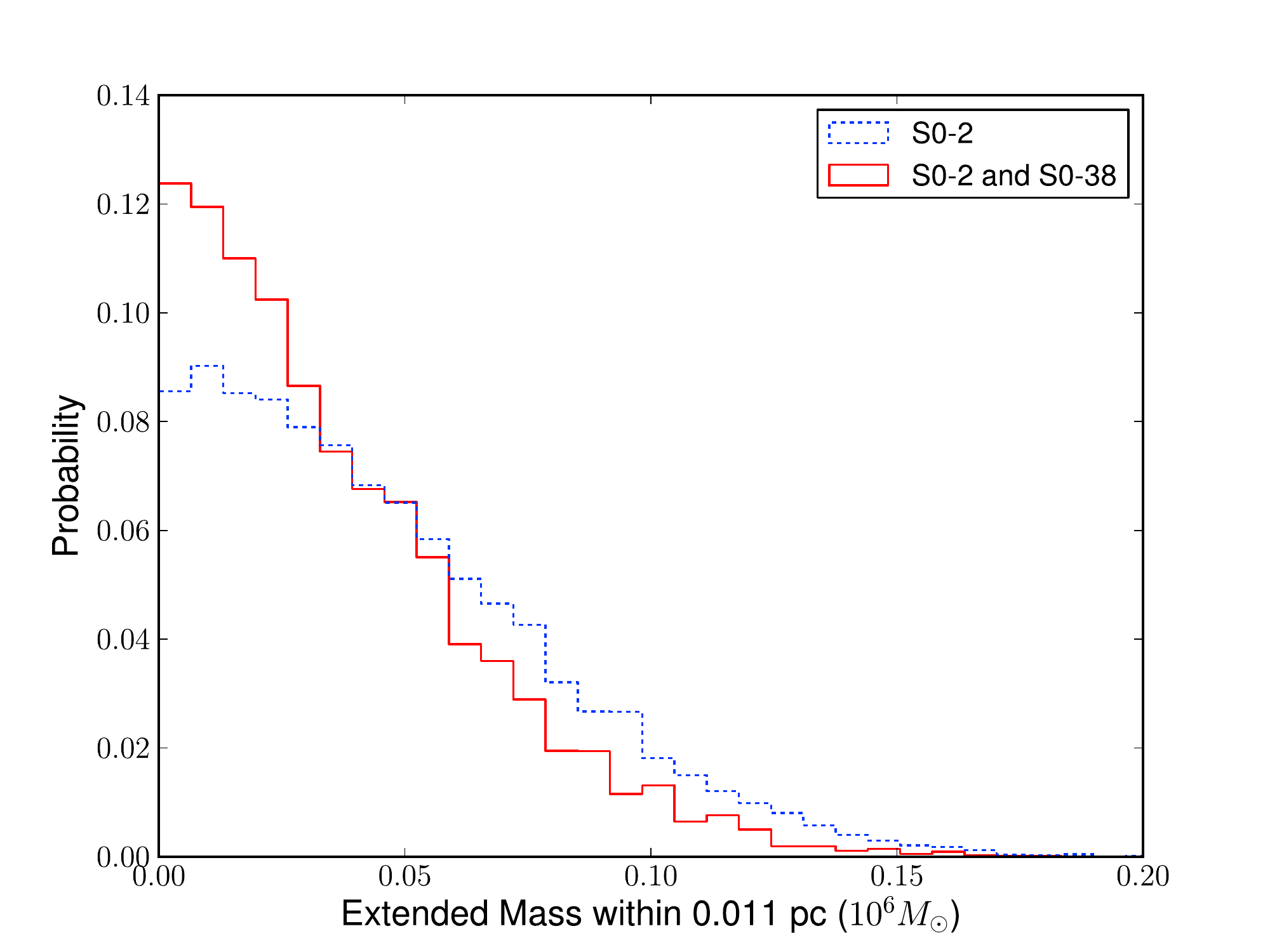}  
 \caption{\label{fig:extmass_compare} The probability distribution of the extended dark mass within 0.011 pc of the supermassive black hole, as determined by the fit of S0-2 alone (blue dotted line) and by the simultaneous fit of S0-2 and S0-38 (red solid line).  The 99.7\% confidence upper limit of the extended mass component decreases by $\sim$10\% when the information contained in S0-38's orbital motion is added.}
\end{figure}

The additional information found in S0-38's orbital motion also gives increased constraints on extended dark mass within the apoapse distances of its and S0-2's orbits.  
In addition to the central supermassive black hole, a central cluster of stellar mass black holes and/or other compact objects has been predicted and theoretically explored (e.g., \citealt{1993ApJ...408..496M}; \citealt{2000ApJ...545..847M}; \citealt{2006ApJ...649...91F}; \citealt{2010ApJ...718..739M}).  The amount of extended dark mass within the orbits of S0-2 and S0-38 influences their orbital motion.  
Figure \ref{fig:extmass_compare} shows the probability distribution of the extended mass contained within this radius for a fit of S0-2 alone and a simultaneous fit of S0-2 and S0-38.  The resulting 68.3\% confidence ($1\sigma$) upper limits on the $M_{ext}$ within 0.011 pc are 0.06 $\times 10^6  M_{\odot}$ for the S0-2 only fit and 0.05 $\times 10^6 M_{\odot}$ for the simultaneous S0-2 and S0-38 fit.  The 99.7\% confidence ($3\sigma$) upper limits for these fits are 0.16 and 0.15 $\times 10^6 M_{\odot}$ respectively.

\section{Discussion}
\label{discussion}

\subsection{Improvements with the Additional Information from S0-38}
\label{discussion1}

We have demonstrated the power of our new methodology of searching for speckle holography detections of S0-38 using the information in the AO detections by applying it to S0-38.  The combination of S0-2 and S0-38 leads to a significant improvement in our knowledge of $R_o$ and the mass of Sgr A* compared to using the information from S0-2 alone.  
The reasons why the addition of S0-38 improves the constraints on $M_{bh}$ and $R_o$ so much are described here.

\begin{figure}[h]
\begin{center}
\includegraphics[width=0.5\columnwidth]{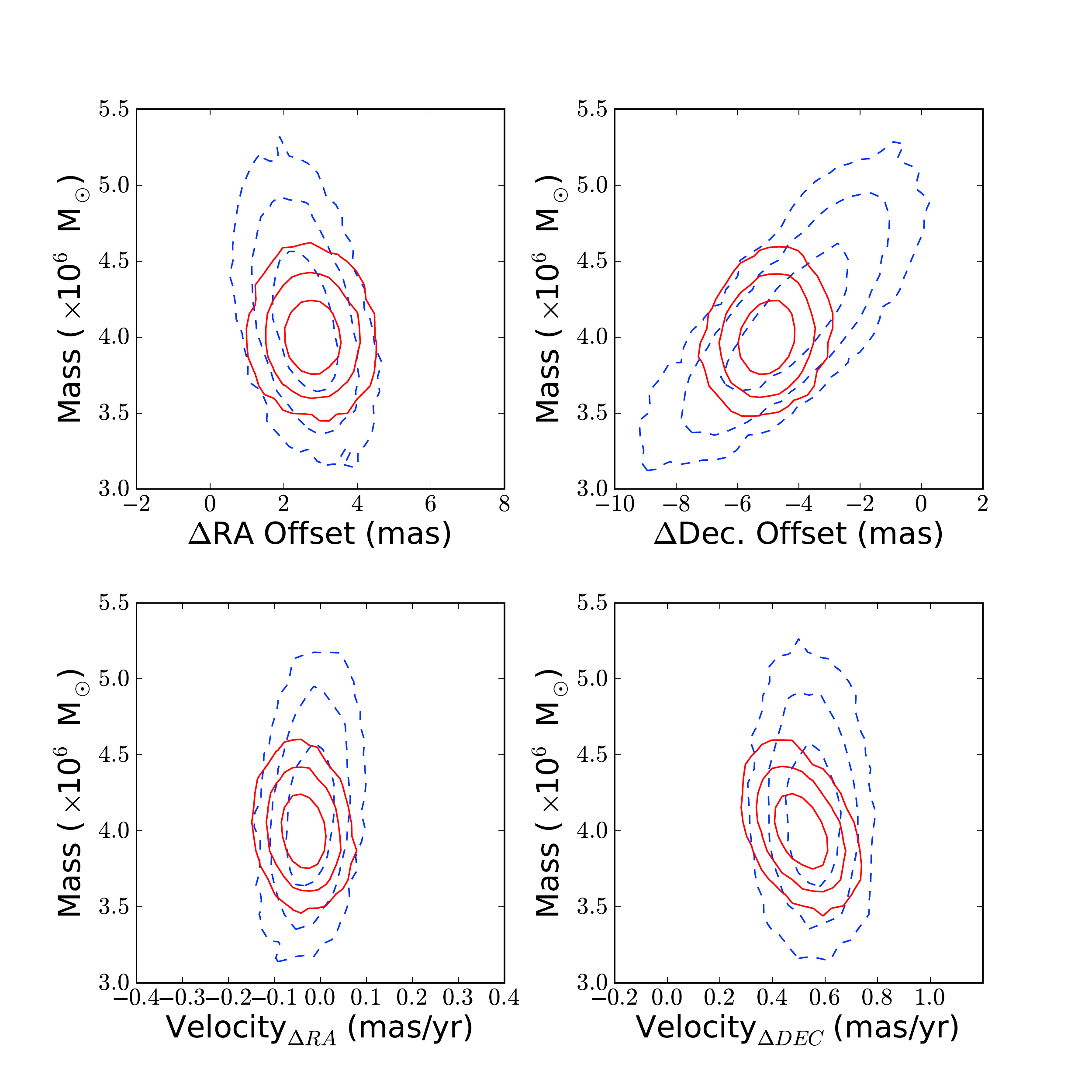}
\includegraphics[width=0.5\columnwidth]{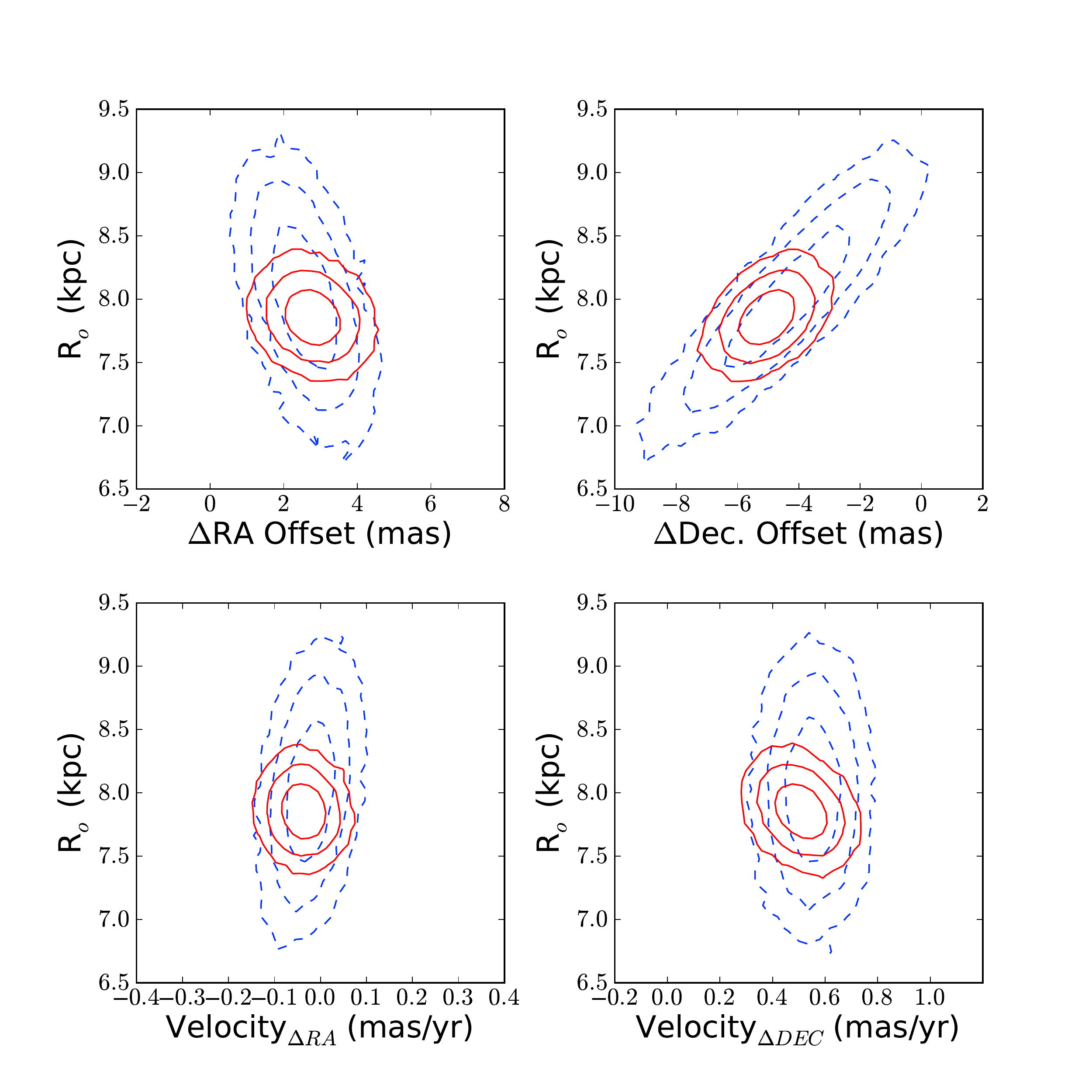}
\caption{\label{fig:MassRo_PosVel}
Joint probability distributions of $M_{bh}$ (\emph{top}) and $R_o$ (\emph{bottom}) and $x_o$, $y_o$, $V_x$, and $V_y$ as determined by the orbital fit of S0-2 alone (blue dotted line) and by the simultaneous orbital fit of S0-2 and S0-38, with all new speckle holography detections included (red solid line).  
These probability distributions are shifted by the bias determined in the jack knife analysis (see Appendix \ref{appendix:jackknife}).  The correlation between these sets of parameters in the case of S0-2 alone shows that our knowledge of $M_{bh}$ and $R_o$ is limited by how well S0-2 constrains $y_o$.  With the addition of the measurements of S0-38's position and RV, $y_o$, and therefore also $M_{bh}$ and $R_o$, are much better determined.  The errors in $M_{bh}$ and $R_o$ in the combined S0-2 and S0-38 fit are $\sim$2 and $\sim$2.5 times smaller than in the S0-2-only fit respectively.}
\end{center}
\end{figure}

The results of fitting S0-38 alone show that S0-38 primarily gives information about the position and velocity of Sgr A* on the plane of the sky ($x_o$, $y_o$, $V_x$, and $V_y$).  
Figure~\ref{fig:MassRo_PosVel} shows the joint probability distributions of these four parameters with $M_{bh}$ and $R_o$ from the fit of S0-2 alone and in the simultaneous fit of S0-2 and S0-38.  
In the fit of S0-2 alone, $M_{bh}$ and $R_o$ are both correlated with $y_o$, demonstrating that a better constraint on $y_o$ would lead to a better constraint on $M_{bh}$ and $R_o$.  In the simultaneous fit, the errors on $y_o$ and thereby $M_{bh}$ and $R_o$ are significantly reduced.  So it is specifically through improved knowledge of $y_o$ that S0-38 decreases the errors on $M_{bh}$ and $R_o$ when simultaneously fit with S0-2.  
The fact that $M_{bh}$ and $R_o$ are not correlated with $V_y$ in Figure~\ref{fig:MassRo_PosVel} additionally shows that the difference in values of this parameter from the separate fits of S0-2 and S0-38 does not affect the $M_{bh}$ and $R_o$ results from the simultaneous fit.  

The reason that the addition of S0-38 gives such an improved constraint on $y_o$ is the orientation of this star's orbit on the plane of the sky.  
Due to S0-2's orientation on the plane of the sky and the fact that we must omit the detections of S0-2 around its periapse position because of confusion with the NIR counterpart of Sgr A*, S0-2 gives a better constraint on $x_o$ than $y_o$.  Figure \ref{fig:XY_S0-2_S0-38} shows the current set of astrometric measurements for S0-2, including those points left out of the fit due to confusion with Sgr A* and other known sources (indicated by open circles), leaving very few unbiased points in the lower third of the ellipse of S0-2's orbit.  
S0-38 is fortunately on an orbit that is nearly perpendicular on the plane of the sky to that of S0-2, also shown in Figure \ref{fig:XY_S0-2_S0-38}.  Therefore, this star can further constrain the $y_o$ position of Sgr A* and in turn the values of $M_{bh}$ and $R_o$. 
The best-fit solutions of $M_{bh}$ and $R_o$ for each case agree well within 1 sigma.  With the addition of the information provided by S0-38, the errors on these parameters decrease by a factor of $\sim$2 and 2.5 respectively.  

The joint probability distribution of $M_{bh}$ and $R_o$ (Figure \ref{fig:MassRo}) also shows that S0-38 contains different information than S0-2 about these parameters.  $M_{bh}$ and $R_o$ are not individually constrained by the orbital fit of S0-38 alone, but their correlation is constrained.  The correlation as determined by S0-38 alone has a different slope than the correlation determined by S0-2 alone.  This different slope leads to the additional constraints on $M_{bh}$ and $R_o$ when information from S0-38 is added.  
In the future, the addition of more stars with high orbital phase coverage and orbits perpendicular to S0-2 will lead to even smaller errors on $M_{bh}$ and $R_o$.

\subsection{Scientific Implications of New Constraints on $M_{bh}$ and $R_o$}

The values of $M_{bh}$ and $R_o$ presented here agree within uncertainties with previous measurements made using stellar orbits.  The most recent previous measurements from Keck and VLT data respectively are $4.1\pm0.4$ (\citealt{Meyer12}) and $4.30\pm0.36$ (\citealt{2010RvMP...82.3121G}) million solar masses 
for the mass of the black hole and $7.7 \pm 0.4$ (\citealt{Meyer12}) and $8.2 \pm 0.34$ (\citealt{2013Gillessen}) kpc for $R_o$.  The uncertainties derived from the orbital fit of S0-2 alone are slightly smaller than the uncertainties of these previous measurements due to the increased orbital phase coverage of this star, but the primary improvement in this work is the added information from the orbit of S0-38.  

The method of using multiple stars' orbits to determine $M_{bh}$ and $R_o$ was also used in \cite{2009Gillessen}.  
In that work, the orbits of 5 stars in addition to S0-2 were simultaneously fit to determine the gravitational potential parameters.  These 5 stars had orbital phase coverages ranging from 12 - 35\% and three of the stars had multiple radial velocity measurements.  
S0-38 was not included in this set of stars, since it had only been observed from 2005 through 2008 at the time.  
With our observations now covering over 80\% of S0-38's orbit, just including this one star in addition to S0-2 has significantly decreased the errors on $M_{bh}$ and $R_o$.  From the arguments presented in the previous section, this seems to mainly be due to S0-38's orientation on the sky, as well as the more than 80\% orbital phase coverage.  Of the 5 other stars used in \cite{2009Gillessen}, two of them have orientations $\sim45$ degrees away from S0-2's vertical orientation, but none has the perpendicular orientation of S0-38.  

The $R_o$ value presented in this work also agrees with other, recent direct measurements of the distance to the center of our galaxy within $\sim2$~sigma.  
VLBI measurements of the trigonometric parallax of H$_2$O masers in the star forming region Sgr B2 give a value of $7.9\pm0.8$ kpc (\citealt{2009ApJ...705.1548R}), while  dynamical modelling of the nuclear star cluster gives a statistical parallax distance to the Galactic Center of $8.27\pm0.13$ kpc (\citealt{2015MNRAS.447..948C}).  
Another recent indirect measurement, using VLBI parallaxes and proper motions of over 100 masers to model the motion and structure of the Milky Way, gives a comparable value and uncertainty in $R_o$ to that of the statistical parallax method: $8.34\pm0.16$ kpc (\citealt{2014ApJ...783..130R}).  
The direct statistical parallax method and indirect modelling method have similar uncertainties in $R_o$, but a value of $\sim2$~sigma higher than presented in this work.  
It remains to be seen whether continued orbital monitoring and further improvement of $R_o$ constraints from the orbits of S0-2 and S0-38 will maintain this mild disagreement.  

The direct measurement of $R_o$ presented here has implications for constraints on the structure and kinematics of our galaxy.  
\cite{2004Reid_propermotion} measured the apparent proper motion of Sgr A*, which is due only to the galactic orbit of the sun if the supermassive black hole is assumed to be at the center of the galaxy.  This measurement gave a ratio of the circular rotation speed at the radius of the Sun ($\Theta_o$) and $R_o$ of $29.45 \pm 0.15$ km s$^{-1}$ kpc$^{-1}$.  Combining this ratio with the $R_o$ measured in this work gives $\Theta_o = 231$ $\pm 4.3$ km s$^{-1}$.  
This value of $\Theta_o$ agrees well with the independently measured value found by \citealt{2014ApJ...783..130R} of $240\pm8$ km s$^{-1}$.  

The new mass value presented here does not significantly change the position of the Milky Way in the observed correlations between mass of the central black hole and host galaxy properties, such as stellar velocity dispersion and mass of the bulge (see \citealt{2013ARA&A..51..511K} for a review).  Sgr A*, along with other central black holes in galaxies with pseudobulges, has a lower mass than expected from the tight correlation seen in ellipticals and galaxies with bulges.  

The new upper limit on the extended dark mass within the orbits of S0-2 and S0-38 is also improved compared to previous work.  To compare the limits presented in Section \ref{results} with previous measurements from \cite{2008Ghez} and \cite{2009Gillessen}, we tranform our upper limits to find the extended mass within the apoapse distance of S0-2: 3.1 $\times 10^{11}$ km $=$ 0.01 pc.  The 1-sigma (3-sigma) upper limits within this radius are 0.05 (0.14) $\times 10^6  M_{\odot}$ for the S0-2 only fit and 0.04 (0.13) $\times 10^6 M_{\odot}$ for the simultaneous S0-2 and S0-38 fit.   \cite{2008Ghez} and \cite{2009Gillessen} found 1-sigma upper limits of 0.12 and 0.17 $\times 10^6 M_{\odot}$ respectively; therefore our new upper limits are a factor of 3-4 lower than previous measurements.  S0-38 does improve the limit compared to S0-2 alone, but the main reason for the lower limit compared to earlier work is due to increased time coverage of the orbit of S0-2.  Our new limit is still over an order of magnitude more than the $\sim$500 - 1000 $M_{\odot}$ of stellar remnants predicted to be within 0.01 pc (e.g., \citealt{2006ApJ...649...91F}), and other models predict even less mass within 0.01 pc (\citealt{2010ApJ...718..739M}).  As observations of the orbits of S0-2 and S0-38 continue, the limits on the extended dark mass within 0.01 pc will continue to decrease.  

The improved constraints on the gravitational potential that come with the addition of S0-38 also impacts future tests of general relativity in the Galactic Center.  
General relativistic deviations from pure Keplerian motion are expected to be detectable when S0-2 goes though its time of closest approach to the black hole in 2018.  The deviations will be observable as the shift of the measured velocity of S0-2 due to the gravitational redshift.  Measuring the deviations from S0-2's Keplerian orbit require as much knowledge of S0-2's Keplerian orbit and the gravitational potential as possible, so the additional constraints from S0-38 are important to this future probe of general relativity.  
Another observable deviation from a Keplerian orbit is the precession of the point of periapse.  This general relativistic precession is confounded by the Newtonian precession due to extended dark mass within stellar orbits.  Measuring the general relativistic precession therefore requires the measurement of precession in at least two stars.  
Additionally, the measurement of the ratio of $M_{bh}$ and $R_o$ from stellar orbits is required to compare the size of the black hole shadow as measured by the upcoming Event Horizon Telescope to theoretical predictions (\citealt{2015Psaltis}).  The theoretical half-opening angle of the shadow of Sgr A* as observed from Earth is predicted to be $(5\pm0.2)GM_{bh}/R_{o}c^2$, regardless of the spin of the black hole.  
With the addition of S0-38 and an increased time baseline of observations, the gravitational radius is now known within $\sim3\%$, which is less than $\pm4\%$ range in the theoretically predicted sizes of the black hole shadow assuming no knowledge of the spin of Sgr A*.  In the future, our knowledge of the gravitational potential in the Galactic Center will increase with more observations of S0-2 and S0-38 as well as with the addition of other short-period stars, thereby also increasing the possibility of measuring the effects of general relativity in this extreme environment.

\acknowledgements{
We thank the staff of the Keck Observatory, especially Randy Campbell, Jason Chin, Scott Dahm, Heather Hershey, Carolyn Jordan, Marc Kassis, Jim Lyke, Gary Puniwai, Julie Renaud-Kim, Luca Rizzi, Terry Stickel, Hien Tran, Peter Wizinowich, and former director Taft Armandroff for all their help in obtaining the new observations.  Support for this work at UCLA was provided by NSF grants AST-0909218 and AST-1412615, the Levine-Leichtman Family Foundation, the Preston Family Graduate Fellowship (held by A. B. and B. N. S.), the Galactic Center Board of Advisors, the Janet Marott Student Travel Awards, the UCLA Graduate Division Dissertation Year Fellowship (held by B. N. S.), and Janet Marott for her support of research on S0-38 through the Galactic Center Group's Stellar Patron Program.  The research by R.S. leading to these results has received funding from the European Research Council under the European Union's Seventh Framework Programme (FP7/2007-2013) / ERC grant agreement n$^\circ$ [614922].  The W. M. Keck Observatory is operated as a scientific partnership among the California Institute of Technology, the University of California, and the National Aeronautics and Space Administration. The authors wish to recognize that the summit of Mauna Kea has always held a very significant cultural role for the indigenous Hawaiian community. We are most fortunate to have the opportunity to observe from this mountain. The Observatory was made possible by the generous financial support of the W. M. Keck Foundation.}
\clearpage

\appendix
\section{Astrometry of Secondary Standards}
\label{app:ref_frame}
The astrometric absolute reference frame is updated using the methods described in \cite{2014Yelda} with the addition of new observations of the masers, which are summarized in Table \ref{tab:maserobs}.  
The results of this analysis, carried out by Sylvana Yelda, are the updated astrometry measurements for the IR secondary astrometric standards, shown in Table \ref{tab:secondary_short}.  

\begin{deluxetable}{lrrrrrrrr}
\tabletypesize{\scriptsize}
\tablewidth{0pt}
\tablecaption{Summary of New Maser Mosaic Observations}
\tablehead{
  \colhead{Date} & 
  \colhead{Date} &  
  \colhead{Frames} & 
  \colhead{Frames} & 
  \colhead{FWHM} & 
  \colhead{Strehl} & 
  \colhead{N$_{stars}$} & 
  \colhead{K$_{lim}\tablenotemark{a}$} & 
  \colhead{$\sigma_{pos}\tablenotemark{b}$} \\
  \colhead{(UT)} & 
  \colhead{(Decimal)} & 
  \colhead{Obtained} & 
  \colhead{Used} & 
  \colhead{(mas)} & 
  \colhead{} & 
  \colhead{} & 
  \colhead{(mag)} & 
  \colhead{mas} \\
}
\startdata
 2011 July 20 & 2011.549 &   65 &   64 & 62 & 0.23 & 2103 & 15.8 & 1.87\\ 
  2012 May 16 & 2012.373 &   54 &   54 & 58 & 0.24 & 2040 & 15.7 & 1.82\\ 
  2013 July 1 & 2013.501 &  175 &  173 & 59 & 0.24 & 2685 & 16.4 & 1.63\\ 
\enddata 
\label{tab:maserobs}

\tablenotetext{a}{K$_{lim}$ is the magnitude at which the cumulative distribution function of the observed K magnitudes reaches 90\% of the total sample size.}
\tablenotetext{b}{Positional error taken as error on the mean from the three sub-images in each epoch and is derived from stars with $K <$ 15.  These positional errors also include distortion error.}

\end{deluxetable}
\clearpage

\begin{deluxetable}{lcccccccccc}
\tabletypesize{\scriptsize}
\tablewidth{0pt}
\tablecaption{Galactic Center Secondary IR Astrometric Standards}
\tablehead{
  \colhead{Name} & 
  \colhead{K'} & 
  \colhead{$T_{0,IR}$} & 
  \colhead{Radius} & 
  \colhead{$\Delta$ R.A.} & 
  \colhead{$\sigma_{R.A.}$\tablenotemark{a}} & 
  \colhead{$\Delta$ Dec.} & 
  \colhead{$\sigma_{Dec}$\tablenotemark{a}} & 
  \colhead{v$_{RA}$\tablenotemark{b}} & 
  \colhead{v$_{Dec}$\tablenotemark{b}} \\  
  \colhead{} & 
  \colhead{(mag)} & 
  \colhead{(year)} & 
  \colhead{(arcsec)} & 
  \colhead{(arcsec)} & 
  \colhead{(mas)} & 
  \colhead{(arcsec)} & 
  \colhead{(mas)} & 
  \colhead{(mas yr$^{-1}$)} & 
  \colhead{(mas yr$^{-1}$)} 
}
\startdata
      S0-6  & 14.2  & 2010.11  & 0.36  &   0.0200  & 1.1  &  -0.3558  & 1.1  &     -5.2 $\pm$      0.1     &      3.5 $\pm$      0.2\\ 
     S0-11  & 15.4  & 2010.42  & 0.50  &   0.4934  & 1.1  &  -0.0607  & 1.2  &     -3.8 $\pm$      0.2     &     -2.4 $\pm$      0.3\\ 
      S0-7  & 15.4  & 2010.08  & 0.52  &   0.5145  & 1.1  &   0.1013  & 1.2  &      5.8 $\pm$      0.2     &      1.0 $\pm$      0.3\\ 
\enddata 
\tablecomments{Table \ref{tab:secondary_short} is published in its entirety in the electronic version of this paper.}
\tablenotetext{a}{Positional errors include centroiding, alignment, and residual distortion (1 mas) errors, but do not include error in position of Sgr A* (2.0 mas, 1.4 mas in RA and Dec, respectively).}
\tablenotetext{b}{Velocity errors do not include error in velocity of Sgr A* (0.13 mas yr$^{-1}$, 0.23 mas yr$^{-1}$ in RA and Dec, respectively).}
\label{tab:secondary_short}

\end{deluxetable}
\clearpage

\section{S0-2 Data and Orbital Analysis}
\label{appendix:S0-2}
Here we report new astrometric and spectroscopic observations of the star S0-2.  New AO astrometric measurements were obtained in the same way as described in Section \ref{AOimagedata} for new observation epochs.  Table \ref{tab:S0-2_astrometry} lists the astrometric measurements of S0-2 used in this work.  The speckle holography detections used in the fit of S0-2 are all from the set of direct detections discussed in Section \ref{speckledata}; i.e., S0-2 was detected in all three subset images as well as the main image.  Speckle holography direct detections in which S0-2 was confused with another known source are not used in the orbital fit, as in \cite{2008Ghez}.  
This is the first paper in which we fit S0-2's orbit with the speckle holography astrometric data.  These data supercede the shift-and-add speckle data reported in previous works.   
Table \ref{tab:S0-2_astrometry} lists the astrometric measurements of S0-2 used in this work.  

New spectroscopic measurements of S0-2 were obtained with OSIRIS on the Keck 1 and Keck 2 telescopes using the Kn3 filter that is centered on the Br-$\gamma$ line at $2.1661 \mu m$.  The previously unreported measurements as well as the derived LSR-corrected radial velocities of S0-2 are summarized in Table \ref{tab:S0-2_spectraobs}.  Orbital fits of S0-2 performed in this work used these new radial velocities as well as previously reported radial velocities from both Keck and VLT (originally published in \citealt{2008Ghez} and \citealt{2009Gillessen};  also see \citealt{2009ApJ...707L.114G} in which both data sets are presented together).

In order to perform the orbital fits of stars in our sample (as detailed in Section \ref{orbitalfits}), S0-2's data must first be fit alone in order to determine the probability distribution of the 7 BH parameters (in addition to S0-2's 6 own Keplerian orbital parameters).  
This 13-dimensional orbital fit was done using the Bayesian multimodal nested sampling algorithm called {\sc MultiNest} (see \citealt{2008Feroz} and \citealt{2009Feroz}).  We also use the results of this orbital fit of S0-2 alone to compare with the results of fitting S0-38 alone and S0-2 and S0-38 simultaneously (Section \ref{combofits}).  
In addition to a purely Keplerian orbital fit, we also fit S0-2's orbit alone with an additional free parameter describing the amount of extended dark mass within a characteristic radius.  This orbital fit also serves as a comparison to the results from fitting S0-2 and S0-38 simultaneously.

\begin{deluxetable}{lllll}
\tabletypesize{\scriptsize}
\tablewidth{0pt}
\tablecaption{S0-2 Astrometric Measurements}
\tablehead{
  \colhead{Date} & 
  \colhead{$\Delta$R.A.} & 
  \colhead{$\Delta$Dec.} & 
  \colhead{$\Delta$R.A. Error} & 
  \colhead{$\Delta$Dec. Error} \\
  \colhead{(Decimal)} & 
  \colhead{(arcsec)} & 
  \colhead{(arcsec)} & 
  \colhead{(arcsec)} & 
  \colhead{(arcsec)} \\
}
\startdata
1995.439 & -0.0401 & 0.16216 & 0.0010 & 0.00086\\ 
1996.485 & -0.0481 & 0.1553 & 0.0044 & 0.0029\\ 
1997.367 & -0.0547 & 0.1388 & 0.0012 & 0.0015\\ 
1999.333 & -0.0682 & 0.09808 & 0.0011 & 0.00074\\ 
1999.559 & -0.0677 & 0.08971 & 0.0011 & 0.00077\\ 
2000.305 & -0.0638 & 0.0747 & 0.0013 & 0.0016\\ 
2000.381 & -0.06411 & 0.0643 & 0.00087 & 0.0019\\ 
2000.548 & -0.0631 & 0.0563 & 0.0014 & 0.0023\\ 
2000.797 & -0.0584 & 0.0519 & 0.0030 & 0.0019\\ 
2001.351 & -0.0520 & 0.0261 & 0.0015 & 0.0013\\ 
2001.572 & -0.0504 & 0.0152 & 0.0010 & 0.0012\\ 
2003.303 & 0.0386 & 0.0726 & 0.0017 & 0.0016\\ 
2003.554 & 0.03863 & 0.08430 & 0.00089 & 0.00082\\ 
2003.682 & 0.0435 & 0.0949 & 0.0024 & 0.0019\\ 
2004.327 & 0.03595 & 0.11539 & 0.00076 & 0.00062\\ 
2004.564 & 0.03261 & 0.12328 & 0.00079 & 0.00053\\ 
2004.567 & 0.0355 & 0.1261 & 0.0050 & 0.0046\\ 
2004.660 & 0.03137 & 0.12584 & 0.00069 & 0.00060\\ 
2005.312 & 0.02483 & 0.14110 & 0.00081 & 0.00059\\ 
2005.495 & 0.0201 & 0.1473 & 0.0053 & 0.0023\\ 
2005.566 & 0.02084 & 0.1492 & 0.00083 & 0.0011\\ 
2005.580 & 0.0235 & 0.1493 & 0.0038 & 0.0010\\
2006.336 & 0.01296 & 0.16191 & 0.00015 & 0.00016\\ 
2008.371 & -0.01016 & 0.18101 & 0.00013 & 0.00014\\ 
2008.562 & -0.01232 & 0.18184 & 0.00016 & 0.00016\\ 
2009.340 & -0.02114 & 0.18266 & 0.00011 & 0.00011\\ 
2009.561 & -0.02343 & 0.18269 & 0.00013 & 0.00020\\ 
2009.689 & -0.02478 & 0.18265 & 0.00018 & 0.00014\\ 
2010.342 & -0.03194 & 0.18059 & 0.00012 & 0.00012\\ 
2010.511 & -0.03382 & 0.17984 & 0.00013 & 0.00012\\ 
2010.620 & -0.03501 & 0.17931 & 0.00016 & 0.00013\\ 
2011.401 & -0.04277 & 0.17377 & 0.00022 & 0.00017\\ 
2011.543 & -0.04423 & 0.17196 & 0.00014 & 0.00014\\ 
2011.642 & -0.04517 & 0.17121 & 0.00025 & 0.00025\\ 
2012.371 & -0.05136 & 0.16326 & 0.00013 & 0.00017\\ 
2012.562 & -0.05312 & 0.16074 & 0.00014 & 0.00020\\ 
2013.318 & -0.05837 & 0.14965 & 0.00021 & 0.00026\\ 
2013.550 & -0.05979 & 0.14541 & 0.00014 & 0.00021\\ 
\enddata 
\label{tab:S0-2_astrometry}

\end{deluxetable}
\clearpage

\begin{deluxetable}{lcrrrcc}
\tabletypesize{\scriptsize}
\tablewidth{0pt}
\tablecaption{Summary of New Spectroscopic Observations and Radial Velocities of S0-2}
\tablehead{
  \multicolumn{2}{c}{Date} & 
  \colhead{$N_{frames} \times t_{int}$} & 
  \colhead{FWHM\tablenotemark{a}} &
  \colhead{$V_z$\tablenotemark{b}}&
  \colhead{$V_z$ Error} &
  \colhead{$V_{LSR}$\tablenotemark{c}}\\
  \colhead{(UT)} & 
  \colhead{(Decimal)} & 
  \colhead{} & 
  \colhead{(mas)} & 
  \colhead{(km s$^{-1}$)} &
  \colhead{(km s$^{-1}$)} &  
  \colhead{(km s$^{-1}$)} 
}
\startdata
2008 May 16&  2008.373 &   11 $\times$ 900 sec &  75 & -417  & 32 & 26\\
2008 July 25&  2008.564 & 10 $\times$ 900 sec& 84  & -380 & 43 & -7\\
2009 May 5 \& 6& 2009.344 & 24 $\times$ 900 sec& 74 & -285 & 32 & 30\\

2010 May 5 \& 8&  2010.349&   16$\times$ 900 sec&  69 &   -123 & 22 & 30\\   %
2011 July 10&  2011.521&  12$\times$ 900 sec &  86  &    23.5  & 22 & -11\\

2012 June 8 - 11&  2012.441& 10 $\times$ 900 sec&  72&    183&49 & 14\\
2012 July 21 \& 22&  2012.556& 11 $\times$ 900 sec &  102 & 187& 33 & -5\\
2012 August 12 \& 13&  2012.616& 26 $\times$ 900 sec &  61  &167& 25 & -13\\
2013 May 14 \& 16&  2013.369& 18 $\times$ 900 sec  & 67 &  328 &66 & 27\\
2013 July 25 - 27&  2013.566&26 $\times$ 900 sec  & 84 &  339& 23 & -18\\
2013 August 10 - 13&  2013.612&31 $\times$ 900 sec & 78 &
  349&  16 & -13\\
\enddata 
\label{tab:S0-2_spectraobs}

\tablenotetext{a}{Average FWHM of S0-2 in the mosaic made of all frames, measured by fitting a two-dimensional Gaussian to the source.}
\tablenotetext{b}{All radial velocities are corrected for the local standard of rest velocity.}
\tablenotetext{c}{The local standard of rest velocity used to convert the observed velocities to the local standard of rest.}
\end{deluxetable}
\clearpage

\section{Jack Knife Approximation to Reference Frame Uncertainties}
\label{appendix:jackknife}

We expect systematic errors in the orbital analysis of stars at the Galactic Center to be dominated by inaccuracies in our distortion solution and unaccounted errors in the construction of the absolute reference frame.  While the former is potentially a significant source of uncertainty, currently its exact contribution is unclear and thus is the subject of future research.  We also expect non-propagated statistical uncertainty in the reference frame to be a source of significant error.  Such issues may arise from rigid model assumptions or improper propagation of error correlations.  In this case, this issue is the general of several hundred secondary astrometric standards that have been treated as if their uncertainties are uncorrelated.  This assumption is incorrect as the secondary standards are derived based on only 7 primary astrometric standards.  If these correlated uncertainties are not incorporated into the full Bayesian orbital analysis, then the underlying statistical variance could produce a systematic bias in the resultant posterior distributions of the black hole parameters.  

We can gain some insight into the effect and magnitude of this bias through various resampling methodologies.  We use a jack knife resampling methodology to infer, to first order, the random statistical bias of an unbiased estimator by utilizing subsets of the full set of SiO maser positions that are used in the construction of the absolute reference frame (see \citealt{2014Yelda} for details on the refernece frame).  
Seven subsets were made by systematically excluding one maser position at a time.  These subsets were then used to construct seven new reference frames and a full orbital analysis of S0-2 was performed.\footnote{Note that the orbital analysis of S0-2 was performed on the speckle shift-and-add astrometry of S0-2 and not the new speckle holography astrometry.  We do not expect this to affect the results of the maser jack knife analysis, since the absolute reference frame construction is the same in the jack knife analysis and in the speckle holography analysis presented in this work.}  The S0-2 astrometric measurements in each reference frame is presented in Table \ref{tab:droponemaser_S02}.  
The results of the orbital fits performed in the seven subset reference frames are then used to infer the statistical bias of some estimator $x$ by \citep{2001math......9002G}
\begin{equation}
\label{eq:jackbias}
 x_{\textrm{bias}} \approx (n-1)(\bar{x}_{n-1} - x_{n})
\end{equation}
with a variance of
\begin{equation}
\label{eq:jackvar}
 \sigma^2(x_n) \leq \frac{n-1}{n} \sum\limits_{i=1}^{n} {\left(x_{n-1, i} - \bar{x}_{n-1}\right)^2}
\end{equation}
where $n$ is the sample size (in this case $7$), $x_n$ is the estimator derived using the full sample, $x_{n-1, i}$ is the estimator derived by the exclusion of the $i$-th maser, and $\bar{x}_{n-1}$ is the average of these subsets.  Table \ref{tab:jackknife_bias_var} lists this bias and the square root of its variance ($\sigma$) on the inferred average value of the 7 black hole parameters considered in this work.  In the black hole parameter probability distribution figures presented in Sections \ref{results} and \ref{discussion} we overlay 68\%, 95\%, and 99\% contours over their respective posteriors.  For the joint posteriors we infer the correlations through a multidimensional form of Equation \ref{eq:jackvar}:
\begin{equation}
\label{eq:multijackvar}
 \sigma^2(x_n, y_n) \leq \frac{n-1}{n} \sum\limits_{i=1, j = 1}^{n} {\left(x_{n-1, i} - \bar{x}_{n-1}\right)\left(y_{n-1, j} - \bar{y}_{n-1}\right)}
\end{equation}

As noted in Section \ref{BHconstraints}, the deviations of the $x_o$, $y_o$, $V_x$, $V_y$ posteriors appear to be explained by previously unaccounted for reference frame uncertainties, while the $M_{bh}$  
and $R_o$ posteriors are not affected by these systematics.

We emphases that the bias estimated by this analysis is only relevant assuming that these uncertainties are not already incorporated into our Bayesian analysis.  The bias define in Equation \ref{eq:jackbias} is interpreted as, to first order, the difference between a statistic inferred from a hypothetical set of infinite masers and the same statistic inferred from our set of seven masers.  Ideally, an uncertainty of this type would be inherently incorporated into any robust statistical methodology.  If, contrary to our assumption, the uncertainties presented in this paper do indeed account for reference frame construction uncertainties, then direct incorporation of the jack knife results would overestimate the uncertainties and thereby hide additional systematic effects.  Thus, here we present our Bayesian uncertainties and jack knife result {\em separately} in Table \ref{tab:bestfit_BHandorbitparams}.  

If our assumption holds and if the Bayesian posteriors are parametrized in the sufficient statistic of $x_n$, then we may directly derive the resultant `total' probability
of a parameter $x$ as:
\begin{equation}
\label{eq:jackcombo}
 \mathscr{P}(x \vert x_{\textrm{bias}}, \sigma^2(x_n)) = \int d \hat{x}_{\textrm{bias}} \mathscr{P}(x \vert x_n + \hat{x}_{\textrm{bias}})\mathscr{P}(\hat{x}_{\textrm{bias}} \vert x_{\textrm{bias}}, \sigma^2(x_n))
\end{equation}
where $\mathscr{P}(x \vert x_n + \hat{x}_{\textrm{bias}})$ is the posterior parameterized in $x_n$ and $\mathscr{P}(\hat{x}_{\textrm{bias}} \vert x_{\textrm{bias}}, \sigma^2(x_n))$ is a normal distribution whose parametrization is determined by Equations \ref{eq:jackbias} and \ref{eq:jackvar}.  In the limit of asymptotic normality, the net effect of incorporating this uncertainty is to shift the posterior by $x_{\textrm{bias}}$ and adding the variance $\sigma(x_n)$ in quadrature with the derived Bayesian uncertainty.  All posteriors plotted with the 68\%, 95\%, and 99\% contours in Sections \ref{results} and \ref{discussion} are therefore shifted by the bias presented in Table \ref{tab:jackknife_bias_var}.  The best-fit values of the black hole parameters presented in Table \ref{tab:bestfit_BHandorbitparams} are also shifted by this bias and include the standard deviation of the bias as an additional uncertainty term.  The most significant effect of the bias shifts and the corresponding additional uncertainties is in the black hole's position and velocity on the plane of the sky ($x_o$, $y_o$, $V_x$, and $V_y$).

\begin{deluxetable}{lllll}
\tabletypesize{\scriptsize}
\tablewidth{0pt}
\tablecaption{S0-2 Shift-and-Add Astrometric Measurements Used in Jack Knife Analysis}
\tablehead{
  \colhead{Date} & 
  \colhead{$\Delta$R.A.} & 
  \colhead{$\Delta$Dec.} & 
  \colhead{$\Delta$R.A. Error} & 
  \colhead{$\Delta$Dec. Error} \\
  \colhead{(Decimal)} & 
  \colhead{(arcsec)} & 
  \colhead{(arcsec)} & 
  \colhead{(arcsec)} & 
  \colhead{(arcsec)} 
}
\startdata
All 7 Masers:& & & & \\ 
1995.439 & -0.0418 & 0.1637 & 0.0011 & 0.0014\\
1996.485 & -0.0520 & 0.1557 & 0.0046 & 0.0035\\ 
1997.367 & -0.0556 & 0.1384 & 0.0012 & 0.0025\\ 
1999.333 & -0.0648 & 0.0923 & 0.0011 & 0.0015\\ 
1999.559 & -0.0659 & 0.08949 & 0.0014 & 0.00093\\ 
2000.381 & -0.06306 & 0.06357 & 0.00097 & 0.00081\\ 
2000.548 & -0.0626 & 0.0591 & 0.0012 & 0.0012\\ 
2000.797 & -0.0615 & 0.0487 & 0.0024 & 0.0032\\ 
2001.351 & -0.0534 & 0.0258 & 0.0015 & 0.0016\\ 
2001.572 & -0.0492 & 0.0142 & 0.0012 & 0.0017\\ 
2003.303 & 0.0397 & 0.0703 & 0.0015 & 0.0021\\ 
2003.554 & 0.0410 & 0.08209 & 0.0013 & 0.00067\\ 
2003.682 & 0.0393 & 0.08902 & 0.0018 & 0.00084\\ 
2004.327 & 0.03691 & 0.1140 & 0.00081 & 0.0012\\ 
2004.564 & 0.0333 & 0.12245 & 0.0011 & 0.00093\\ 
2004.567 & 0.0352 & 0.1265 & 0.0050 & 0.0046\\ 
2004.660 & 0.03083 & 0.1255 & 0.00098 & 0.0020\\ 
2005.312 & 0.02412 & 0.1430 & 0.00081 & 0.0011\\ 
2005.495 & 0.0199 & 0.1477 & 0.0053 & 0.0023\\ 
2005.566 & 0.0216 & 0.1493 & 0.0021 & 0.0012\\ 
2005.580 & 0.0233 & 0.1497 & 0.0038 & 0.0010\\ 
2006.336 & 0.01280 & 0.16237 & 0.00013 & 0.00013\\ 
2008.371 & -0.01039 & 0.18146 & 0.00015 & 0.00015\\ 
2008.562 & -0.01247 & 0.18227 & 0.00017 & 0.00017\\ 
2009.340 & -0.02141 & 0.18308 & 0.00016 & 0.00016\\ 
2009.561 & -0.02368 & 0.18315 & 0.00016 & 0.00022\\ 
2009.689 & -0.02491 & 0.18311 & 0.00018 & 0.00018\\ 
2010.342 & -0.03216 & 0.18103 & 0.00015 & 0.00018\\ 
2010.511 & -0.03415 & 0.18025 & 0.00021 & 0.00020\\ 
2010.620 & -0.03527 & 0.17972 & 0.00019 & 0.00018\\ 
2011.401 & -0.04295 & 0.17417 & 0.00024 & 0.00022\\ 
2011.543 & -0.04441 & 0.17241 & 0.00017 & 0.00022\\ 
2011.642 & -0.04519 & 0.17152 & 0.00026 & 0.00023\\ 
2012.371 & -0.05162 & 0.16365 & 0.00016 & 0.00026\\ 
2012.562 & -0.05335 & 0.16112 & 0.00018 & 0.00029\\ 
2013.318 & -0.05847 & 0.15042 & 0.00024 & 0.00041\\ 
2013.550 & -0.05995 & 0.14586 & 0.00018 & 0.00032\\ 
\\ 
 
No IRS 10EE:& & & & \\ 
1995.439 & -0.0424 & 0.1639 & 0.0013 & 0.0014\\ 
1996.485 & -0.0526 & 0.1558 & 0.0047 & 0.0035\\ 
1997.367 & -0.0561 & 0.1385 & 0.0011 & 0.0026\\ 
1999.333 & -0.0653 & 0.0925 & 0.0013 & 0.0015\\ 
1999.559 & -0.0663 & 0.08967 & 0.0014 & 0.00099\\ 
2000.381 & -0.06335 & 0.06360 & 0.00090 & 0.00084\\ 
2000.548 & -0.0629 & 0.0594 & 0.0011 & 0.0013\\ 
2000.797 & -0.0618 & 0.0489 & 0.0024 & 0.0033\\ 
2001.351 & -0.0536 & 0.0261 & 0.0014 & 0.0016\\ 
2001.572 & -0.0495 & 0.0145 & 0.0012 & 0.0018\\ 
2003.303 & 0.0395 & 0.0705 & 0.0015 & 0.0022\\ 
2003.554 & 0.0408 & 0.08232 & 0.0013 & 0.00071\\ 
2003.682 & 0.0392 & 0.08920 & 0.0018 & 0.00084\\ 
2004.327 & 0.03679 & 0.1142 & 0.00086 & 0.0011\\ 
2004.564 & 0.0332 & 0.12270 & 0.0011 & 0.00089\\ 
2004.567 & 0.0352 & 0.1268 & 0.0050 & 0.0046\\ 
2004.660 & 0.03072 & 0.1258 & 0.00099 & 0.0020\\ 
2005.312 & 0.02406 & 0.1433 & 0.00083 & 0.0012\\ 
2005.495 & 0.0198 & 0.1480 & 0.0053 & 0.0023\\ 
2005.566 & 0.0215 & 0.1495 & 0.0022 & 0.0012\\ 
2005.580 & 0.0231 & 0.1501 & 0.0038 & 0.0010\\ 
2006.336 & 0.01280 & 0.16265 & 0.00013 & 0.00012\\ 
2008.371 & -0.01028 & 0.18174 & 0.00015 & 0.00015\\ 
2008.562 & -0.01235 & 0.18255 & 0.00017 & 0.00017\\ 
2009.340 & -0.02125 & 0.18337 & 0.00016 & 0.00016\\ 
2009.561 & -0.02351 & 0.18342 & 0.00015 & 0.00021\\ 
2009.689 & -0.02473 & 0.18340 & 0.00018 & 0.00016\\ 
2010.342 & -0.03195 & 0.18130 & 0.00014 & 0.00016\\ 
2010.511 & -0.03393 & 0.18053 & 0.00020 & 0.00017\\ 
2010.620 & -0.03504 & 0.18004 & 0.00019 & 0.00018\\ 
2011.401 & -0.04267 & 0.17447 & 0.00024 & 0.00022\\ 
2011.543 & -0.04413 & 0.17276 & 0.00017 & 0.00021\\ 
2011.642 & -0.04492 & 0.17183 & 0.00025 & 0.00023\\ 
2012.371 & -0.05128 & 0.16391 & 0.00015 & 0.00025\\ 
2012.562 & -0.05301 & 0.16143 & 0.00017 & 0.00027\\ 
2013.318 & -0.05811 & 0.15079 & 0.00023 & 0.00039\\ 
2013.550 & -0.05958 & 0.14618 & 0.00017 & 0.00030\\ 
\\ 
 
No IRS 12N:& & & & \\ 
1995.439 & -0.0426 & 0.1612 & 0.0017 & 0.0017\\ 
1996.485 & -0.0544 & 0.1538 & 0.0049 & 0.0041\\ 
1997.367 & -0.0579 & 0.1367 & 0.0017 & 0.0030\\ 
1999.333 & -0.0655 & 0.0904 & 0.0012 & 0.0018\\ 
1999.559 & -0.0669 & 0.0880 & 0.0014 & 0.0018\\ 
2000.381 & -0.0651 & 0.0617 & 0.0012 & 0.0019\\ 
2000.548 & -0.0640 & 0.0584 & 0.0014 & 0.0015\\ 
2000.797 & -0.0639 & 0.0486 & 0.0026 & 0.0041\\ 
2001.351 & -0.0552 & 0.0249 & 0.0017 & 0.0020\\ 
2001.572 & -0.0505 & 0.0145 & 0.0020 & 0.0021\\ 
2003.303 & 0.0385 & 0.0702 & 0.0016 & 0.0023\\ 
2003.554 & 0.0391 & 0.0812 & 0.0014 & 0.0013\\ 
2003.682 & 0.0385 & 0.0880 & 0.0012 & 0.0010\\ 
2004.327 & 0.03550 & 0.1135 & 0.00074 & 0.0012\\ 
2004.564 & 0.0320 & 0.1217 & 0.0012 & 0.0011\\ 
2004.567 & 0.0340 & 0.1261 & 0.0050 & 0.0046\\ 
2004.660 & 0.0308 & 0.1251 & 0.0014 & 0.0023\\ 
2005.312 & 0.0227 & 0.1425 & 0.0012 & 0.0015\\ 
2005.495 & 0.0185 & 0.1473 & 0.0053 & 0.0023\\ 
2005.566 & 0.0198 & 0.1488 & 0.0022 & 0.0020\\ 
2005.580 & 0.0218 & 0.1493 & 0.0039 & 0.0011\\
2006.336 & 0.01160 & 0.16187 & 0.00013 & 0.00013\\ 
2008.371 & -0.01139 & 0.18136 & 0.00015 & 0.00016\\ 
2008.562 & -0.01355 & 0.18213 & 0.00016 & 0.00017\\ 
2009.340 & -0.02251 & 0.18318 & 0.00016 & 0.00019\\ 
2009.561 & -0.02491 & 0.18320 & 0.00021 & 0.00026\\ 
2009.689 & -0.02618 & 0.18310 & 0.00020 & 0.00021\\ 
2010.342 & -0.03339 & 0.18122 & 0.00018 & 0.00023\\ 
2010.511 & -0.03514 & 0.18023 & 0.00022 & 0.00028\\ 
2010.620 & -0.03637 & 0.17981 & 0.00021 & 0.00020\\ 
2011.401 & -0.04404 & 0.17436 & 0.00027 & 0.00030\\ 
2011.543 & -0.04541 & 0.17245 & 0.00021 & 0.00030\\ 
2011.642 & -0.04616 & 0.17169 & 0.00034 & 0.00029\\ 
2012.371 & -0.05271 & 0.16394 & 0.00021 & 0.00029\\ 
2012.562 & -0.05437 & 0.16143 & 0.00022 & 0.00030\\ 
2013.318 & -0.05946 & 0.15118 & 0.00027 & 0.00053\\ 
2013.550 & -0.06103 & 0.14638 & 0.00022 & 0.00038\\ 
\\ 
 
No IRS 15NE:& & & & \\ 
1995.439 & -0.0407 & 0.1636 & 0.0012 & 0.0014\\ 
1996.485 & -0.0509 & 0.1557 & 0.0047 & 0.0035\\ 
1997.367 & -0.0546 & 0.1384 & 0.0012 & 0.0025\\ 
1999.333 & -0.0641 & 0.0926 & 0.0012 & 0.0015\\ 
1999.559 & -0.0652 & 0.08971 & 0.0014 & 0.00091\\ 
2000.381 & -0.06237 & 0.06384 & 0.00092 & 0.00083\\ 
2000.548 & -0.0619 & 0.0596 & 0.0012 & 0.0012\\ 
2000.797 & -0.0608 & 0.0492 & 0.0024 & 0.0032\\ 
2001.351 & -0.0527 & 0.0264 & 0.0015 & 0.0016\\ 
2001.572 & -0.0487 & 0.0148 & 0.0013 & 0.0017\\ 
2003.303 & 0.0401 & 0.0709 & 0.0015 & 0.0021\\ 
2003.554 & 0.0415 & 0.08275 & 0.0014 & 0.00082\\ 
2003.682 & 0.0397 & 0.08961 & 0.0018 & 0.00088\\ 
2004.327 & 0.03729 & 0.1147 & 0.00081 & 0.0012\\ 
2004.564 & 0.0336 & 0.12320 & 0.0011 & 0.00095\\ 
2004.567 & 0.0356 & 0.1272 & 0.0050 & 0.0046\\ 
2004.660 & 0.03120 & 0.1263 & 0.00096 & 0.0020\\ 
2005.312 & 0.02439 & 0.1439 & 0.00086 & 0.0011\\ 
2005.495 & 0.0202 & 0.1486 & 0.0053 & 0.0023\\ 
2005.566 & 0.0219 & 0.1501 & 0.0022 & 0.0012\\ 
2005.580 & 0.0235 & 0.1506 & 0.0038 & 0.0010\\
2006.336 & 0.01304 & 0.16336 & 0.00012 & 0.00013\\ 
2008.371 & -0.01032 & 0.18266 & 0.00014 & 0.00015\\ 
2008.562 & -0.01240 & 0.18346 & 0.00017 & 0.00017\\ 
2009.340 & -0.02143 & 0.18436 & 0.00016 & 0.00016\\ 
2009.561 & -0.02369 & 0.18449 & 0.00015 & 0.00021\\ 
2009.689 & -0.02495 & 0.18445 & 0.00018 & 0.00016\\ 
2010.342 & -0.03224 & 0.18243 & 0.00014 & 0.00016\\ 
2010.511 & -0.03424 & 0.18167 & 0.00019 & 0.00016\\ 
2010.620 & -0.03542 & 0.18119 & 0.00019 & 0.00018\\ 
2011.401 & -0.04312 & 0.17567 & 0.00023 & 0.00022\\ 
2011.543 & -0.04465 & 0.17396 & 0.00016 & 0.00020\\ 
2011.642 & -0.04542 & 0.17307 & 0.00025 & 0.00024\\ 
2012.371 & -0.05186 & 0.16529 & 0.00015 & 0.00023\\ 
2012.562 & -0.05365 & 0.16282 & 0.00016 & 0.00025\\ 
2013.318 & -0.05887 & 0.15217 & 0.00024 & 0.00037\\ 
2013.550 & -0.06035 & 0.14767 & 0.00017 & 0.00029\\ 
\\ 
 
No IRS 17:& & & & \\ 
1995.439 & -0.0414 & 0.1643 & 0.0012 & 0.0014\\ 
1996.485 & -0.0516 & 0.1562 & 0.0048 & 0.0035\\ 
1997.367 & -0.0553 & 0.1388 & 0.0012 & 0.0025\\ 
1999.333 & -0.0647 & 0.0927 & 0.0010 & 0.0014\\ 
1999.559 & -0.0657 & 0.08990 & 0.0014 & 0.00095\\ 
2000.381 & -0.06292 & 0.06384 & 0.00093 & 0.00081\\ 
2000.548 & -0.0624 & 0.0595 & 0.0012 & 0.0012\\ 
2000.797 & -0.0614 & 0.0491 & 0.0025 & 0.0034\\ 
2001.351 & -0.0532 & 0.0262 & 0.0015 & 0.0016\\ 
2001.572 & -0.0491 & 0.0146 & 0.0011 & 0.0017\\ 
2003.303 & 0.0397 & 0.0705 & 0.0014 & 0.0021\\ 
2003.554 & 0.0410 & 0.08229 & 0.0014 & 0.00068\\ 
2003.682 & 0.0394 & 0.08915 & 0.0018 & 0.00082\\ 
2004.327 & 0.03696 & 0.1142 & 0.00079 & 0.0011\\ 
2004.564 & 0.0333 & 0.12264 & 0.0011 & 0.00092\\ 
2004.567 & 0.0353 & 0.1267 & 0.0050 & 0.0046\\ 
2004.660 & 0.0309 & 0.1257 & 0.0010 & 0.0020\\ 
2005.312 & 0.02412 & 0.1432 & 0.00083 & 0.0011\\ 
2005.495 & 0.0199 & 0.1479 & 0.0053 & 0.0023\\ 
2005.566 & 0.0216 & 0.1494 & 0.0021 & 0.0011\\ 
2005.580 & 0.0233 & 0.1500 & 0.0038 & 0.0010\\ 
2006.336 & 0.01275 & 0.16250 & 0.00013 & 0.00013\\ 
2008.371 & -0.01054 & 0.18153 & 0.00014 & 0.00015\\ 
2008.562 & -0.01263 & 0.18232 & 0.00017 & 0.00017\\ 
2009.340 & -0.02161 & 0.18309 & 0.00016 & 0.00016\\ 
2009.561 & -0.02389 & 0.18315 & 0.00014 & 0.00022\\ 
2009.689 & -0.02513 & 0.18311 & 0.00018 & 0.00016\\ 
2010.342 & -0.03243 & 0.18099 & 0.00014 & 0.00017\\ 
2010.511 & -0.03441 & 0.18022 & 0.00019 & 0.00018\\ 
2010.620 & -0.03555 & 0.17970 & 0.00019 & 0.00018\\ 
2011.401 & -0.04326 & 0.17409 & 0.00023 & 0.00023\\ 
2011.543 & -0.04474 & 0.17235 & 0.00016 & 0.00020\\ 
2011.642 & -0.04551 & 0.17144 & 0.00026 & 0.00023\\ 
2012.371 & -0.05199 & 0.16352 & 0.00015 & 0.00024\\ 
2012.562 & -0.05373 & 0.16100 & 0.00016 & 0.00026\\ 
2013.318 & -0.05886 & 0.15030 & 0.00023 & 0.00038\\ 
2013.550 & -0.06038 & 0.14573 & 0.00017 & 0.00030\\ 
\\ 
 
No IRS 28:& & & & \\ 
1995.439 & -0.0413 & 0.1646 & 0.0014 & 0.0014\\ 
1996.485 & -0.0516 & 0.1565 & 0.0047 & 0.0035\\ 
1997.367 & -0.0552 & 0.1391 & 0.0012 & 0.0024\\ 
1999.333 & -0.0645 & 0.0930 & 0.0012 & 0.0014\\ 
1999.559 & -0.0656 & 0.09016 & 0.0015 & 0.00091\\ 
2000.381 & -0.06277 & 0.06402 & 0.00093 & 0.00078\\ 
2000.548 & -0.0623 & 0.0598 & 0.0012 & 0.0012\\ 
2000.797 & -0.0613 & 0.0493 & 0.0024 & 0.0034\\ 
2001.351 & -0.0531 & 0.0265 & 0.0015 & 0.0016\\ 
2001.572 & -0.0490 & 0.0148 & 0.0012 & 0.0017\\ 
2003.303 & 0.0398 & 0.0707 & 0.0014 & 0.0021\\ 
2003.554 & 0.0411 & 0.08252 & 0.0013 & 0.00075\\ 
2003.682 & 0.0394 & 0.08940 & 0.0018 & 0.00084\\ 
2004.327 & 0.03703 & 0.1144 & 0.00081 & 0.0011\\ 
2004.564 & 0.0334 & 0.12284 & 0.0011 & 0.00092\\ 
2004.567 & 0.0354 & 0.1269 & 0.0050 & 0.0046\\ 
2004.660 & 0.0309 & 0.1259 & 0.0010 & 0.0020\\ 
2005.312 & 0.02422 & 0.1434 & 0.00079 & 0.0011\\ 
2005.495 & 0.0200 & 0.1481 & 0.0053 & 0.0023\\ 
2005.566 & 0.0217 & 0.1496 & 0.0021 & 0.0013\\ 
2005.580 & 0.0233 & 0.1502 & 0.0038 & 0.0010\\ 
2006.336 & 0.01286 & 0.16268 & 0.00012 & 0.00013\\ 
2008.371 & -0.01040 & 0.18163 & 0.00015 & 0.00015\\ 
2008.562 & -0.01249 & 0.18243 & 0.00017 & 0.00017\\ 
2009.340 & -0.02147 & 0.18318 & 0.00016 & 0.00016\\ 
2009.561 & -0.02374 & 0.18325 & 0.00015 & 0.00021\\ 
2009.689 & -0.02496 & 0.18321 & 0.00018 & 0.00016\\ 
2010.342 & -0.03224 & 0.18109 & 0.00014 & 0.00017\\ 
2010.511 & -0.03423 & 0.18030 & 0.00021 & 0.00018\\ 
2010.620 & -0.03540 & 0.17974 & 0.00019 & 0.00018\\ 
2011.401 & -0.04307 & 0.17419 & 0.00023 & 0.00021\\ 
2011.543 & -0.04455 & 0.17244 & 0.00017 & 0.00023\\ 
2011.642 & -0.04532 & 0.17151 & 0.00026 & 0.00023\\ 
2012.371 & -0.05177 & 0.16357 & 0.00016 & 0.00024\\ 
2012.562 & -0.05350 & 0.16107 & 0.00018 & 0.00028\\ 
2013.318 & -0.05867 & 0.15029 & 0.00024 & 0.00039\\ 
2013.550 & -0.06016 & 0.14575 & 0.00018 & 0.00031\\ 
\\ 
 
No IRS 7:& & & & \\ 
1995.439 & -0.0399 & 0.1631 & 0.0017 & 0.0021\\ 
1996.485 & -0.0498 & 0.1561 & 0.0055 & 0.0044\\ 
1997.367 & -0.0545 & 0.1386 & 0.0013 & 0.0028\\ 
1999.333 & -0.0634 & 0.0920 & 0.0012 & 0.0016\\ 
1999.559 & -0.0644 & 0.0892 & 0.0017 & 0.0012\\ 
2000.381 & -0.0619 & 0.06333 & 0.0012 & 0.00089\\ 
2000.548 & -0.0617 & 0.0587 & 0.0012 & 0.0013\\ 
2000.797 & -0.0621 & 0.0483 & 0.0033 & 0.0044\\ 
2001.351 & -0.0526 & 0.0254 & 0.0015 & 0.0016\\ 
2001.572 & -0.0482 & 0.0140 & 0.0012 & 0.0018\\ 
2003.303 & 0.0414 & 0.0694 & 0.0016 & 0.0021\\ 
2003.554 & 0.0420 & 0.0812 & 0.0015 & 0.0010\\ 
2003.682 & 0.0411 & 0.08802 & 0.0020 & 0.00086\\ 
2004.327 & 0.03773 & 0.1133 & 0.00086 & 0.0013\\ 
2004.564 & 0.0341 & 0.1213 & 0.0012 & 0.0010\\ 
2004.567 & 0.0363 & 0.1255 & 0.0050 & 0.0046\\ 
2004.660 & 0.0322 & 0.1248 & 0.0014 & 0.0020\\ 
2005.312 & 0.0253 & 0.1419 & 0.0010 & 0.0012\\ 
2005.495 & 0.0209 & 0.1465 & 0.0053 & 0.0023\\ 
2005.566 & 0.0230 & 0.1483 & 0.0022 & 0.0013\\ 
2005.580 & 0.0240 & 0.1485 & 0.0038 & 0.0010\\ 
2006.336 & 0.01372 & 0.16106 & 0.00013 & 0.00013\\ 
2008.371 & -0.00944 & 0.17984 & 0.00016 & 0.00015\\ 
2008.562 & -0.01155 & 0.18067 & 0.00017 & 0.00017\\ 
2009.340 & -0.02057 & 0.18139 & 0.00018 & 0.00016\\ 
2009.561 & -0.02281 & 0.18135 & 0.00017 & 0.00021\\ 
2009.689 & -0.02406 & 0.18129 & 0.00018 & 0.00016\\ 
2010.342 & -0.03134 & 0.17919 & 0.00016 & 0.00018\\ 
2010.511 & -0.03339 & 0.17839 & 0.00023 & 0.00018\\ 
2010.620 & -0.03447 & 0.17777 & 0.00022 & 0.00018\\ 
2011.401 & -0.04205 & 0.17215 & 0.00023 & 0.00023\\ 
2011.543 & -0.04357 & 0.17041 & 0.00017 & 0.00020\\ 
2011.642 & -0.04430 & 0.16950 & 0.00031 & 0.00028\\ 
2012.371 & -0.05081 & 0.16147 & 0.00016 & 0.00028\\ 
2012.562 & -0.05254 & 0.15899 & 0.00016 & 0.00029\\ 
2013.318 & -0.05768 & 0.14818 & 0.00026 & 0.00046\\ 
2013.550 & -0.05916 & 0.14347 & 0.00019 & 0.00034\\ 
\\ 
 
No IRS 9:& & & & \\ 
1995.439 & -0.0426 & 0.1637 & 0.0013 & 0.0014\\ 
1996.485 & -0.0527 & 0.1556 & 0.0047 & 0.0036\\ 
1997.367 & -0.0563 & 0.1383 & 0.0013 & 0.0026\\ 
1999.333 & -0.0655 & 0.0923 & 0.0013 & 0.0015\\ 
1999.559 & -0.0664 & 0.08948 & 0.0015 & 0.00095\\ 
2000.381 & -0.06354 & 0.06341 & 0.00095 & 0.00086\\ 
2000.548 & -0.0631 & 0.0592 & 0.0012 & 0.0012\\ 
2000.797 & -0.0620 & 0.0488 & 0.0024 & 0.0032\\ 
2001.351 & -0.0538 & 0.0260 & 0.0015 & 0.0016\\ 
2001.572 & -0.0498 & 0.0144 & 0.0013 & 0.0017\\ 
2003.303 & 0.0393 & 0.0703 & 0.0014 & 0.0021\\ 
2003.554 & 0.0406 & 0.08216 & 0.0014 & 0.00071\\ 
2003.682 & 0.0390 & 0.08905 & 0.0018 & 0.00082\\ 
2004.327 & 0.03666 & 0.1141 & 0.00083 & 0.0012\\ 
2004.564 & 0.0330 & 0.12256 & 0.0011 & 0.00090\\ 
2004.567 & 0.0350 & 0.1266 & 0.0050 & 0.0046\\ 
2004.660 & 0.03062 & 0.1256 & 0.00096 & 0.0020\\ 
2005.312 & 0.02396 & 0.1431 & 0.00078 & 0.0011\\ 
2005.495 & 0.0197 & 0.1478 & 0.0053 & 0.0023\\ 
2005.566 & 0.0214 & 0.1494 & 0.0021 & 0.0012\\ 
2005.580 & 0.02302 & 0.14978 & 0.0038 & 0.00099\\ 
2006.336 & 0.01269 & 0.16255 & 0.00013 & 0.00013\\ 
2008.371 & -0.01037 & 0.18165 & 0.00015 & 0.00015\\ 
2008.562 & -0.01245 & 0.18246 & 0.00017 & 0.00017\\ 
2009.340 & -0.02135 & 0.18327 & 0.00016 & 0.00015\\ 
2009.561 & -0.02359 & 0.18336 & 0.00014 & 0.00021\\ 
2009.689 & -0.02482 & 0.18333 & 0.00018 & 0.00016\\ 
2010.342 & -0.03202 & 0.18127 & 0.00014 & 0.00016\\ 
2010.511 & -0.03401 & 0.18048 & 0.00020 & 0.00017\\ 
2010.620 & -0.03515 & 0.17995 & 0.00020 & 0.00018\\ 
2011.401 & -0.04275 & 0.17444 & 0.00023 & 0.00021\\ 
2011.543 & -0.04422 & 0.17271 & 0.00016 & 0.00021\\ 
2011.642 & -0.04499 & 0.17179 & 0.00025 & 0.00022\\ 
2012.371 & -0.05135 & 0.16390 & 0.00015 & 0.00024\\ 
2012.562 & -0.05306 & 0.16143 & 0.00017 & 0.00025\\ 
2013.318 & -0.05818 & 0.15075 & 0.00023 & 0.00038\\ 
2013.550 & -0.05963 & 0.14620 & 0.00018 & 0.00028\\ 
\\ 
 
\enddata 
\label{tab:droponemaser_S02}

\end{deluxetable}
\clearpage

\begin{deluxetable}{lcccc}
\tabletypesize{\scriptsize}
\tablewidth{0pt}
\tablecaption{Jack Knife Bias and Variance}
\tablehead{
  \colhead{Black Hole Parameter (units)} & 
  \colhead{Bias} & 
  \colhead{Standard Deviation of Bias ($\sigma$)} & 
}
\startdata
                     Distance (kpc)   & 0.01   & 0.04\\ 
             Mass ($10^6M_{\odot}$)   & 0.09   & 0.04\\ 
       $X$ Position of Sgr A* (mas)   & -0.50   & 1.90\\ 
       $Y$ Position of Sgr A* (mas)   & -0.73   & 1.23\\ 
              $X$ Velocity (mas/yr)   & -0.08   & 0.13\\ 
              $Y$ Velocity (mas/yr)   & -0.17   & 0.22\\ 
              $Z$ Velocity (km/sec)   & 4.07   & 4.28\\ 
\enddata 
\label{tab:jackknife_bias_var}

\end{deluxetable}
\clearpage

\bibliographystyle{apj}


\begin{thebibliography}{}
\expandafter\ifx\csname natexlab\endcsname\relax\def\natexlab#1{#1}\fi

\bibitem[{{Bates} {et~al.}(1973){Bates}, {Gough}, \&
  {Napier}}]{1973A&A....22..319B}
{Bates}, R.~H.~T., {Gough}, P.~T., \& {Napier}, P.~J. 1973, \aap, 22, 319

\bibitem[{{Chatzopoulos} {et~al.}(2015){Chatzopoulos}, {Fritz}, {Gerhard},
  {Gillessen}, {Wegg}, {Genzel}, \& {Pfuhl}}]{2015MNRAS.447..948C}
{Chatzopoulos}, S., {Fritz}, T.~K., {Gerhard}, O., {et~al.} 2015, \mnras, 447,
  948

\bibitem[{{Christou}(1991)}]{1991PASP..103.1040C}
{Christou}, J.~C. 1991, \pasp, 103, 1040

\bibitem[{{Diolaiti} {et~al.}(2000){Diolaiti}, {Bendinelli}, {Bonaccini},
  {Close}, {Currie}, \& {Parmeggiani}}]{2000SPIE.4007..879D}
{Diolaiti}, E., {Bendinelli}, O., {Bonaccini}, D., {et~al.} 2000, in Presented
  at the Society of Photo-Optical Instrumentation Engineers (SPIE) Conference,
  Vol. 4007, Proc. SPIE Vol. 4007, p. 879-888, Adaptive Optical Systems
  Technology, Peter L. Wizinowich; Ed., ed. P.~L. {Wizinowich}, 879--888

\bibitem[{{Do} {et~al.}(2013){Do}, {Lu}, {Ghez}, {Morris}, {Yelda}, {Martinez},
  {Wright}, \& {Matthews}}]{2013Do}
{Do}, T., {Lu}, J.~R., {Ghez}, A.~M., {et~al.} 2013, \apj, 764, 154

\bibitem[{{Eckart} \& {Genzel}(1997)}]{1997MNRAS.284..576E}
{Eckart}, A., \& {Genzel}, R. 1997, \mnras, 284, 576

\bibitem[{{Eckart} {et~al.}(2002){Eckart}, {Genzel}, {Ott}, \&
  {Sch{\"o}del}}]{2002MNRAS.331..917E}
{Eckart}, A., {Genzel}, R., {Ott}, T., \& {Sch{\"o}del}, R. 2002, \mnras, 331,
  917

\bibitem[{{Eisenhauer} {et~al.}(2003){Eisenhauer}, {Sch{\"o}del}, {Genzel},
  {Ott}, {Tecza}, {Abuter}, {Eckart}, \& {Alexander}}]{2003ApJ...597L.121E}
{Eisenhauer}, F., {Sch{\"o}del}, R., {Genzel}, R., {et~al.} 2003, \apjl, 597,
  L121

\bibitem[{{Feroz} \& {Hobson}(2008)}]{2008Feroz}
{Feroz}, F., \& {Hobson}, M.~P. 2008, \mnras, 384, 449

\bibitem[{{Feroz} {et~al.}(2009){Feroz}, {Hobson}, \& {Bridges}}]{2009Feroz}
{Feroz}, F., {Hobson}, M.~P., \& {Bridges}, M. 2009, \mnras, 398, 1601

\bibitem[{{Ferrarese} \& {Merritt}(2000)}]{2000ApJ...539L...9F}
{Ferrarese}, L., \& {Merritt}, D. 2000, \apjl, 539, L9

\bibitem[{{Freitag} {et~al.}(2006){Freitag}, {Amaro-Seoane}, \&
  {Kalogera}}]{2006ApJ...649...91F}
{Freitag}, M., {Amaro-Seoane}, P., \& {Kalogera}, V. 2006, \apj, 649, 91

\bibitem[{{Genzel} {et~al.}(2010){Genzel}, {Eisenhauer}, \&
  {Gillessen}}]{2010RvMP...82.3121G}
{Genzel}, R., {Eisenhauer}, F., \& {Gillessen}, S. 2010, Reviews of Modern
  Physics, 82, 3121

\bibitem[{{Ghez} {et~al.}(1998){Ghez}, {Klein}, {Morris}, \&
  {Becklin}}]{1998Ghez}
{Ghez}, A.~M., {Klein}, B.~L., {Morris}, M., \& {Becklin}, E.~E. 1998, \apj,
  509, 678

\bibitem[{{Ghez} {et~al.}(2000){Ghez}, {Morris}, {Becklin}, {Tanner}, \&
  {Kremenek}}]{2000Ghez}
{Ghez}, A.~M., {Morris}, M., {Becklin}, E.~E., {Tanner}, A., \& {Kremenek}, T.
  2000, \nat, 407, 349

\bibitem[{{Ghez} {et~al.}(2005{\natexlab{a}}){Ghez}, {Salim}, {Hornstein},
  {Tanner}, {Lu}, {Morris}, {Becklin}, \& {Duch{\^e}ne}}]{2005aGhez}
{Ghez}, A.~M., {Salim}, S., {Hornstein}, S.~D., {et~al.} 2005{\natexlab{a}},
  \apj, 620, 744

\bibitem[{{Ghez} {et~al.}(2003){Ghez}, {Duch{\^e}ne}, {Matthews}, {Hornstein},
  {Tanner}, {Larkin}, {Morris}, {Becklin}, {Salim}, {Kremenek}, {Thompson},
  {Soifer}, {Neugebauer}, \& {McLean}}]{2003ApJ...586L.127G}
{Ghez}, A.~M., {Duch{\^e}ne}, G., {Matthews}, K., {et~al.} 2003, \apjl, 586,
  L127

\bibitem[{{Ghez} {et~al.}(2005{\natexlab{b}}){Ghez}, {Hornstein}, {Lu},
  {Bouchez}, {Le Mignant}, {van Dam}, {Wizinowich}, {Matthews}, {Morris},
  {Becklin}, {Campbell}, {Chin}, {Hartman}, {Johansson}, {Lafon}, {Stomski}, \&
  {Summers}}]{2005bGhez}
{Ghez}, A.~M., {Hornstein}, S.~D., {Lu}, J.~R., {et~al.} 2005{\natexlab{b}},
  \apj, 635, 1087

\bibitem[{{Ghez} {et~al.}(2008){Ghez}, {Salim}, {Weinberg}, {Lu}, {Do}, {Dunn},
  {Matthews}, {Morris}, {Yelda}, {Becklin}, {Kremenek}, {Milosavljevic}, \&
  {Naiman}}]{2008Ghez}
{Ghez}, A.~M., {Salim}, S., {Weinberg}, N.~N., {et~al.} 2008, \apj, 689, 1044

\bibitem[{{Gillessen} {et~al.}(2009{\natexlab{a}}){Gillessen}, {Eisenhauer},
  {Fritz}, {Bartko}, {Dodds-Eden}, {Pfuhl}, {Ott}, \&
  {Genzel}}]{2009ApJ...707L.114G}
{Gillessen}, S., {Eisenhauer}, F., {Fritz}, T.~K., {et~al.} 2009{\natexlab{a}},
  \apjl, 707, L114

\bibitem[{{Gillessen} {et~al.}(2013){Gillessen}, {Eisenhauer}, {Fritz},
  {Pfuhl}, {Ott}, \& {Genzel}}]{2013Gillessen}
{Gillessen}, S., {Eisenhauer}, F., {Fritz}, T.~K., {et~al.} 2013, in IAU
  Symposium, Vol. 289, IAU Symposium, ed. R.~{de Grijs}, 29--35

\bibitem[{{Gillessen} {et~al.}(2009{\natexlab{b}}){Gillessen}, {Eisenhauer},
  {Trippe}, {Alexander}, {Genzel}, {Martins}, \& {Ott}}]{2009Gillessen}
{Gillessen}, S., {Eisenhauer}, F., {Trippe}, S., {et~al.} 2009{\natexlab{b}},
  \apj, 692, 1075
 %
\bibitem[{{Gottlieb}(2001)}]{2001math......9002G}
{Gottlieb}, A.~D. 2001, ArXiv Mathematics e-prints, math/0109002

\bibitem[{{Hornstein} {et~al.}(2002){Hornstein}, {Ghez}, {Tanner}, {Morris},
  {Becklin}, \& {Wizinowich}}]{2002ApJ...577L...9H}
{Hornstein}, S.~D., {Ghez}, A.~M., {Tanner}, A., {et~al.} 2002, \apjl, 577, L9

\bibitem[{{Kormendy} \& {Ho}(2013)}]{2013ARA&A..51..511K}
{Kormendy}, J., \& {Ho}, L.~C. 2013, \araa, 51, 511

\bibitem[{{Larkin} {et~al.}(2006){Larkin}, {Barczys}, {Krabbe}, {Adkins},
  {Aliado}, {Amico}, {Brims}, {Campbell}, {Canfield}, {Gasaway}, {Honey},
  {Iserlohe}, {Johnson}, {Kress}, {LaFreniere}, {Lyke}, {Magnone}, {Magnone},
  {McElwain}, {Moon}, {Quirrenbach}, {Skulason}, {Song}, {Spencer}, {Weiss}, \&
  {Wright}}]{2006SPIE.6269E..42L}
{Larkin}, J., {Barczys}, M., {Krabbe}, A., {et~al.} 2006, in Society of
  Photo-Optical Instrumentation Engineers (SPIE) Conference Series, Vol. 6269,
  Society of Photo-Optical Instrumentation Engineers (SPIE) Conference Series

\bibitem[{{Lu} {et~al.}(2005){Lu}, {Ghez}, {Hornstein}, {Morris}, \&
  {Becklin}}]{2005Lu}
{Lu}, J.~R., {Ghez}, A.~M., {Hornstein}, S.~D., {Morris}, M., \& {Becklin},
  E.~E. 2005, \apjl, 625, L51

\bibitem[{{Lu} {et~al.}(2009){Lu}, {Ghez}, {Hornstein}, {Morris}, {Becklin}, \&
  {Matthews}}]{2009Lu}
{Lu}, J.~R., {Ghez}, A.~M., {Hornstein}, S.~D., {et~al.} 2009, \apj, 690, 1463

\bibitem[{{Lucy}(2014)}]{2014Lucy}
{Lucy}, L.~B. 2014, \aap, 563, A126

\bibitem[{{Matthews} {et~al.}(1996){Matthews}, {Ghez}, {Weinberger}, \&
  {Neugebauer}}]{1996PASP..108..615M}
{Matthews}, K., {Ghez}, A.~M., {Weinberger}, A.~J., \& {Neugebauer}, G. 1996,
  \pasp, 108, 615

\bibitem[{{Matthews} \& {Soifer}(1994)}]{1994ExA.....3...77M}
{Matthews}, K., \& {Soifer}, B.~T. 1994, Experimental Astronomy, 3, 77

\bibitem[{{Merritt}(2010)}]{2010ApJ...718..739M}
{Merritt}, D. 2010, \apj, 718, 739

\bibitem[{{Meyer} {et~al.}(2012){Meyer}, {Ghez}, {Sch{\"o}del}, {Yelda},
  {Boehle}, {Lu}, {Do}, {Morris}, {Becklin}, \& {Matthews}}]{Meyer12}
{Meyer}, L., {Ghez}, A.~M., {Sch{\"o}del}, R., {et~al.} 2012, Science, 338, 84

\bibitem[{{Miralda-Escud{\'e}} \& {Gould}(2000)}]{2000ApJ...545..847M}
{Miralda-Escud{\'e}}, J., \& {Gould}, A. 2000, \apj, 545, 847

\bibitem[{{Morris}(1993)}]{1993ApJ...408..496M}
{Morris}, M. 1993, \apj, 408, 496

\bibitem[{{Olling} \& {Merrifield}(2000)}]{2000MNRAS.311..361O}
{Olling}, R.~P., \& {Merrifield}, M.~R. 2000, \mnras, 311, 361

\bibitem[{{Petr} {et~al.}(1998){Petr}, {Coud{\'e} du Foresto}, {Beckwith},
  {Richichi}, \& {McCaughrean}}]{1998ApJ...500..825P}
{Petr}, M.~G., {Coud{\'e} du Foresto}, V., {Beckwith}, S.~V.~W., {Richichi},
  A., \& {McCaughrean}, M.~J. 1998, \apj, 500, 825

\bibitem[{{Primot} {et~al.}(1990){Primot}, {Rousset}, \&
  {Fontanella}}]{1990JOSAA...7.1598P}
{Primot}, J., {Rousset}, G., \& {Fontanella}, J.~C. 1990, Journal of the
  Optical Society of America A, 7, 1598

\bibitem[{{Psaltis} {et~al.}(2015){Psaltis}, {{\"O}zel}, {Chan}, \&
  {Marrone}}]{2015Psaltis}
{Psaltis}, D., {{\"O}zel}, F., {Chan}, C.-K., \& {Marrone}, D.~P. 2015, \apj,
  814, 115

\bibitem[{{Rafelski} {et~al.}(2007){Rafelski}, {Ghez}, {Hornstein}, {Lu}, \&
  {Morris}}]{2007Rafelski}
{Rafelski}, M., {Ghez}, A.~M., {Hornstein}, S.~D., {Lu}, J.~R., \& {Morris}, M.
  2007, \apj, 659, 1241

\bibitem[{{Rayner} {et~al.}(2009){Rayner}, {Cushing}, \&
  {Vacca}}]{2009ApJS..185..289R}
{Rayner}, J.~T., {Cushing}, M.~C., \& {Vacca}, W.~D. 2009, \apjs, 185, 289

\bibitem[{{Reid}(1993)}]{1993ARA&A..31..345R}
{Reid}, M.~J. 1993, \araa, 31, 345

\bibitem[{{Reid} \& {Brunthaler}(2004)}]{2004Reid_propermotion}
{Reid}, M.~J., \& {Brunthaler}, A. 2004, \apj, 616, 872

\bibitem[{{Reid} {et~al.}(2007){Reid}, {Menten}, {Trippe}, {Ott}, \&
  {Genzel}}]{2007ApJ...659..378R}
{Reid}, M.~J., {Menten}, K.~M., {Trippe}, S., {Ott}, T., \& {Genzel}, R. 2007,
  \apj, 659, 378

\bibitem[{{Reid} {et~al.}(2009){Reid}, {Menten}, {Zheng}, {Brunthaler}, \&
  {Xu}}]{2009ApJ...705.1548R}
{Reid}, M.~J., {Menten}, K.~M., {Zheng}, X.~W., {Brunthaler}, A., \& {Xu}, Y.
  2009, \apj, 705, 1548

\bibitem[{{Reid} {et~al.}(2014){Reid}, {Menten}, {Brunthaler}, {Zheng}, {Dame},
  {Xu}, {Wu}, {Zhang}, {Sanna}, {Sato}, {Hachisuka}, {Choi}, {Immer},
  {Moscadelli}, {Rygl}, \& {Bartkiewicz}}]{2014ApJ...783..130R}
{Reid}, M.~J., {Menten}, K.~M., {Brunthaler}, A., {et~al.} 2014, \apj, 783, 130

\bibitem[{{Sch{\"o}del} {et~al.}(2010){Sch{\"o}del}, {Najarro}, {Muzic}, \&
  {Eckart}}]{2010A&A...511A..18S}
{Sch{\"o}del}, R., {Najarro}, F., {Muzic}, K., \& {Eckart}, A. 2010, \aap, 511,
  A18

\bibitem[{{Sch{\"o}del} {et~al.}(2013){Sch{\"o}del}, {Yelda}, {Ghez}, {Girard},
  {Labadie}, {Rebolo}, {P{\'e}rez-Garrido}, \& {Morris}}]{2013SchoedelHolo}
{Sch{\"o}del}, R., {Yelda}, S., {Ghez}, A., {et~al.} 2013, \mnras, 429, 1367

\bibitem[{{Sch{\"o}del} {et~al.}(2002){Sch{\"o}del}, {Ott}, {Genzel},
  {Hofmann}, {Lehnert}, {Eckart}, {Mouawad}, {Alexander}, {Reid}, {Lenzen},
  {Hartung}, {Lacombe}, {Rouan}, {Gendron}, {Rousset}, {Lagrange}, {Brandner},
  {Ageorges}, {Lidman}, {Moorwood}, {Spyromilio}, {Hubin}, \&
  {Menten}}]{2002Natur.419..694S}
{Sch{\"o}del}, R., {Ott}, T., {Genzel}, R., {et~al.} 2002, \nat, 419, 694

\bibitem[{{Tremaine} {et~al.}(2002){Tremaine}, {Gebhardt}, {Bender}, {Bower},
  {Dressler}, {Faber}, {Filippenko}, {Green}, {Grillmair}, {Ho}, {Kormendy},
  {Lauer}, {Magorrian}, {Pinkney}, \& {Richstone}}]{2002ApJ...574..740T}
{Tremaine}, S., {Gebhardt}, K., {Bender}, R., {et~al.} 2002, \apj, 574, 740

\bibitem[{{van Dam} {et~al.}(2006){van Dam}, {Bouchez}, {Le Mignant},
  {Johansson}, {Wizinowich}, {Campbell}, {Chin}, {Hartman}, {Lafon}, {Stomski},
  \& {Summers}}]{2006PASP..118..310V}
{van Dam}, M.~A., {Bouchez}, A.~H., {Le Mignant}, D., {et~al.} 2006, \pasp,
  118, 310

\bibitem[{{Wizinowich} {et~al.}(2006){Wizinowich}, {Le Mignant}, {Bouchez},
  {Campbell}, {Chin}, {Contos}, {van Dam}, {Hartman}, {Johansson}, {Lafon},
  {Lewis}, {Stomski}, {Summers}, {Brown}, {Danforth}, {Max}, \&
  {Pennington}}]{2006PASP..118..297W}
{Wizinowich}, P.~L., {Le Mignant}, D., {Bouchez}, A.~H., {et~al.} 2006, \pasp,
  118, 297

\bibitem[{{Yelda}(2012)}]{2012Yelda}
{Yelda}, S. 2012, PhD thesis, University of California, Los Angeles

\bibitem[{{Yelda} {et~al.}(2014){Yelda}, {Ghez}, {Lu}, {Do}, {Meyer}, {Morris},
  \& {Matthews}}]{2014Yelda}
{Yelda}, S., {Ghez}, A.~M., {Lu}, J.~R., {et~al.} 2014, ArXiv e-prints,
  arXiv:1401.7354

\bibitem[{{Yelda} {et~al.}(2010){Yelda}, {Lu}, {Ghez}, {Clarkson}, {Anderson},
  {Do}, \& {Matthews}}]{2010Yelda}
{Yelda}, S., {Lu}, J.~R., {Ghez}, A.~M., {et~al.} 2010, \apj, 725, 331

\end{thebibliography}

\end{document}